\documentclass[12pt,twoside,openright]{book}
\usepackage[pdftex]{graphicx}
\usepackage[a4paper, tmargin=3.5cm, bmargin=2.5cm, lmargin=3.75cm, rmargin=2.5cm, twoside]{geometry}
\usepackage{latexsym,amsmath,amssymb,amsfonts,bm,bbm}             % AMS Math
\usepackage[toc,page]{appendix}
\usepackage{physics}
\usepackage{rotating}                    % Sideways of figures & tables
\usepackage{natbib}          % Cross-reference package (Natural BiB)
\usepackage{soul}                                         % Put References at the end of each chapter
\usepackage{xcolor}
\usepackage{float}
\usepackage{comment}
\usepackage{fancyhdr}                    % Fancy Header and Footer
\usepackage{setspace}
\usepackage[english]{babel}
\doublespace          % Line spacing
\usepackage{tocloft}
\usepackage{lscape}

\usepackage[small,hang]{caption}
\usepackage[compact]{titlesec}
\numberwithin{equation}{section}
\usepackage[T1]{fontenc}                    % Public Times New Roman text & math font
\makeatletter
\def\cleardoublepage{\clearpage\if@twoside \ifodd\c@page\else%
    \hbox{}%
    \thispagestyle{empty}%              % Empty header styles
    \newpage%
    \if@twocolumn\hbox{}\newpage\fi\fi\fi}
\makeatother
\clearpage{\pagestyle{plain}\cleardoublepage}
\newcommand{\be}{\begin{equation}}
\newcommand{\ee}{\end{equation}}
\newcommand{\bea}{\begin{eqnarray}}
\newcommand{\eea}{\end{eqnarray}}
\newcommand{\ba}{\begin{array}}
\newcommand{\ea}{\end{array}}
\newcommand{\bi}{\begin{itemize}}
\newcommand{\ei}{\end{itemize}}
\newcommand{\bc}{\begin{center}}
\newcommand{\ec}{\end{center}}
\newcommand{\bfr}{\begin{flushright}}
\newcommand{\efr}{\end{flushright}}

\usepackage{color}

\begin{document}
\thispagestyle{empty}

{ \renewcommand{\baselinestretch}{1.5}
\begin{center}
%\vspace*{1.0cm}
\begin{spacing}{2}
\noindent{\Large \bf INDUCING AND PROBING CHARGE MIGRATION IN MOLECULAR SYSTEMS}
\end{spacing}
\end{center}

\vspace{18cm}
\begin{flushright}
{ {\LARGE\em  \textbf{Sucharita Giri} ~~}}
%\noindent \vskip
%-1.0\baselineskip \noindent \rule[-2.5 mm]{\textwidth}{3 pt}\\}
\end{flushright}}

	\cleardoublepage
	\frontmatter
	\addcontentsline{toc}{chapter}{Title Page}
	\newpage
\thispagestyle{empty}

% ******* Title page *******
% **************************
\begin{titlepage}
\begin{center}

{
%% Ici je  lis  logoensem.jpg
%%  commenter si vous voulez faire du DVI :
%

%\vskip3cm

\textbf{\Large INDUCING AND PROBING CHARGE MIGRATION IN MOLECULAR SYSTEMS} \vskip1.0cm %\textbf{\large\emph{Thesis
%submitted to the\\  Indian Institute of Technology Bombay}}
{\large\emph{Submitted in partial fulfillment of the requirements}} \vskip 0.03cm 
{\large\emph{of the degree of}}
\singlespacing 
%\textbf{\large\emph{For award of the
%degree\\\vspace{5.0cm}of}}
%\textbf{\large\emph{For award of the
%degree}}\vskip 0.03cm
%\textbf{\large\emph{of}}
% \singlespacing
\textbf {\large Doctor of Philosophy}\\
\singlespacing
 {\large\emph{by}}\\
\singlespacing \textbf{\large Sucharita Giri} \vskip0.2cm
%{\large (Roll No. 164120007)}\\\
\vskip1cm
{\large Supervisor:}\\
\singlespacing \textbf{\large{Prof. Gopal Dixit}} \vskip1cm

\includegraphics[height=42mm]{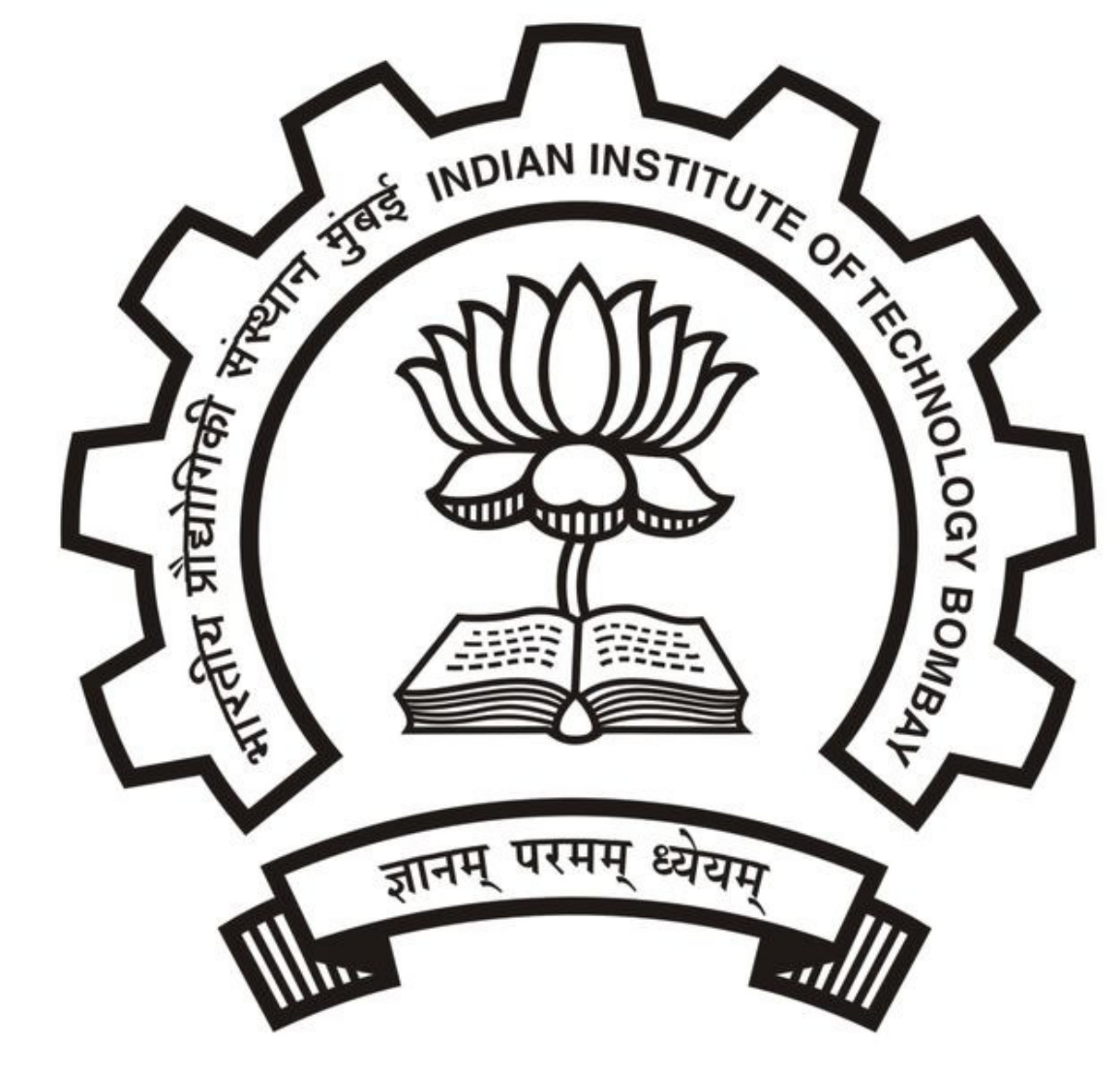}
\doublespacing
\begin{center}
\textbf{\large{DEPARTMENT OF PHYSICS\\
INDIAN INSTITUTE OF TECHNOLOGY BOMBAY\\\singlespacing
2023}}\\\singlespacing \copyright~2023 Sucharita Giri All rights reserved.
\end{center}

\fboxsep6mm
\fboxrule1.3pt
}

\end{center}
\end{titlepage}
\pagestyle{fancy}

	\cleardoublepage
	\newpage
\thispagestyle{empty}
\setlength{\baselineskip}{32pt}
\bigskip
\bigskip
\bigskip
\bigskip
\bigskip
\vspace*{5cm}
\begin{center}
%\end{flushright}
\vspace*{0.5cm}
%\begin{flushright}
{\Large  \bf {Dedicated to my parents, and my sisters.} } \\
\end{center}
%\vspace*{5cm}
%\bigskip
%{\parindent 5pt \centerline{\Large \bf \it Dedicated To }}
%{\parindent 0pt \centerline{\huge \bf \it \texttt{My Mother}}
%
%\bigskip
%\bigskip

\setlength{\baselineskip}{18pt}

	%\cleardoublepage \addcontentsline{toc}{chapter}{Thesis Approval}
	%\include{HeadTail/certi_appro}
	\cleardoublepage
	\addcontentsline{toc}{chapter}{Declaration}
	\newpage
\thispagestyle{empty}
 \vspace*{2cm}
\begin{center}
\textbf{\LARGE{Declaration}}
\end{center}
\vspace{1cm}
\begin{spacing}{1.3}
I declare that this written submission represents my ideas in my own words and where
others' ideas or words have been included, I have adequately cited and referenced the original
sources. I also declare that I have adhered to all principles of academic honesty and integrity
and have not misrepresented or fabricated or falsified any idea/data/fact/source in my
submission. I understand that any violation of the above will be cause for disciplinary action
by the Institute and can also evoke penal action from the sources which have thus not been
properly cited or from whom proper permission has not been taken when needed.
\vspace{2.5cm}
\begin{flushright}
Sucharita Giri \\
(Roll No. 174123002)
\end{flushright}
%\begin{flushleft}
%	Date: 27/10/2021
%\end{flushleft}
\end{spacing} 
	\cleardoublepage
	\addcontentsline{toc}{chapter}{Acknowledgements}
	\newpage
\thispagestyle{empty}
\begin{center}
\vspace*{-0.4cm}
{\LARGE {\textbf{Acknowledgements}}}
\end{center}

{\setlength{\baselineskip}{8pt} \setlength{\parskip}{2pt}
\begin{spacing}{1.5}
%A journey is easier when you travel together. Interdependence is certainly more valuable than independence.
%This thesis is the result of four and half years of work whereby I have been accompanied and supported by many
%people. It is a pleasant aspect that I have now the opportunity to express my gratitude for all of them.
%First and foremost, I would like to thank my greatest teacher of all: God.
%I know that I am here and that I am able to write all of this for a reason.
%I will do my best in never forgetting what a great fortune
%I have had in just being here, and that it comes with a lesson
%and a responsibility. I hope I am doing the work you have planned me to do.

I would like to acknowledge  everyone who has helped me grow personally and academically throughout this Ph.D. journey.
I can't possibly include everyone, but I'll try to mention the ones who have made a significant impact. 

First and foremost, I would like to thank my supervisor, Prof. Gopal Dixit, for his guidance, patience, and support throughout the course of my research. His valuable advice and feedback helped me to complete this thesis.

I would like to express my sincerest gratitude to Prof. Jean Christophe Tremblay, for his guidance throughout the entire process and for inviting me to Freie University Berlin as a guest researcher with financial support. I could not have completed my Ph.D. without his help. Thank you for being such a great collaborator.

I would like to express my gratitude to the collaborators from IIT KGP, Prof. Partha Pratim Jana and Dr. Nilanjan Roy. I have really enjoyed learning the new concepts during the pandemic.

I would also like to thank the RPC members, Prof.  Sumiran Pujari and Prof.  Amber Jain, for their guidance and advice, and for taking the time to read my reports and be present in my annual progress seminar (APS) presentations.

I am thankful to all my charming and caring friends.
A special thanks to Adhip, Monish, Manisha, Supriti and Suman for all the good moments shared throughout my campus journey. 
Thank you Sovan for being there through all the ups and downs. 
You are always a constant support in my life.
I want to thank Avishek, who was a help to me throughout the writing of my thesis.
Thank you Aneesha for the friendly discussions and all the encouragement.

Most importantly, I would like to thank my family for their unwavering love and support. 
They have understood throughout this entire process. 
Maa and Baba, thank you for being my heroes all my life. 
My little sisters, thank you for all your support and sacrifices.
I could not have asked for better people to have in my corner.

I want to thank Irfana, Mrudul, Adhip, Navdeep, and Amar for being such friendly and supportive labmates.
Furthermore, I would like to record my thanks to all the staff members of IITB for making life inside IITB much easier.  
Finally, I would like to thank my institution for providing the resources and infrastructure.
\vskip2cm
%\hspace{4.58in}(Gopal Dixit)
%
%\hspace{2.65in}Department of Physics and Meteorology
%

\begin{flushright}
Sucharita Giri\\
Department of Physics\\
IITB
\end{flushright}
%\begin{flushleft}
%	Date: 19/12/2022\\
%\end{flushleft}
\end{spacing}

	\cleardoublepage

	\addcontentsline{toc}{chapter}{List of Symbols and Abbreviations} %%%%%%%%
	\markboth{List of Symbols}{List of Symbols and Abbreviations}  %%%%%%%%%%%
	\cleardoublepage
	\addcontentsline{toc}{chapter}{Abstract}
	\newpage
\thispagestyle{empty}
\begin{center}
\vspace*{-0.4cm}
{\LARGE {\textbf{Abstract}}}
\end{center}

{\setlength{\baselineskip}{8pt}} \setlength{\parskip}{2pt}
\begin{spacing}{1.5}
Technological advancements in generation of ultrafast and intense laser pulses have enabled the real-time observation and control of charge migration in molecules on their natural timescale, which  
ranges from few femtoseconds to several hundreds of attoseconds.
Present thesis discusses the effect of symmetry on the adiabatic attosecond charge migration in different molecular systems. 
The spatial representation of the charge migration is documented  by 
time-dependent electronic charge and flux  densities.
Furthermore, the induced charge migration is imaged via time resolved x-ray diffraction (TRXD) 
with atomic-scale spatiotemporal resolution in few cases. 

We have studied how the charge migration will be modified if one reduces the
symmetry of the molecule by transiting from six-membered benzene  
to heterocyclic five-membered ring molecules.
For this purpose, the effects of symmetry reduction and electronegativity are 
investigated by a comparative study 
of the charge migration in pyrrole, furan, and  oxazole with the help of the time-dependent  electronic charge and flux  densities are used.
In the next case, we have studied the role of symmetry reduction on the  charge migration in 
copper corrole -- a non-planer molecule with saddled geometry.
Correlation of the flux densities and TRXD signals indicates that the signature of the structural saddling during charge migration is imprinted in TRXD signals.
After studying charge migration in molecules with certain symmetries, we have explored charge migration in molecules without any symmetry -- chiral molecules.
Control over charge migration in R- and S-epoxypropane is achieved by choosing the different orientations of the linearly polarized pulse.
For oriented and floppy molecules, we find that the charge migration is different for enantiomers when the polarization of the pulse lies in the mirror plane defining the enantiomer pair.
It is found that the total TRXD signals are significantly different for both enantiomers. Furthermore,  a asymmetry parameter is introduced to discern the concentration of an enantiomer in the racemic mixture. 
%\vspace{0.5in}

\textbf{Key words: Charge migration, Molecular symmetry, Electronic flux density, TRXD, Chirality} 

\end{spacing}

	\cleardoublepage
	\renewcommand{\contentsname}{Contents}  % Original name = Contents
	\begin{spacing}{1.2}  % Environment for 1.2 line spacing for contents and lists
		\tableofcontents
		\cleardoublepage
		\addcontentsline{toc}{chapter}{List of Figures} %%%%%%%%
		\listoffigures
		\cleardoublepage
		\addcontentsline{toc}{chapter}{List of Symbols and Abbreviations} %%%%%%%%
		\markboth{List of Symbols}{List of Symbols and Abbreviations}  %%%%%%%%%%%
		\chapter*{List of Symbols and Abbreviations}
\noindent {\bf Symbols}
\begin{tabbing}
aaaaaaaaaaaa \= abababababab \kill
$\mathcal{H}_{0}$\> Time-independent molecular Hamiltonian\\
$\mathcal{H}(t)$\> Time-dependent Hamiltonian\\
$ \mathcal{H}_{\textrm{int}}(t)$\>Time-dependent interaction Hamiltonian \\
$| \Phi_{k} (\mathbf{r}) \rangle$\> Time-independent multi-electronic wave function\\
$ | \Psi (\textbf{r}, t) \rangle$\> Laser-induced electronic wavepacket \\
%$C_{k}(t)$\> Time-dependent  expansion coefficient \\
%$T_{p}$\> Pulse duration of laser pulse\\
$\rho (\textbf{r}, t)$ \> Time-dependent electronic charge density \\
$\textbf{j} (\textbf{r}, t)$\> Time-dependent electronic flux density\\
${dP}/{d\Omega}$ \> Differential scattering probability\\
${dP_{e}}/{d\Omega}$ \> Differential scattering probability of a free electron\\
{\sffamily{T}} \> Pump-probe time delay \\
$\tau$ \> Characteristic timescale  of the charge migration \\
\end{tabbing}
\noindent {\bf Abbreviations}
 \begin{tabbing}
aaaaaaaaaaaa \= abababababab  \kill
TDSE \> Time-dependent Schr\"odinger equation\\
HF \> Hartree-Fock\\
CIS \> Configuration interaction singles\\
DFT \> Density functional theory\\
TDDFT \> Time-dependent density functional theory\\
TDDFT/CI \> Hybrid time-dependent density functional theory-configuration interaction \\
TRXD \> Time-resolved x-ray diffraction\\
EFD \> Electronic flux density\\
DSP \> Differential scattering probability\\
IRREP \> Irreducible representation\\
NTOs \> Natural transition orbitals\\
%CI \> Configuration interaction\\
%CID \> Configuration interaction doubles\\
%KS \> Kohn-Sham\\
%TDKS \> Time-dependent Kohn-Sham\\
%LvN \> Liouville-von Neumann  \\
%QED \> Quantum electrodynamics\\
\end{tabbing}
%
%\noindent {\bf Superscript}
% \begin{tabbing}
%aaaaaaaaaaaa \= abababababab  \kill
% $*$ \>  dimensional quantity
%\end{tabbing}

		\cleardoublepage

	\end{spacing}
	
%	\addcontentsline{toc}{chapter}{List of Symbols} %%%%%%%%
%	\markboth{List of Symbols}{List of Symbols}  %%%%%%%%%%%
%	%\include{HeadTail/listofsymbol}
%	\cleardoublepage
\mainmatter
\begin{spacing}{1.5}
\chapter{Introduction}\label{Chaper1}

The motion of atoms within molecules associated with light-induced exciton dynamics, conformational changes, bond formation and breakage during chemical reactions occurs on the femtosecond 
(1 fs = 10$^{-15}$ s) timescale. Thus, understanding  the  dynamics of atoms (nuclei)  during complex chemical and biological processes has started an era of femtochemistry~\citep{zewail2000femtochemistry}.  
Electrons are the glue that keeps the atoms within molecules together and 
their dynamics play a crucial role in the function and transformation of molecules.  
The distinctive timescale of electronic motion, responsible for electron-hole dynamics and charge migration processes in molecules can be even faster, ranges from a few femtoseconds  
to several hundreds of attoseconds (1 as = 10$^{-18}$ s).

When a molecule is exposed to an ultrashort laser pulse, 
electrons  are the first to respond to the  action of the incident laser. 
Depending on the energy and pulse duration of a coherent laser pulse, 
several interesting processes can take place in a molecule during laser-molecule interaction.   
A superposition of several electronic states in the form of 
an electronic wavepacket can be created
if the energy of the laser is smaller than the ionization potential of the valence electron in a molecule and 
the energy bandwidth  is of the order of few electron-volts. 
The characteristic timescale of the created wavepacket can be estimated from $\tau = \hbar/ \Delta E$, 
where $\Delta E$ is the energy difference between electronic states involved in the wavepacket. 
In general, the electronic motion associated with the wavepacket occurs on attosecond timescale and 
not impacted by nuclear motion. 
Thus, a coherent electronic motion can be described as a coherent superposition of multiple electronic states with  fixed nuclei configuration.
Observations of the attosecond electronic motion was not possible until the realization of 
the coherent attosecond laser pulses. 
Before we discuss the main theme of this thesis, let us briefly recap the key mechanisms and early 
findings in attosecond physics, an essential ingredient for charge migration, in the next section. 

\section{Attosecond Physics }\label{attosecondpulse}
Recent developments in coherent light sources made the generation of attosecond pulses possible 
and gave a birth to a new era of ultrafast physics -- attosecond physics.
Attosecond pulses can be generated from a high-harmonic generation, which is a strong-field driven nonlinear process. 
In last two decades, attosecond pulses have been used to study a number of interesting  
ultrafast processes in atoms, molecules and solids. 
Pump-probe technique is the most common approach in attosecond experiments in which 
an ultrashort ``pump'' pulse induces an  attosecond electronic motion, which is subsequently captured by the 
second ``probe'' pulse. 
The induced motion can be interrogated at different instances  during the dynamics by varying the delay between the pump and probe pulses. Time-resolved photoionization (emission) spectroscopy, attosecond transient-absorption spectroscopy, two-colour high-harmonic spectroscopy are few commonly used  spectroscopy methods on attosecond timescale. 
Not only generating attosecond pulses but its characterization is essential before 
performing any attosecond experiment.   
Attosecond streaking and RABBITT (reconstruction of attosecond beating by interference of two-photon transitions) are two commonly used attosecond pulse characterization techniques. 
Moreover, these two methods are also used to interrogate attosedond processes in atoms, molecules and solids 
as both the methods are based on  time-resolved photoionization concept. 
Attosecond streaking is closely related to RABBITT in which isolated attosecond pulse is used 
instead of attosecond pulse train for photoionization~\citep{krausz2009attosecond}. 

Soon after the realization of attosecond pulses in 2001~\citep{hentschel2001attosecond,paul2001observation}, 
time-resolved spectroscopy of Auger relaxation in krypton is demonstrated~\citep{drescher2002time}. 
Over the years, several fundamental attosecond processes in different atomic systems  
have been experimentally observed such as real-time observation of valence electron in krypton via 
attosecond transient absorption spectroscopy~\citep{goulielmakis2010real}, 
characterization of electron wavepacket in helium~\citep{mauritsson2010attosecond}, 
measurement of photoemission time-delay in neon~\citep{schultze2010delay}, 
real-time electron dynamics in the vicinity of Fano resonance in helium~\citep{gruson2016attosecond},
tunnelling dynamics of electron in neon~\citep{uiberacker2007attosecond}, to name but a few. 

Not only ultrafast electronic motion in atoms but also in molecules and solids has been investigated by attosecond spectroscopy.  Few selected examples include attosecond time-delay in 
photoelectrons emitted from localized core and delocalized conduction-band states in crystalline tungsten~\citep{cavalieri2007attosecond}, light-driven  insulator-conductor phase transition
in a dielectric~\citep{schiffrin2013optical}, 
electron dynamics between valence and conduction energy bands in silicon using attosecond transient absorption spectroscopy~\citep{schultze2014attosecond}, nonlinear polarization and light-matter energy transfer in solids~\citep{sommer2016attosecond}.
Moreover, attosecond pulse driven dissociative dynamics in diatomic molecules have been explored in 
H$_{2}$~\citep{sansone2010electron,kelkensberg2011attosecond} and O$_{2}$~\citep{siu2011attosecond}.
In later part of this chapter, we will discuss attosecond spectroscopy of ``more'' complicated molecules. 
 
Attosecond pulses, produced by high-harmonic generation process,  have two serious limitations: 
Relatively low intensity (fluence) and limited range of available photon energies. In the last decade, alternate source of coherent ultrashort pulses, based on particle acceleration technology, came into existence.  
Thanks to technological  advances that have made possible to 
generate intense ultrashort pulses from recently operational  novel x-ray source: X-ray free-electron laser (XFEL), which offers to circumvent these limitations upto certain  
extent~\citep{ishikawa2012compact, pellegrini2016physics, emma2010first}. 
Laser pulses with few femtoseconds pulse duration in energy regime ranges from extreme ultraviolet to hard x-ray
are routinely employed to perform various interesting experiments at different 
XFEL sources such as LCLS, SACLA, and European XFEL. 
Moreover, feasibility to generate  attosecond x-ray pulses was demonstrated experimentally  in recent years~\citep{hartmann2018attosecond, duris2020tunable}.

\subsection{Time-Resolved X-ray Diffraction}

Since the discovery of x-rays, diffraction of x-rays from matter has been used to unveil the arrangement of atoms inside a matter with atomic-scale spatial resolution. 
Also, insights about the excited electronic states of atoms and molecules can be obtained via  scattering of x-rays from a matter.  
Thus, x-ray diffraction has become a well-established method in several areas of science to access 
atomic-scale structural information of complex materials, ranging from molecules to biological complexes. Moreover, utilizing the Fourier relationship between the electron density of an object and the diffraction  intensity (i.e., elastic x-ray scattering), coherent diffractive imaging  became 
a powerful lensless technique to obtain three-dimensional structural information of nonperiodic and periodic samples. 

The availability of ultrashort x-ray pulses  from XFELs 
have enabled to extend  
static x-ray diffraction into time domain with 
exceptional temporal resolution~\citep{lindroth2019challenges}.
Time-resolved x-ray diffraction (TRXD) within pump-probe configuration  has become a method of choice   
to obtain snapshots of a
temporarily-evolving electronic charge distribution in matter with spatiotemporal resolution on atomic and electronic scales~\citep{peplow2017next}.  
The concept of ``molecular movies'', which track the motion of atoms on femtosecond
timescale, can be extended with the availability of the ultrashort, tunable, and high-energy x-ray pulses from XFEL, which  promises to provide ``electronic movies'' that take place on few  femtosecond to attosecond timescale~\citep{dixit2012imaging, vrakking2012x}.
A series of diffraction signals, obtained at different instants of the motion at various pump-probe time delay,  may be stitched together to make a movie of the electronic motion with unprecedented spatiotemporal resolution. 
In this context, a straightforward extension of x-ray diffraction  from the static to the time domain would seem to suggest that the diffraction signal  encodes information related to the instantaneous electron density on attosecond timescale.  However, it 
has been established that the TRXD signal is not related to the instantaneous electron density and different formalisms of TRXD have been developed over the years~\citep{henriksen2008theory, dixit2012imaging, dixit2013role, bennett2018monitoring,  simmermacher2019electronic, dixit2013proposed, dixit2014theory, santra2014comment, dixit2017time, kowalewski2017monitoring}.

To demonstrate the capabilities of x-rays from XFELs, aligned 2,5-diiodobenzo nitrile is tested as a first  molecule from which static x-ray diffraction has been performed at LCLS~\citep{kupper2014x}.
Frequency-resolved TRXD was employed to disentangle dissociative and bound electronic states during vibrational motion in iodine~\citep{ware2019characterizing}; and vibrational motion in  iodine was imaged by TRXD on femtosecond timescale~\citep{glownia2016self}. 
Anisotropic TRXD measurements have been performed  to determine transition dipole moment and assign excited electronics states in molecules~\citep{yong2018determining}.
Light-induced cis-trans photochemical structural changes in photoactive yellow protein~\citep{pande2016femtosecond}, and
structural changes during ring opening electrocyclic chemical reaction in cyclohexadiene~\citep{minitti2015imaging, minitti2014toward} were imaged  by TRXD. 
Furthermore, the changes in the electron density during ring opening in 1,3-cyclohexadiene  were discussed  experimentally and theoretically using TRXD ~\citep{yong2020observation}.
In this thesis, we will employ TRXD to image the dynamics of an electronic wavepacket in selected molecules, which we will discuss in later chapter. 

\section{Charge Migration}
As discussed earlier, an interaction of the coherent laser pulse with a molecule leads the creation of a coherent superposition of several electronic states. 
If the energy of the coherent laser pulse is high enough to remove an electron from an inner-shell state by photoionization, a hole can be  created and the ionized molecule is driven to a non-equilibrium state.  
Owing to the large  density of ionic states, the formation of a hole can be  
seen as a coherent superposition of several ionic electronic states. 
The hole is spatially localized in the beginning. As the time progresses, the coherent superposition evolves 
in time, which leads the migration of the hole around the atoms within the molecular scaffold~\citep{kuleff2014ultrafast}. 
Cederbaum and Zobeley were first to introduce the notion of  charge migration on attosecond timescale 
associated with the hole dynamics~\citep{cederbaum1999ultrafast}.   
It is assumed that the nuclei are frozen during the charge migration, which is completely different from 
nuclei-driven electron transfer in molecules~\citep{may2008charge}. 
The first theoretical prediction of charge migration in difluoropropadienone has triggered a number of theoretical study in variety of molecules with the emphasis of multielectron effects such as  2-propyn-1-ol, N-methyl acetamide, oligopeptide Gly-Gly-NH-CH$_3$, 4-methylphenol, 
2-Phenylethyl-N,N-dimethylamine (PENNA), tetrapeptides, glycine to name but a few~\citep{breidbach2003migration, breidbach2005universal, hennig2005electron, remacle2006electronic, kuleff2005multielectron, kuleff2014ultrafast}.  
Two experimental proposals with appropriate theoretical description on charge migration in polypeptide molecules have been discussed~\citep{lepine2014attosecond}.
Predictions made by Cederbaum and others have not been 
verified till 2014  when attosecond spectroscopy reached to an advance level. 

As mentioned earlier, attosecond spectroscopy during the initial phase was limited to simple diatomic 
molecules like H$_2$ and O$_2$. 
However, as time progressed,  attosecond spectroscopy 
steadily advanced toward the investigation of more complicated molecules 
with the ability to initiate and interrogate coherent charge migration. 
In 2014, Calegari and co-workers have performed first experiment to probe charge migration in 
amino acid phenylalanine using pump-probe method in which an attosecond pump pulse was used 
to create a hole wavepacket and a near infrared/visible probe pulse was applied to probe the charge migration~\citep{calegari2014ultrafast}. 
%Similar experimental approach was applied to probe charge migration in amino acid phenylalanine~\citep{belshaw2012observation}.   
Subsequently, two-colour high-harmonic spectroscopy was used to observe the charge migration in iodoacetylene spatiotemporally~\citep{kraus2015measurement}. In this experiment, an intense infrared pulse was used for a strong-field ionization of iodoacetylene and a resultant hole wavepacket was probed by high-harmonic spectroscopy. 
It was observed that the orientation of iodoacetylene with respect to the ionizing pulse has a significant impact on the charge migration~\citep{kraus2015measurement}.
By combined experimental and theoretical investigation of the charge migration in tryptophan, it has been  established
that the  nuclear dynamics do not significantly alter the coherence induced by the attosecond pulse during the
 early stages of the wavepacket dynamics~\citep{lara2018attosecond}.  
Later,  time-resolved mapping  of the correlation-driven charge migration in the nucleic-acid base adenine 
was performed  by double ionization pump-probe method~\citep{maansson2021real}. 
Recently,  high-harmonic spectroscopy was applied to probe charge migration in N$_{2}$ and CO$_{2}$ 
with  50 attoseconds temporal resolution~\citep{he2022filming}.
The influences of the length,
bond order, and halogenation  on  the charge migration in halogenated hydrocarbon chains have been explored theoretically~\citep{folorunso2021molecular}. 
Detailed reviews on charge migration in molecules of chemical and biological interest,  and relevant  
state-of-the-art attosecond spectroscopy and theoretical methods  have been documented in Refs.~\citep{nisoli2017attosecond, worner2017charge, merritt2021attochemistry}. 

All the reported observations of charge migrations in different molecules are performed using various spectroscopic methods within pump-probe configuration. 
Direct imaging of charge migration in real-time and in real-space (four-dimensional) is still missing --
one of the main themes of this thesis.   

\section{Aim of This Thesis}\label{MolecularSymmetry}
There are  numerous theoretical works  on charge migration in  different molecules. 
However, there are still several open questions such as
sensitivity of the charge migration with respect to the molecular structural and electronic symmetries. 
In other words, 
whether symmetry as a fundamental property of a molecule  will be imprinted on the charge migration or
can one deduce generic rules on charge migration  based on the symmetries of the molecules?  
This thesis aims to address such  questions. 
To investigate the role of molecular symmetries on charge migration, 
an overview about the choice of molecules is presented in Fig.~\ref{fig11}.
In this thesis, we will explore charge migration corresponding to valence electron excitation in  
neutral molecules. 

Benzene as one of most symmetric molecules has been investigated from charge migration perspective in detail. 
Control over aromaticity in benzene is understood by  analysing attosecond charge migration ~\citep{ulusoy2011correlated}.
A pincer-type flow during charge migration in benzene has been established, which indicates inactive role of a significant fraction of electrons during the charge migration.
Furthermore, it has been found that the nature and direction of the charge migration can be controlled by selective excitation using optimized laser parameters~\citep{hermann2016multidirectional, jia2017quantum, hermann2017attosecond, hermann2020probing}.
At this juncture, it is natural to envisage how the charge migration  will alter if  one reduces the symmetry of the molecule by transiting from six-membered benzene ($\mathcal{D}_{\textrm{6h}}$ point group) to five-membered ring-shaped molecules.
This will be  the starting point of the present thesis. 
Both cyclopentene and cyclopentadiene are homocyclic and nonaromatic  pentagon molecules of carbon atoms. 
Moreover, both molecules are ionic, making them less attractive for the present purpose. 
On the other hand, pyrrole, furan and oxazole are five-membered ring-shaped  neutral molecules.
Thus, these three molecules with $\mathcal{C}_{\textrm{2v}}$ and $\mathcal{C}_{\textrm{s}}$ point group symmetries will serve our first purpose to explore the role of symmetry reduction on charge migration.

To explore the role of the symmetry on charge migration, 
we will transit from a planer to non-planer molecule as a second example. 
Magnesium porphyrin is another  most symmetric and planer molecule in which 
attosecond charge migration has been   
studied in detail~\citep{nam2020monitoring, tremblay2021time, barth2006periodic, barth2006unidirectional, koksal2017effect}. 
Corroles have out-of-plane atoms which is a reason for the saddling in their molecular structure, whereas 
porphyrins are planar in nature.    
Corroles and porphyrins are macrocyclic organic molecules as they sustain a conjugation channel bearing a large number (18) of $\pi$ electrons. Due to a slightly smaller number of atoms,  
corroles are commonly known as reduced dimensional models for porphyrins~\citep{johnson1965corroles}, with whom they share similar electronic properties. 
The chemistry of porphyrins and corroles is a never-ending field of exciting problems, for both fundamental and practical reasons~\citep{mingos2012molecular}.
Over the years, many significant research works have been performed 
to understand ultrafast light-induced 
processes in porphyrins from a theoretical perspective~\citep{tremblay2021time, nam2020monitoring, koksal2017effect, barth2006periodic, barth2006unidirectional}.  
In contrast, corroles have not received similar attention in the context of light-induced ultrafast processes~\citep{lemon2020corrole}. 
As a part of this thesis, we have taken a first step towards exploring ultrafast charge migration in metal-coordinated corroles where we will explore  the role of structural 
saddling on the charge migration in copper corrole.

After studying charge migration in molecules with certain symmetries, 
we will explore charge migration in molecules without any symmetry. 
For this purpose, chiral molecules are  best suited.  
Chirality is a general property observed in nature, and can be defined as the geometric property of a molecule of being non-superposable on its mirror image. 
Enantiomers -- a pair of chiral molecules --  possess similar physical properties but they show strong enantiomeric preference during chemical reactions. 
Distinguishing and understanding the chirality of molecules is essential in a broad range of sciences. One example is the important role that homochirality plays in life on earth~\citep{bada1995origins, blackmond2010origin, meierhenrich2013amino}. Therefore,  detecting and measuring the enantiomeric excess and handedness of chiral molecules play a crucial role in chemistry, biology and pharmacy~\citep{mori2011bioactive, fischer2005nonlinear, inoue2004chiral}. 
As a result, development of evermore reliable methods to discern enantiomers is an ongoing quest, which has received considerable attention in recent years. 
%~\citep{castiglioni2011experimental}.
 In this part of the thesis, we will explore how chiral control over coherent electronic  motion can be 
achieved by choosing the different orientations of an ultrashort linearly polarized pulse, which is
 used to drive an ultrafast charge migration process by the excitation of a small number of low-lying excited states from the ground electronic state of R- and S-epoxypropane.
Moreover, we will explore 
the conditions required to understand charge migration in oriented and floppy chiral molecules.

\begin{figure}[h!]
\includegraphics[width= \linewidth]{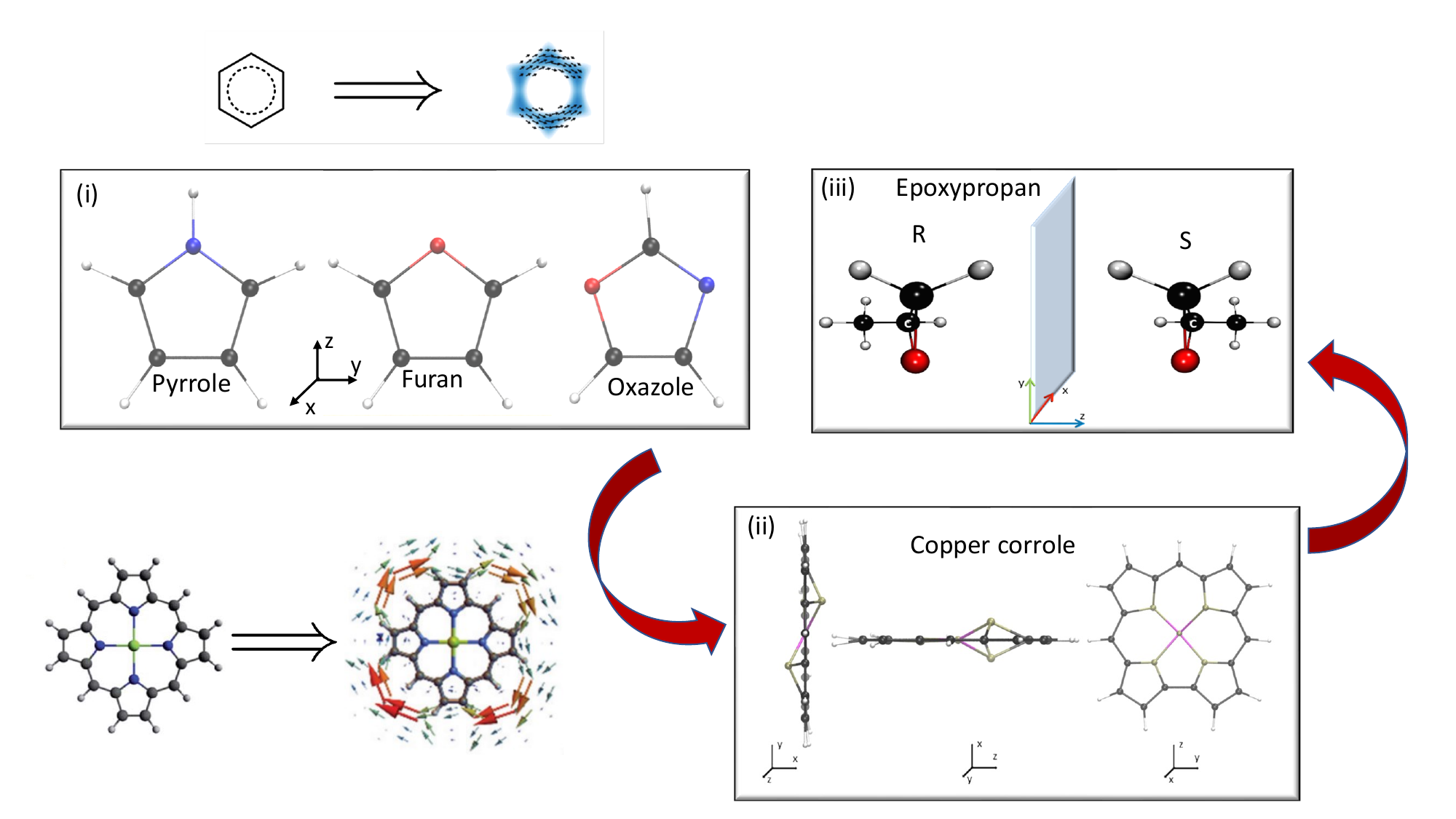}
\caption{An illustration of the overview of the impact of molecular symmetry during laser-induced 
charge migration in different molecules. (i) Planar molecules pyrrole and furan with $\mathcal{C}_{\textrm{2v}}$ and oxazole with $\mathcal{C}_{\textrm{s}}$ point group symmetries, (ii) structural saddling in copper corrole with $\mathcal{C}_{2}$ symmetry 
and (iii) R- and S-epoxypropane with  $\mathcal{C}_{1}$ symmetry. 
Results of  benzene and magnesium porphyrin are adapted from ~\citep{hermann2020probing} and~\citep{nam2020monitoring}, respectively.} \label{fig11}
\end{figure}

Another important aspect of this thesis is a four-dimensional imaging of charge migration. 
For this purpose, spatial representation of charge migration is essential.
As the coherent electronic wavepacket associated with the charge migration evolves in time, the
electronic charge distribution migrates from one point in space to another point within a molecule. 
Following the quantum version of the continuity equation, time-dependent electronic charge distribution and thus the charge migration 
is accompanied by time-dependent  electronic flux density (EFD)~\citep{sakurai2006advanced}. 
Not only the EFD provides mechanistic information about ultrafast process but also  a directional representation of the spatial distribution  of the charge migration along the direction of flow.  
The  EFD is a vector quantity, whereas   the charge distribution is a scalar quantity. 
Over the years, EFD has emerged as an important method to gain detailed insights about the underlying mechanism of ultrafast charge migration in systems ranging from  molecules to nano-materials~\citep{barth2006periodic,barth2006unidirectional,nagashima2009electron, giri2020time, okuyama2009electron, giri2021imaging, giri2022probing, shao2020electronic, hermann2016ultrafast, pohl2019imaging, shao2021local, sobottka2020tuning,diestler2012coupled,takatsuka2011exploring,patchkovskii2012electronic,okuyama2012dynamical,diestler2013computation,takatsuka2014chemical,hermann2014electronic,yamamoto2015electron,hermann2016multidirectional, bredtmann2014x}. 
In 2020,  it was established that TRXD from an electronic wavepacket is related to the EFD associated with the charge migration. Laser-induced charge migration in benzene was used to establish this connection~\citep{hermann2020probing}. 
Following this work, charge migration and ring current associated with the hole wavepacket in  oxazole was 
discussed in the context of  TRXD~\citep{carrascosa2021mapping}. 
Also, stimulated resonant x-ray Raman process was used to prepare a hole wavepacket in oxazole and the resultant charge migration was imaged by TRXD~\citep{yong2021ultrafast}.
Recently, it was demonstrated that the combination of  time-resolved electron  and x-ray homodyne diffraction has potential to provide a real-space picture of charge migration in molecule without the influence of nuclear motion~\citep{yong2022attosecond}.

In this thesis, we will use time-dependent electronic charge distribution  and associated EFD to unravel the influences of the molecular symmetries  in the charge migration and 
its underlying mechanism in different  molecules. 

\section{Outline}\label{Outline}
The thesis is primarily divided into six main chapters and bibliography.

A brief overview of ultrafast processes in matter is discussed in \textbf{Chapter 1}. 
The recent developments  of attosecond physics with the generation of the coherent 
attosecond pulses and their key applications in atoms and solids  are introduced.  
The concept of TRXD using ultrashort x-ray pulses within pump-probe configuration  
and the recent results of TRXD are discussed. 
Following the attosecond physics and TRXD, the notion of the charge migration with recent theoretical and experimental findings  are presented.  
A detailed aim of the present thesis with an overview is offered at the end of this chapter. 

\textbf{Chapter 2} is dedicated for the theoretical framework of inducing and four-dimensional imaging of the charge migration in molecules.
A brief overview of the electronic structure theory, based on wave function and density functional methods, is presented.  
Concepts of time-dependent density functional theory  and time-dependent configuration interaction are introduced to  construct the laser-induced electronic wavepacket associated with the charge migration.    
 A theory of electronic continuity equation is introduced for spatiotemporal analysis of the charge migration. 
This chapter ends with a brief theory of TRXD and technical details of the numerical simulations performed in this thesis .

In \textbf{Chapter 3}, we investigate  a connection between the charge migration and molecular symmetry. For this purpose, charge migration in three planar molecules, pyrrole, furan and oxazole, is presented.  
The effects of  the electronegative of the foreign atoms are discussed by comparing the charge migration in pyrrole and furan. 
Moreover, the effect of symmetry reduction from $\mathcal{C}_{\textrm{2v}}$ to $\mathcal{C}_{\textrm{s}}$ is analysed by a comparative study of charge migration in oxazole with pyrrole and furan. 

\textbf{Chapter 4} contains a detailed discussion on the effect of the molecular structural saddling on charge migration in copper corrole.  
A linearly polarized pulse is used to trigger  the charge migration in copper corrole.  
Time-resolved difference  electronic charge  and flux densities are used to quantify 
the role of the saddling on charge migration, which is subsequently imaged by TRXD. 
The simulated results are compared for  
planes containing the saddling and plane without saddling  at different instances of the charge migration. 
To ensure the impact  of the saddling on the charge migration, 
we have also compared the static x-ray diffraction signal corresponding to ground states of copper corrole  and copper porphyrin in which the structural saddling  is absent and its structure is planner. 

Charge migration  in molecules without any symmetry, i.e., chiral molecules is investigated 
in \textbf{chapter 5}.
The control over charge migration in space-oriented R- and S-enantiomers of epoxypropane is discussed by tuning the parameters of linearly polarized laser pulse.
In the first case, we have applied a  linearly polarized pulse along an axis to induce 
charge migration in both enantiomers. 
In the second case, we tune the direction of the polarization of the linear  pulse in a plane containing normal axis.  
This results drastically different  charge migration in both enantiomers.
We have also investigated the charge migration in loosely-oriented floppy chiral molecules. 
Time-resolved difference  electronic charge  and flux densities are employed  to validate our findings. 
Furthermore, TRXD is employed to image charge migration in both enantiomers. 
This chapter is concluded with the introduction of an asymmetry parameter associated with TRXD 
to discern the enantiomers.

We conclude our findings in this thesis with future direction in \textbf{Chapter 6}.

\cleardoublepage
\chapter{Theoretical Framework  for Laser-Induced Charge Migration}\label{Chapter2}

Atomic units are used throughout in this chapter unless stated otherwise.
The  time evolution of an electronic charge distribution in a molecule within 
non-relativistic framework  can be described by time-dependent Schr\"{o}dinger equation (TDSE) as
\begin{equation}{\label{eq:TDSE}}
    i \frac{d}{dt} | \Psi_{\textrm{tot}} (\textbf{r}, \textbf{R}, t) \rangle =
    \mathcal{H}(t) | \Psi_{\textrm{tot}} (\textbf{r}, \textbf{R}, t) \rangle,
\end{equation}
where \textbf{r} and \textbf{R} represent the electronic and nuclear coordinates, respectively. 
The total time-dependent Hamiltonian, $\mathcal{H}(t)$, can be written as a sum of  
time-independent Hamiltonian for a molecule, $\mathcal{H}_{0}$, and the time-dependent interaction Hamiltonian between a molecule and a laser, $ \mathcal{H}_{\textrm{int}}(t)$, as
\begin{equation}{\label{eq:Tot_Hamiltonian}}
 \mathcal{H}(t) = \mathcal{H}_{0} + \mathcal{H}_{\textrm{int}}(t).    
\end{equation}
The molecular Hamiltonian within Born-Oppenheimer approximation can be written as 
\begin{equation}\label{eq:BO_Mol_Hamiltonian}
   \mathcal{H}_{0} = - \sum_{i} \frac{1}{2} \nabla_{i}^{2}  - \sum_{A,i} \frac{Z_{A}}{ |\textbf{r}_{iA}|} + \frac{1}{2}\sum_{i \neq j} \frac{1}{ |\textbf{r}_{ij}|},
\end{equation}
where the first term stands as  the kinetic energy for electrons followed by the electron-nucleus and electron-electron interaction terms with  $Z_{A}$  as the nuclear charge of the $A^{\textrm{th}}$ atom~\citep{tannor2007introduction}. 
The applicability of the Born-Oppenheimer approximation is justified in this thesis 
as we are interested in the laser-induced charge migration taking place on attosecond timescale. 
Within Born-Oppenheimer approximation,  TDSE  for electrons with a fixed nuclear configuration 
is  written as  
\begin{equation}{\label{eq:el_TDSE}}
    i \frac{d}{dt} | \Psi (\textbf{r}, t) \rangle =
    \mathcal{H} (t) | \Psi (\textbf{r}, t) \rangle, 
\end{equation}
where time-dependent wave function, $ | \Psi (\textbf{r}, t) \rangle$, is expanded as 
\begin{equation}{\label{eq:TD_wavefunction}}
| \Psi (\mathbf{r}, t) \rangle =  \sum_{k} C_{k} (t) | \Phi_{k} (\mathbf{r}) \rangle = \sum_{k} | \Psi_{k} (\mathbf{r}, t) \rangle.   
\end{equation}
Here, $C_{k}(t)$ is a time-dependent  expansion coefficient and $| \Phi_{k} (\mathbf{r}) \rangle$ is a time-independent multi-electronic wave function. Solving the TDSE for electrons in the presence of a laser is a challenging task. 
In the following, we will first  discuss theoretical methods to obtain  $| \Phi_{k} (\mathbf{r}) \rangle$, which will be followed by techniques to  construct   time-dependent coefficient $C_{k}(t)$ in the presence of a laser. 

\section{Electronic Structure Theory}{\label{Sec:Stationary_wave_function}}
It is well-known that the time-independent molecular Hamiltonian, $\mathcal{H}_{0}$, is not 
exactly solvable except for few systems such as hydrogen~\citep{szabo2012modern}.
The difficulty in solving  $\mathcal{H}_{0}$ 
is a drastic increase in computational cost toward the exact solution as the number of electrons increases.  
A set  of appropriate  approximations and a large number of mathematical operations are employed to obtain 
solutions of $\mathcal{H}_{0}$  as given in Eq.~(\ref{eq:BO_Mol_Hamiltonian}),  upto some level of accuracy. In the following, we will discuss two approximate methods to solve $\mathcal{H}_{0}$:  
the first one relies  on wave function based approach, whereas  the other is density-based approach.  

\subsection{Wave Function Based Methods}
Independent-electron model is the simplest approach to solve $\mathcal{H}_{0}$ in which electrons are 
considered as independent and do not interact with each other. Within this model, electron-electron interaction is 
completely ignored, i.e., the last term of Eq.~(\ref{eq:BO_Mol_Hamiltonian}) is omitted. 
 Going beyond the independent-electron model is Hartree-Fock method,  which accounts average interaction among the electrons and the multi-electronic wave function is expressed   in the form of a Slater determinant.  
Numerous theoretical methods are developed over the years by considering different levels of the electron-electron correction to improve the results obtained by Hartree-Fock method.  
These methods are known as post Hartree-Fock method. 
Configuration Interaction (CI) method is one of the most popular and robust post Hartree-Fock methods~\citep{szabo2012modern, jensen2017introduction, helgaker2014molecular}. 

Total electronic wave function within CI method  is expressed in terms of the linear combination of several Slater determinants as
\begin{equation}{\label{eq:CI_wavefunction}}
| \Phi^{(\lambda)}_{\textrm{CI}} \rangle = D^{(\lambda)}_{0} | \Phi^{(0)}_{\textrm{HF}} \rangle + \sum_{ar} D^{r(\lambda)}_{a} | \Phi^{(0)}_{\textrm{HF}} \rangle^{r}_{a} + \sum_{abrs} D^{rs(\lambda)}_{ab} | \Phi^{(0)}_{\textrm{HF}} \rangle^{rs}_{ab} + \cdots,
\end{equation}
where $| \Phi^{(0)}_{\textrm{HF}} \rangle$ is the Hartree-Fock wave function,  which is treated as  
a reference wave function,  $D's$ are coefficients of the configuration interaction. 
Electrons are excited from occupied $\{ a, b, \cdots\}$ to virtual $\{ r, s, \cdots\}$ molecular orbitals  to 
construct the excited-state Slater determinants from a ground-state Slater determinant.  
An exact solution of the molecular Hamiltonian can be obtained 
by considering all possible excitations. 
However, the number of possible excitations increases drastically as 
the number of electrons in a molecular system increases, which 
makes full CI  numerically impractical to implement for a moderate size molecular system. 
To make a delicate balance between the computational cost and accuracy of the wave function, 
either the rank of the excitation can be restricted to CI singles (CIS), or CI doubles (CID) and so on, 
or the excitation can be restricted to a specific configuration  space. 
We have considered only single excitations to all the results presented 
in this thesis.  

Hartree-Fock  and post Hartree-Fock methods are computationally  expensive 
as they have to keep track of all the  spatial coordinates of all the electrons in a molecule explicitly. 
To reduce computational cost,  an alternate density-based method becomes popular and practical over the years. 
In the following section, let us briefly discuss the mathematical framework of Density Functional Theory (DFT). 

\subsection{Density Based Methods}
Let us start our discussion about DFT by writing time-independent Schr\"{o}dinger equation corresponding to $\mathcal{H}_{0}$ as 
\begin{equation}
	\left[ \mathcal{T}_e + \mathcal{V}_{ei} + \mathcal{W}_{ee} \right] | \Psi_0(\mathbf{r}) \rangle =  \mathcal{E}_{0} | \Psi_0(\mathbf{r}) \rangle,
\end{equation}  
where $\mathcal{T}_e$ is the  kinetic energy operator, $\mathcal{V}_{ei}$ is the attraction between electron and nucleus, and $\mathcal{W}_{ee}$ is the electron-electron interaction terms.  $| \Psi_0(\mathbf{r}) \rangle$ and $ \mathcal{E}_{0}$ 
are the ground state wave function and corresponding energy, respectively. 
$\mathcal{V}_{ei}$ characterize the molecular system within non-relativistic framework,  whereas 
 the form of $\mathcal{T}_e$ and $\mathcal{W}_{ei}$ are universal. 

Hohenberg and Kohn demonstrated in their seminal work that any observable of a 
molecular system can be deduced from their ground-state density, $\rho_0(\textbf{r})$, without the requirement of a multi-electronic wave function 
$| \Psi_0(\mathbf{r}) \rangle$~\citep{hohenberg1964inhomogeneous}. 
Moreover, a one-to-one correspond between $v_{ei}(\textbf{r})$ and $\rho_0(\textbf{r})$ has been established with 
$| \Psi_0(\mathbf{r}) \rangle$ is a unique functional of $\rho_0(\textbf{r})$. 
This leads to the conclusion that  the ground-state density contains all the relevant information of a molecular  system. This is the foundation of DFT.  
The ground-state energy functional within DFT is  defined as
\begin{equation}
	E[\rho_0] = \left\langle \Psi[\rho_0] \right| \mathcal{{H}}_0 \left| \Psi[\rho_0] \right\rangle.
\end{equation}
However, the exact form of the energy functional can be obtained variationally. 
The energy functional can be expressed  in terms of single-particle orbitals, $\phi_\mu$, as 
\begin{equation}\label{eq:Efun}
	E[\rho] =   T_s[\rho] + W_H[\rho] + E_{xc}[\rho] + \int d \mathbf{r}~\rho(\mathbf{r})v_{ei}(\mathbf{r}).
\end{equation}
Here,  $T_s$ is the kinetic energy of a non-interacting system with electronic density, $\rho$, and can be written as 
\begin{equation}
	T_s[\rho] = -\frac{1}{2} \sum_{i=\mu}^N \int d {\mathbf r}~\phi^*_\mu(\mathbf{r})\nabla^2 \phi_\mu(\mathbf{r}),
\end{equation} 
and the classical Hartree term $W_H$ is expressed as 
\begin{equation}
	W_H[\rho] = \frac{1}{2}\int d {\mathbf r}\int d {\mathbf r}^{\prime}~\frac{\rho(\mathbf{r})\rho(\mathbf{r}^{\prime})}{|\mathbf{r}-\mathbf{r}^{\prime}|}.
\end{equation}
The exchange-correlation energy functional, $E_{xc}$,  contains all 
information about a molecular system  by construction. 
Kohn-Sham (KS) equation  is obtained by 
minimizing the energy functional in Eq.~(\ref{eq:Efun}) using variational principle as~\citep{kohn1965self}
\begin{equation}
	\left(-\frac{1}{2}\nabla^2 + v_s[\rho_0](\textbf{r})\right) | \phi_\mu(\textbf{r}) \rangle = \epsilon_\mu| \phi_\mu(\textbf{r}) \rangle.
\end{equation}
Here, the effective single-particle potential is defined as
\begin{equation}{\label{eq:KS_potential}}
	v_s[\rho](\textbf{r}) = v_{ei}[\rho](\mathbf{r}) + v_H[\rho](\mathbf{r}) + v_{xc}[\rho](\mathbf{r}).
\end{equation}
Here, the Hartree and exchange-correlation  potentials are defined as 
$v_H[\rho](\mathbf{r}) = \partial W_H[\rho]/\partial \rho(\mathbf{r})$, and  $v_{xc}[\rho](\mathbf{r}) = \partial E_{xc}[\rho]/\partial \rho(\mathbf{r})$, respectively. 
KS equation is equivalent to single-electron Schr\"odinger equation, and its  solution is obtained self consistently.

\subsubsection{Time-Dependent Density Functional Theory}
Basic idea of the ground-state DFT is extended to 
Time-Dependent DFT (TDDFT)
to calculate  the excitation energies of the excited states, frequency-dependent response properties, and other   time-dependent properties~\citep{gross1996density, marques2004time, marques2012fundamentals, casida2009time}. 
The building blocks of the TDDFT are two theorems: Runge-Gross theorem~\citep{runge1984density}, which is a time-dependent extension of the Hohenberg-Kohn theorem~\citep{hohenberg1964inhomogeneous}, and the Van Leeuwen theorem~\citep{van1999mapping}. 
Runge and Gross showed that, for a multi-electronic  molecular system evolving 
from a ground-state $| \Psi(t = t_0) \rangle = | \Psi_0 \rangle$, 
there is a one-to-one mapping between external potential $v_{ext}(\textbf{r},t)$ and time-dependent density $\rho(\textbf{r}, t)$ with $\mathcal{V}_{ext}(t) = \sum_{i=\mu}^N v_{ext}(\textbf{r},t)$ as the scalar external potential. 
Time-dependent KS (TDKS) equation analogous to single-electron TDSE can be written  as 
\begin{equation}{\label{eq:TDKS}}
	i \frac{\partial}{\partial t} | \phi_\mu(\textbf{r},t) \rangle =\left(-\frac{1}{2}\nabla^2 + v_s(\textbf{r},t)\right) 
	|\phi_\mu(\textbf{r},t) \rangle.
\end{equation}
Here,  TDKS orbitals of non-interacting electrons  are used to obtain 
the time-dependent density as 
\begin{equation}
	\rho(\textbf{r},t) =  \sum_{i=\mu}^N |\phi_\mu(\textbf{r},t)|^2.
\end{equation}
The time-dependent density  evolves in the presence of the TDKS potential. 
In general, the exact expression of the time-dependent exchange-correlation potential is not known. Thus, 
it is not very straightforward to solve TDKS equation as given in Eq.~(\ref{eq:TDKS}). 
In this thesis, we have adopted first-order  linear response approach  to solve Eq.~(\ref{eq:TDKS}). 

Linear-Response TDDFT (LR-TDDFT) is a first-order perturbative solution of the TDKS equation, which  
describes the molecular system in the presence of a weak 
perturbation~\citep{gross1996density, marques2004time, marques2012fundamentals}. 
The first-order  correction to density within LR-TDDFT approach can be written using Taylor series expansion  as~\citep{ernzerhof1994taylor}  
\begin{equation}
\rho^{(1)} (\mathbf{r}, t) = \int \int dt^{\prime} d \mathbf{r}^{\prime} \chi_{\textrm{eff}} (\mathbf{r}, t, \mathbf{r^{\prime}}, t^{\prime}) \hat{v}_{\textrm{eff}}^{(1)} (\mathbf{r^{\prime}}, t^{\prime}).
\end{equation}
Here, $\hat{v}_{\textrm{eff}}^{(1)}(\mathbf{r^{\prime}}, t^{\prime})$ is the first-order correction of the TDKS potential and consists of three terms  same as mentioned in Eq.~(\ref{eq:KS_potential}). 
The last term due to non-trivial multi-electronic effects, i.e., the exchange correlation term, has the following form 
\begin{equation}
\hat{v}_{\textrm{XC}}^{(1)} (\mathbf{r}, t) = \int \int dt^{\prime} d\mathbf{r}^{\prime} \hat{f}_{\textrm{XC}} (\mathbf{r}, t, \mathbf{r^{\prime}}, t^{\prime}) \rho^{(1)} (\mathbf{r^{\prime}}, t^{\prime}),
\end{equation}
where the time-dependent exchange-correlation kernel, $\hat{f}_{\textrm{XC}} (\mathbf{r}, t, \mathbf{r^{\prime}}, t^{\prime})$, can be written as 
\begin{equation}
\hat{f}_{\textrm{XC}} (\mathbf{r}, t, \mathbf{r^{\prime}}, t^{\prime}) = \left. \frac{\delta \hat{v}^{(1)}_{\textrm{XC}} [\rho] (\mathbf{r}, t)}{\delta \rho (\mathbf{r^{\prime}}, t^{\prime})}\right|_{\rho (\mathbf{r}, t = 0)}. 
\end{equation}
The interacting and non-interacting response functions are connected using the notion of the Fourier space. 
The density-density response function in the Fourier space is 
\begin{equation}
\chi_{\textrm{eff}} (\mathbf{r}, \mathbf{r^{\prime}}, \omega) = \sum_{ar} \left[ \frac{\rho_{ar}(\mathbf{r^{\prime}}) \rho_{ra}(\mathbf{r})}{\omega - \omega_{ra} + i \eta} - \frac{\rho_{ra}(\mathbf{r^{\prime}}) \rho_{ar}(\mathbf{r})}{\omega + \omega_{ra} + i \eta}\right],
\end{equation}
where a transition from an  occupied molecular orbital ``$a$''  to a virtual molecular orbital ``$r$'' is considered. 
To solve the above equation self consistently, it can be modified like a Dyson-type equation as 
\begin{eqnarray}
\nonumber
\chi (\mathbf{r}, \mathbf{r^{\prime}}, \omega) &=& \chi_{\textrm{eff}} (\mathbf{r}, \mathbf{r^{\prime}}, \omega)  \\ \nonumber
&& + \int \int d\mathbf{r}^{\prime \prime} d\mathbf{r}^{\prime \prime \prime} \chi (\mathbf{r}, \mathbf{r}^{\prime \prime}, \omega) \left[ \frac{1}{|\mathbf{r}^{\prime \prime} - \mathbf{r}^{\prime \prime \prime|}}  + \hat{f}_{\textrm{XC}} (\mathbf{r}^{\prime \prime}, \mathbf{r}^{\prime \prime \prime}, \omega) \right] \chi_{\textrm{eff}} (\mathbf{r}^{\prime \prime \prime}, \mathbf{r}^{\prime}, \omega).
\end{eqnarray}
The energies of the excited electronic  states can be extracted from the poles of $\chi (\mathbf{r}, \mathbf{r^{\prime}}, \omega)$ with transition frequencies $\omega_{ra} = (\epsilon_{r} - \epsilon_{a})$. 
A non-Hermitian eigenvalue equation, known as Casida equation, is solved to obtain the 
excitation energies of electronic  states. 
Moreover, by eliminating the de-excitation process using Tamm-Doncoff approximation, 
we arrive the following equation in compact form as~\citep{hirata1999time}
\begin{equation}
    \textbf{AX}_{\lambda} = \Omega_{\lambda} \textbf{X}_{\lambda}. 
\end{equation}
The above equation is equivalent to the CIS method. 
Within Tamm-Doncoff approximation, a CIS-type wave function can be expressed explicitly for each excited state. 
This method thus works as a bridge between the CIS and TDDFT methods and termed as hybrid TDDFT-CI approach.
 
So far, we have discussed methods to solve $\mathcal{H}_{0}$  within Born-Oppenheimer approximation. 
The stationary part of the time-dependent wave function $| \Phi_{k} (\mathbf{r}) \rangle$  can be obtained using   
hybrid TDDFT-CI approach. 
In this thesis, we have expressed these time-independent wave function as a linear combination of the CIS functions as given in Eq.~(\ref{eq:CI_wavefunction}). 
In the following,  we will present a brief discussion  about the theory of laser-induced charge migration -- 
the main  focus of this thesis.

\section{Laser-Induced Charge Migration}{\label{Sec:Dynamics}}
 To characterize the charge migration in a molecular system, 
density operator is the key quantity and can be written as 
\begin{equation}
 \hat{\varrho} (\mathbf{r}, t) = |\Psi (\mathbf{r}, t) \rangle \langle \Psi (\mathbf{r}, t)|.
\end{equation}
The time evolution of an electronic  charge distribution can be described in terms of  the reduced density matrix formalism where the dissipation acts as the effect of the environment. 
Let us introduce the dissipative Liouville-von Neumann (LvN) equation to 
characterize such type of dynamics in a molecule as
\begin{equation}{\label{eq:LvN}}
\frac{\partial}{\partial t} \hat{\varrho} ( \textbf{r}, t) =  
- i \left[ \mathcal{H}_{\textrm{int}}(t), \hat{\varrho} (\textbf{r}, t) \right] + \mathcal{L}_{D} \hat{\varrho} (\textbf{r}, t),
\end{equation}
where $\mathcal{L}_{D}$ is the dissipative Liouvillian. 
The population of the electronic states can be extracted from the diagonal elements in the CIS eigenstate basis. 
Thus, the above equation is expressed in terms of Lindblad form. 
The dissipative Liouvillian within Lindblad formalism can be written as 
\begin{equation}{\label{eq:Dissipative_Liouvillian}}
    \mathcal{L}_{\textrm{D}} \hat{\varrho} (\textbf{r}, t) = \frac{1}{2} \sum_{d} \Bigl\{ \left[ \hat{L}_{d} \hat{\varrho} (\textbf{r}, t), \hat{L}^{\dag}_{d} \right] + \left[ \hat{L}_{d}, \hat{\varrho} (\textbf{r}, t) \hat{L}^{\dag}_{d} \right]\Bigr\}.
\end{equation}
Here, $\hat{L}_{d}$ is the Lindbland operator and has the following form 
\begin{equation}{\label{eq:Lindbland_operator}}
    \hat{L}_{d} = \sqrt{\Gamma_{\nu \to \lambda}} | \Phi_{\lambda} (\mathbf{r})\rangle \langle \Phi_{\nu} (\mathbf{r}) |,
\end{equation}
where $\Gamma_{\nu \to \lambda}$ is the transition rate between 
$| \Phi_{\nu} \rangle $ and $| \Phi_{\lambda} \rangle $ states, 
and can be estimated using Fermi's Golden rule.

The density operator  can be expressed in  a matrix form using a orthonormal basis 
$\{ | \Phi_{k} (\mathbf{r}) \rangle\}$,  and the matrix elements can be written as
 \begin{equation}
    \varrho_{kl} (t) = \langle \Phi_{k} (\mathbf{r})|\hat{\varrho} (\mathbf{r}, t)| \Phi_{l} (\mathbf{r}) \rangle.
 \end{equation}
 Before we discuss the solution of Eq.~(\ref{eq:LvN}) in terms of the density matrix, let us write the 
 interaction Hamiltonian  responsible for the charge migration as 
 \begin{equation}{\label{eq:Interaction_Hamiltonian}}
     \mathcal{H}_{\textrm{int}}(t) = - \hat{\mu} \cdot \mathbf{E} (t),
 \end{equation}
where $\hat{\mu}$ is the dipole operator and $\mathbf{E} (t)$ is the electric field of the laser pulse. 

The time evolution of the direct terms of the density matrix in Eq.~(\ref{eq:LvN}) can be written as 
\begin{equation}{\label{eq:direct_density_operator}}
\frac{d\varrho_{nn} (t)}{dt} =  -i E_{j}(t) \sum_{i} (\mu_{ni} \hat{\varrho}_{in} - \hat{\varrho}_{ni} \mu_{in} )  + \sum_{i} (\Gamma_{i \to n} \hat{\varrho}_{ii} - \Gamma_{n \to i} \hat{\varrho}_{nn}).
\end{equation}
Here, $E_{j}$ is the $j^{\textrm{th}}$-component of $\mathbf{E} (t)$, which 
depends on the polarization of the laser pulse, and $\mu_{ni} = \langle \Phi_{n} | \hat{\mu} |  \Phi_{i} \rangle$. 
Similarly, the time-evolution of the cross terms of the density matrix is written as 
\begin{equation}{\label{eq:indirect_density_operator}}
\frac{d\varrho_{mn} (t)}{dt} = -i \omega_{mn} \hat{\varrho}_{mn} -i E_{j}(t) \sum_{i} (\mu_{mi} \hat{\varrho}_{in} - \hat{\varrho}_{mi} \mu_{in} ) - \gamma_{mn} \hat{\varrho}_{mn},
\end{equation}
where the dephasing rate is $\gamma_{mn} = \gamma_{nm} = (1/2) \sum_{l} (\Gamma_{m \to l} + \Gamma_{n \to l})$. 
The diagonal elements of the density matrix, $\varrho_{kk} (t)$, 
are the probability of finding a molecular system in the basis $\{| \Phi_{k} \rangle \}$ and known  
as population of the $k^{\textrm{th}}$ electronic state. 
The time-dependent expansion coefficient in  Eq.~(\ref{eq:TD_wavefunction}) can be written as  
$C_{k}(t) = \varrho_{kk}(T_{p}) \exp (-i E_{k} t) $  with $T_{p}$ as the pulse duration of laser pulse 
and $E_{k}$ is the energy of the $k^{\textrm{th}}$ electronic state. 

After constructing the  time-dependent expansion coefficient and the wave function, let us 
proceed to  understand  laser-driven charge migration in terms of electronic charge and flux densities 
in the following section.

\subsection{Electronic Continuity Equation}
The flow of an  electronic charge distribution after the end of the laser pulse can be visualized in terms of 
time-dependent electron density, which  can be calculated from the expectation value of the density 
operator after solving Eq.~(\ref{eq:LvN}). 
The time-dependent electron density
using Eq.~(\ref{eq:TD_wavefunction})  can be written as
\begin{equation}{\label{eq:TD_rho}}
    \rho (\mathbf{r}, t) = \sum_{kl} C^{*}_{k} (t) C_{l} (t) \langle \Phi_{k} (\mathbf{r})|\hat{\rho} (\mathbf{r})|\Phi_{l} (\mathbf{r}) \rangle.
\end{equation}
Here, the one-electron density operator, $\hat{\rho}(\mathbf{r})$, has the following form
\begin{equation}{\label{eq:rho_operator}}
    \hat{\rho} (\mathbf{r}) = \sum_{k} \delta (\mathbf{r} - \mathbf{r}_{k}).
\end{equation}
The time-dependent electron density allows us to visualize  the temporal evolution of an electronic charge distribution in space.  Moreover,  
 the direction of the flow of the charge distribution can be obtained by  time-dependent electronic flux density, which is  written as 
\begin{equation}
    \mathbf{j} (\mathbf{r}, t) = \sum_{kl} C^{*}_{k} (t) C_{l} (t) \langle \Phi_{k} (\mathbf{r})|\hat{j} (\mathbf{r})|\Phi_{l} (\mathbf{r}) \rangle.
\end{equation}
Here, the flux density operator, $\hat{j}$, has the following form 
\begin{equation}{\label{eq:j_operator}}
    \hat{j} (\mathbf{r}) = \frac{1}{2} \sum_{k} [\delta (\mathbf{r} - \mathbf{r}_{k}) \hat{p}_{k} - \hat{p}^{\dag}_{k} \delta (\mathbf{r} - \mathbf{r}_{k})], 
\end{equation}
where the momentum of the $k^{\textrm{th}}$ electron is given by the momentum operator as $\hat{p}_{k} = -i\mathbf{\nabla}_{k}$.

The connection between the time-dependent electronic charge  and flux densities can be established by taking a time-derivative of the density matrix, and is expressed  as   
\begin{equation}{\label{eq:continuity}}
    \frac{\partial}{\partial t} \rho (\textbf{r}, t) = - \mathbf{\nabla} \cdot \textbf{j} (\textbf{r}, t).
\end{equation}
This is known as the electronic continuity equation that maintains the charge conservation of a molecular  
system in terms of $\rho (\textbf{r}, t)$ and $\textbf{j} (\textbf{r}, t)$.

In general, transition dipole moments in length and velocity gauges  are compared to ensure the accuracy 
of a wave function obtained via ``high-level" quantum chemistry calculations. 
The dipole moments in length and velocity gauges can be obtained from electronic  charge and electronic current densities, respectively.

Till now, we have discussed how TDSE can be solved and laser-induced 
charge migration can be studied using electronic charge and  flux densities.   
In the following section,  we will briefly discuss the theoretical framework of time-resolved x-ray diffraction (TRXD) 
to image the laser-induced charge migration. 

\section{Time-Resolved X-ray Diffraction}{\label{Sec:TRXD}}
In this thesis, we are interested to employe non-resonant time-resolved x-ray diffraction to capture temporarily-evolving charge distribution. Moreover, we have applied quantum electrodynamics (QED) treatment of  x-ray-molecule interaction.  The corresponding  interaction Hamiltonian for TRXD within QED framework can be written as~\citep{craig1998molecular}
\begin{equation}{\label{eq:int_Hamiltonian_probe}}
    \mathcal{H}^{\textrm{x}}_{\textrm{int}} =  \frac{\alpha^{2}}{2} \int d \textbf{r}~\psi^{\dag} (\textbf{r})~\mathcal{A}^{2} (\textbf{r})~\psi (\textbf{r}),
\end{equation}
where $\alpha$ is the fine-structure constant,  $\psi^{\dag} (\textbf{r}) [\psi (\textbf{r})]$ is electron creation (annihilation) field operator and $\mathcal{A}(\textbf{r})$ is a vector potential of x-ray, which has the following form \begin{equation}{\label{eq:field_vec_2nd}}
\mathcal{A} (\textbf{r}) = \sum_{\textbf{k}, s} \sqrt{\frac{2 \pi}{V \omega_{\textbf{k}} \alpha^{2}}} \{ \hat{\textbf{a}}_{\textbf{k}, s} \mathbf{\epsilon}_{\textbf{k}, s} e^{i \textbf{k} \cdot \textbf{r}} + \hat{\textbf{a}}^{\dag}_{\textbf{k}, s} \mathbf{\epsilon}^{*}_{\textbf{k}, s} e^{-i \textbf{k} \cdot \textbf{r}} \}.    
\end{equation}
Here, $\mathbf{\epsilon}_{\textbf{k}, s}$ is the polarization vector with $s$ as the polarization index and 
$\omega_{\textbf{k}}$ is the photon energy  of the $\textbf{k}^{\textrm{th}}$ mode. 
$\hat{\textbf{a}}_{\textbf{k}, s}  (\hat{\textbf{a}}^{\dag}_{\textbf{k}, s})$ is the photon creation (annihilation) operator and $V$ is the quantization volume.

Differential Scattering Probability (DSP) is the key quantity in TRXD, which  represents the number of diffracted  photons per unit solid angle and is written as
\begin{equation}{\label{eq:DSP}}
    \frac{dP}{d\Omega} = \frac{V\alpha^{3}}{(2\pi)^{3}} \int^{\infty}_{0} d\omega_{\textrm{k}_{s}} \omega^{2}_{\textrm{k}_{s}} \sum_{j} \sum_{\{ n^{\prime} \}}  \langle \Phi_{j} (\mathbf{r}) ; \{ n^{\prime} \} | \hat{\varrho}_{f} (t)| \Phi_{j} (\mathbf{r}) ; \{ n^{\prime} \} \rangle.
 \end{equation}
Here, $\hat{\varrho}_{f}$ is the final density operator of the entire system, i.e., x-ray and molecule. 
$| \{ n^{\prime} \} \rangle$ represents a complete set of numbers that specify the number of photons in all field modes.  The total density operator at the time of measurement can be written as
\begin{equation}
    \hat{\varrho}_{f} (t) = \lim_{t_{f} \to \infty} \lim_{t_{0} \to -\infty} \hat{U}_{\textrm{tot}} (t_{f}, t_{0}) \hat{\varrho}_{\textrm{in}}^{s} \hat{U}^{\dag}_{\textrm{tot}} (t_{f}, t_{0}), 
 \end{equation}
 where $\hat{U}_{\textrm{tot}} (t_{f}, t_{0})$ is the total time-evolution operator. 
 First-order time-dependent perturbation theory is employed to evaluate the first-order correction to the density operator as
\begin{eqnarray}{\label{eq:density1}}
 \nonumber \hat{\rho}^{(1)} &=& \lim_{t_{f} \to \infty} \lim_{t_{0} \to -\infty} \sum_{\{ n \}, \{ \Bar{n} \}} \hat{\rho}^{\textrm{x}}_{\{ n^{\prime} \}, \{ \Bar{n} \}}\\ \nonumber
&& \times \int^{t_{f}}_{t_{0}} \int^{t_{f}}_{t_{0}} dt_{1} dt_{2} \times \left[ \hat{U}_{\textrm{tot}} (t_{f}, t_{1}) \mathcal{H}^{\textrm{x}}_{\textrm{int}}  \hat{U}_{\textrm{tot}} (t_{1}, t_{0}) | \Psi_{\{ n\}}^{(0)} (\mathbf{r}, t_{0}) \rangle \right. \\ %\nonumber
&& \left. \times \langle \Psi_{\{ \Bar{n}\}}^{(0)} (\mathbf{r}, t_{0}) | \hat{U}^{\dag}_{\textrm{tot}} (t_{2}, t_{0})  \mathcal{H}^{\textrm{x} \dagger}_{\textrm{int}}  \hat{U}^{\dag}_{\textrm{tot}} (t_{f}, t_{2}) \right].
\end{eqnarray}
Here, $\hat{\rho}^{\textrm{x}}_{\{ n^{\prime} \}, \{ \Bar{n} \}}$ representes 
the populations and coherences of all the
occupied field modes associated with the incident x-ray pulse. 

After substituting  the expressions from Eqs.~(\ref{eq:int_Hamiltonian_probe}), (\ref{eq:field_vec_2nd}),   and (\ref{eq:density1}) to Eq.~(\ref{eq:DSP}), the expression of DSP can be written as 
\begin{eqnarray}
\nonumber    \frac{dP}{d\Omega} &=& \sum_{\textbf{k}_{1}, s_{1}, \textbf{k}_{2}, s_{2}} \int^{\infty}_{- \infty} \int^{\infty}_{- \infty}  dt_{1} dt_{2} \int^{\infty}_{0} d\omega_{\textbf{k}_{s}}\\ \nonumber
&& \times \frac{\omega_{\textbf{k}_{s}} \alpha^{3}}{2 \pi V \sqrt{\omega_{\textbf{k}_{1}}, \omega_{\textbf{k}_{1}}}} \left( \epsilon_{\textbf{k}_{1}, s_{1}} \cdot \epsilon^{\star}_{\textbf{k}_{s}, s_{s}} \right) \left( \epsilon^{\star}_{\textbf{k}_{2}, s_{2}} \cdot \epsilon_{\textbf{k}_{s}, s_{s}} \right)   \\ \nonumber
&& \times \left[ \int d \textbf{r} \int d \textbf{r}^{\prime} \left\langle \Psi(\mathbf{r}, t_{2}) | \hat{\rho} (\textbf{r}^{\prime})  \hat{U}(t_{2}, t_{1}) \hat{\rho} (\textbf{r}) | \Psi(\mathbf{r}, t_{1})  \right\rangle \right. \\ \nonumber
&& \left. \times e^{-i(\textbf{k}_{2} - \textbf{k}_{s}) \cdot \textbf{r}^{\prime}} e^{i(\textbf{k}_{1} - \textbf{k}_{s}) \cdot \textbf{r}} \ \textrm{Tr} [\hat{\rho}_{\textrm{in}}^{x} \hat{a}^{\dag}_{\textbf{k}_{2}, s_{2}} \hat{a}_{\textbf{k}_{1}, s_{1}}]  e^{-i \omega_{\textbf{k}_{1}} t_{1}} e^{i \omega_{\textbf{k}_{2}} t_{2}} e^{-i \omega_{\textbf{k}_{s}} (t_{2} - t_{1})}   \right], \nonumber
\end{eqnarray}
where 
\begin{equation}
    \textrm{Tr} [\hat{\rho}_{\textrm{in}}^{X} \hat{a}^{\dag}_{\textbf{k}_{2}, s_{2}} \hat{a}_{\textbf{k}_{1}, s_{1}}]=\sum_{\{ m \}} \sum_{\{ n\}, \{ \Bar{n}\}} \rho^{x}_{\{ n\}, \{ \Bar{n}\}} \langle \{ m \} | \hat{a}_{\textbf{k}_{1}, s_{1}} |  \{ n\} \rangle \times \langle \{ \Bar{n}\} | \hat{a}^{\dag}_{\textbf{k}_{2}, s_{2}} | \{ m \} \rangle.
\end{equation}
Here, $ \{ m \} \rangle = \hat{a}_{\textbf{k}_{1 (2)}, s_{1 (2)}} |  \{ n\} \rangle$.

Let us introduced two new time variables as $\gamma = \frac{t_{1} + t_{2}}{2}$ and $\delta = t_{2} - t_{1}$ to simplify 
the expression of DSP. 
Moreover, we can also replace $\epsilon_{\textbf{k}_{1}, s_{1}}$ and $\epsilon_{\textbf{k}_{2}, s_{2}}$ with $\epsilon_{\textbf{k}_{\textrm{in}}, s_{\textrm{in}}}$ as x-ray pulse has small bandwidth and angular spread. 
Using these above approximations the expression for DSP can be written as
\begin{eqnarray}
\nonumber    \frac{dP}{d\Omega} &=& \frac{d\sigma_{\textrm{th}}}{d\Omega}  \int^{\infty}_{- \infty} d \gamma \int^{\infty}_{- \infty}  d \delta \int^{\infty}_{0} d\omega_{\textbf{k}_{s}} \frac{\omega_{\textbf{k}_{s}}}{(2 \pi \omega_{\textbf{k}_{\textrm{in}}})^{2} \alpha} e^{- i \omega_{\textbf{k}_{s}} \delta}  \\ \nonumber
&& \times \int d \textbf{r} \int d \textbf{r}^{\prime} \left\langle \Psi \left(\mathbf{r}, \gamma + \frac{\delta}{2} \right) \left| \hat{\rho} (\textbf{r}^{\prime})~\hat{U}\left(\gamma + \frac{\delta}{2}, \gamma - \frac{\delta}{2}\right)~\hat{\rho} (\textbf{r}) \right| \Psi \left(\mathbf{r}, \gamma - \frac{\delta}{2}\right)  \right\rangle \\ 
&& \times e^{-i \textbf{k}_{s} \cdot (\textbf{r} - \textbf{r}^{\prime})} G^{(1)} \left(  \textbf{r}^{\prime}, \gamma + \frac{\delta}{2}; \textbf{r}, \gamma - \frac{\delta}{2} \right),
\end{eqnarray}
where $d\sigma_{\textrm{th}}/d\Omega$ is the Thomson scattering cross section and $G^{(1)} \left(  \textbf{r}^{\prime}, \gamma + \frac{\delta}{2}; \textbf{r}, \gamma - \frac{\delta}{2} \right)$ is the first-order correlation function for x-ray pulse and has the following form
\begin{eqnarray*}
G^{(1)} \left(  \textbf{r}^{\prime}, \gamma + \frac{\delta}{2}; \textbf{r}, \gamma - \frac{\delta}{2} \right) & = & 
 \frac{2 \pi \omega_{\textbf{k}_{in}}}{V} \sum_{\textbf{k}_{1}, s_{1}, \textbf{k}_{2}, s_{2}}  \textrm{Tr} [\hat{\rho}_{\textrm{in}}^{x} \hat{a}^{\dag}_{\textbf{k}_{2}, s_{2}} \hat{a}_{\textbf{k}_{1}, s_{1}}]  \\ \nonumber 
 && \times e^{i (\omega_{\textbf{k}_{2}} - \omega_{\textbf{k}_{1}}) \gamma} e^{i (\omega_{\textbf{k}_{2}} + \omega_{\textbf{k}_{1}}) \frac{\delta}{2}} e^{-i \textbf{k}_{2} \cdot \textbf{r}^{\prime}} e^{i \textbf{k}_{1} \cdot \textbf{r}}. 
\end{eqnarray*}

For a coherent ensemble of x-ray pulses with a coherence length larger than the size of the object, $\textbf{k}_{1}$ and $\textbf{k}_{2}$ can be  approximated to $\textbf{k}_{in}$. Under this approximation, the expression for DSP can be rewritten as
\begin{eqnarray}
\nonumber    \frac{dP}{d\Omega} &=& \frac{d\sigma_{\textrm{th}}}{d\Omega} \int^{\infty}_{0} d\omega_{\textbf{k}_{s}} \frac{\omega_{\textbf{k}_{s}}}{\omega_{\textbf{k}_{in}}}  \int^{\infty}_{- \infty} d \gamma \frac{I(\textbf{r}_{0}, \gamma)}{\omega_{\textbf{k}_{in}}} \int^{\infty}_{- \infty}  \frac{d \delta}{2 \pi}  C(\delta) e^{- i (\omega_{\textbf{k}_{s}} - \omega_{\textbf{k}_{in}}) \delta}  \\ \nonumber
&& \times \int d \textbf{r} \int d \textbf{r}^{\prime} \left\langle \Psi \left(\mathbf{r}, \gamma + \frac{\delta}{2}\right) \left| \hat{\rho} (\textbf{r}^{\prime}) ~e^{- i \mathcal{H}_{0} \delta}~\hat{\rho} (\textbf{r}) \right| \Psi \left( \mathbf{r}, \gamma - \frac{\delta}{2}\right)  \right\rangle  e^{-i \textbf{Q} \cdot (\textbf{r} - \textbf{r}^{\prime})},
\end{eqnarray}
where the pulse duration, $\tau$, dependent term is included in $C(\delta) = \exp(- \frac{2 \ln{2\delta^{2}}}{\tau^{2}})$ and $\textbf{Q} = \textbf{k}_{in} - \textbf{k}_{s}$ is the momentum transfer. 
The above equation is considered as the key equation for TRXD. 
Let us assume  $|E_{i} - \langle \mathcal{H}_{0} \rangle_{t} | \ll 1/\tau$
holds, where $E_{i}$ is  eigen-energy  
corresponding to $i^{\textrm{th}}$ eigenstate of the wavepacket, and $\langle \mathcal{H}_{0} \rangle_{t}$ 
is the mean energy of the wavepacket.  After performing the $\delta$-integral, the expression of the DSP for a short x-ray  pulse  reduces 
as~\citep{dixit2014theory}
\begin{eqnarray}
\nonumber    \frac{dP}{d\Omega} &=& \frac{dP_{e}}{d\Omega}\int^{\infty}_{0} d\omega_{\textbf{k}_{s}} \frac{\omega_{\textbf{k}_{s}}}{\omega_{\textbf{k}_{in}}} \sum_{f} \frac{\tau_{l}}{\sqrt{8 \pi \ln{2}}} e^{-(\tau^{2}_{l}/ 8 \ln{2})(\omega_{\textbf{k}_{in}} - \omega_{\textbf{k}_{s}} + \Tilde{E} - E_{f})^{2}}  \\ \nonumber
&& \times \int d \textbf{r} \int d \textbf{r}^{\prime} \left\langle \Psi(\mathbf{r}, {\mathsf{T}}) | \hat{\rho} (\textbf{r}^{\prime}) | \Phi_{f} \rangle \langle \Phi_{f} |\hat{\rho} (\textbf{r})| \Psi (\mathbf{r}, {\mathsf{T}}) \right\rangle \times e^{-i \textbf{Q} \cdot (\textbf{r} - \textbf{r}^{\prime})}.
\end{eqnarray}
Here, $ \frac{dP_{e}}{d\Omega} =   \frac{d\sigma_{\textrm{th}}}{d\Omega}  \int^{\infty}_{- \infty} d \gamma \frac{I(\textbf{r}_{0}, \gamma)}{\omega_{\textbf{k}_{in}}}$ is the DSP for a free electron and 
{\sffamily{T}} is the pump-probe time delay.  
For a small energy-transfer within the bandwidth of the x-ray pulse during the diffraction, 
$\omega_{\textbf{k}_{s}}$ can be approximated as  $\omega_{\textbf{k}_{in}}$.  Within this assumption,  
the final expression of DSP can be simplified as  
\begin{equation}{\label{eq:DSP_final}}
    \frac{dP}{d\Omega} = \frac{dP_{e}}{d\Omega} \sum_{f} \left| \int d\textbf{r} \langle \Phi_{f} | \hat{\rho} (\textbf{r}) | \Psi ({\mathsf{T}}) \rangle e^{-i \mathbf{Q} \cdot \mathbf{r}}\right|^{2}.
\end{equation}
Here, the summation over $f$ includes all zeroth-order states included in the dynamics, which amounts
to detect all the scattered photons within a given solid angle, irrespective of the mean photon energy.
In this thesis, we have used the above expression for DSP  to simulate the time-resolved diffraction signals  for different molecular systems.

Refs.\,\citep{kowalewski2017monitoring, simmermacher2019theory} discussed 
different  roles of the time-independent 
and time-dependent diffraction signals  to  the total diffraction signal. Moreover,  a detailed discussion about probing electronic coherences in the  wavepacket have been documented in Refs.~\citep{dixit2012imaging, simmermacher2019electronic}.  Furthermore, the time resolution in TRXD depends on several factors like temporal duration on x-ray pulse, jitter between pump and probe pulse, pump-probe time delay etc.

\section{Technical Details}{\label{Sec:tech}}
In this thesis, ultrafast laser-induced charge migration is simulated using the hybrid  TDDFT/CI 
methodology \citep{klinkusch2016resolution, hermann2016ultrafast}.
This combination of methods provides a good balance between accuracy and computational efficiency.
A generic $N$-electron wavepacket, $| \Psi(\mathbf{r},t) \rangle$, is represented at any given time $t$ as 
a linear combination of the ground state Slater determinant and singly excited many-body excited states as given in Eq.~(\ref{eq:TD_wavefunction}).

In the hybrid TDDFT/CI method, all expansion coefficients $\{D_{0,k},D_{a,k}^r\}$ are obtained from LR-TDDFT. This step is performed using  standard quantum chemistry programs.
The information from the quantum chemistry package is post-processed using the open-source toolbox detCI@ORBKIT.
The contributions of the different Slater determinants extracted from the quantum chemistry programs are first 
pruned by removing all contributions below some numerical threshold (here chosen as 0.001), and then renormalized. In the basis of pseudo-CI eigenfunctions chosen for the propagation, the Hamiltonian is 
considered diagonal. The matrix elements of the dipole moment operator in Eq.\,\eqref{eq:Interaction_Hamiltonian} are 
computed by numerical integration. Finally, by using Eqs.\,\eqref{eq:Tot_Hamiltonian}, ~\eqref{eq:Interaction_Hamiltonian}, and ~\eqref{eq:TD_wavefunction}, the time-evolution of the coefficients $C_k(t)$ in Eq.\,\eqref{eq:TD_wavefunction} is performed by direct numerical integration of Eq.\,\eqref{eq:TDSE}
using a preconditioned adaptive step size Runge-Kutta algorithm \citep{tremblay2004using}.

Integrating the time-dependent many-electron wavepacket in Eq.\,\eqref{eq:TD_wavefunction} 
leads to the one-electron quantum continuity equation as given in Eq.\,\eqref{eq:continuity}.
From the many-body wave function, the expectation value of the one-electron density operator yields the one-electron electronic density [see Eq.\,\eqref{eq:rho_operator}].
The expectation value of the operator $\hat{j} \left( \mathbf{r}\right)$ in Eq.\,\eqref{eq:j_operator} yields the flux density.
The molecular orbitals to compute the integrals are expressed as linear combinations of atom-centered Gaussian functions, and all integrals are computed analytically. 
The electronic charge and flux densities  thus obtained satisfy very accurately the continuity equation, Eq.\,\eqref{eq:continuity}. 
Details on the computation of the one-electron integrals from the many-body wave function using the detCI@ORBKIT
toolbox are given in Refs.~\citep{hermann2016orbkit,pohl2017open, hermann2017open}.

It is assumed that the the probe x-ray pulse duration is much shorter than the timescale of the induced 
charge  migration. 
Thus, TRXD signal is simulated using the expression for  DSP as given in 
Eq.\,\eqref{eq:DSP_final}~\citep{dixit2014theory}.
The signal probes the coherences between eigenstates of the unperturbed Hamiltonian, $| \Phi_{k}  \rangle$,
contained in an electronic wavepacket.  

All many-electron dynamical simulations are performed using in-house 
codes~\citep{tremblay2011dissipative, tremblay2004using, tremblay2008time, tremblay2008guided, tremblay2008time}.
The integrals required to compute the TRXD signals and electronic flux densities are calculated using ORBKIT~\citep{OKgit}.
Mayavi \citep{ramachandran2011mayavi} and Matplotlib \citep{hunter2007matplotlib} are used to visualize the electronic flux densities  and the time-resolved x-ray signals, respectively.

\cleardoublepage
\chapter{Charge Migration in Heterocyclic Five-Membered Planner Molecules}

Recent developments in laser technologies facilitate  to generate intense ultrashort  laser pulses. 
The availability of such laser pulses  allows us to study ultrafast  processes in molecules on their natural timescale.  
During laser-molecule interaction, electrons are the first to respond the action of an ultrashort laser 
pulse. 
As a result of this interaction,  
an electronic wavepacket is created, which sets an electronic charge distribution into motion. 
The migration of  electronic charge around the atoms of the molecular scaffold 
affects the outcomes of chemical reactions and biological functions~\citep{cederbaum1999ultrafast,  kuleff2005multielectron, breidbach2003migration, remacle2006electronic, breidbach2005universal, nisoli2017attosecond, bredtmann2014x}. 
Thus, inducing and controlling the charge migration in molecules  have emerged as one of the cornerstone topics of attosecond science in recent years~\citep{folorunso2021molecular, kraus2015measurement, calegari2014ultrafast, he2022filming}.  

%%%%%

Charge migration of valence electrons and corresponding electronic flux density (EFD) in ring-shaped molecules have received ample 
attention in recent years.  Attosecond charge migration in magnesium porphyrin has been   
studied by means of EFD~\citep{nam2020monitoring, tremblay2021time, barth2006periodic, barth2006unidirectional, koksal2017effect}. Moreover, 
multidirectional angular EFD has been investigated by selective excitation in benzene~\citep{hermann2016multidirectional, jia2017quantum, hermann2017attosecond, hermann2020probing}. 
At this juncture, it is natural to envisage how the nature of the charge migration will alter 
by transiting from six-membered benzene to five-membered 
ring-shaped molecules. This is the main aim of this chapter.  
For this purpose, pyrrole, furan and oxazole, which are  five-membered neutral ring-shaped aromatic molecules, will serve our purpose. 
These three heterocyclic aromatic molecules are present in many important biological systems as a fundamental unit and have enormous applications~\citep{demingos2021first}. 
Series of detailed calculations for electronic properties of pyrrole, furan and oxazole have been 
performed~\citep{christiansen1999electronic, roos2002theoretical, lee1996molecular, serrano1993theoretical, wan2000electronic, burcl2002study, palmer1995electronic, christiansen1998electronic}. Also, few non-equilibrium dynamics in three molecules have been documented in Refs.~\citep{cao2016excited, geng2020time, carrascosa2021mapping, barbatti2010non, barbatti2006nonadiabatic, yong2021ultrafast}. 

This chapter provides a detailed and comparative study of adiabatic attosecond charge migration of valence electrons in 
pyrrole, furan and oxazole. 
In the following, we will discuss  how the charge migration is sensitive to the symmetry reduction, i.e., progressing 
from benzene with $\mathcal{D}_{\textrm{6h}}$ symmetry to these five-membered ring-shaped  
molecules with $\mathcal{C}_{\textrm{2v}}$ and $\mathcal{C}_{\textrm{s}}$ symmetries. 
Furthermore, we will investigate how the nature of electronic charge  and flux densities  during charge migration
is altered by the presence of foreign atoms, i.e.,  nitrogen in pyrrole, oxygen in furan; and nitrogen and oxygen in oxazole.   

In this chapter, to achieve the convergence of the laser-induced charge migration simulations, 
$N_\textrm{states} = $ 24, 22, and 32  lowest-lying bound excited states are used for pyrrole, furan and oxazole, respectively. 
The CAM-B3LYP functional and aug-cc-pVQZ basis sets on all atoms are used to compute these electronic states~\citep{yanai2004new, dunning1989gaussian}. All electronic structure calculations are preformed with the Gaussian16 program~\citep{frisch2016gaussian}.

\section{Results and Discussion}
To compare the charge migration in pyrrole, furan, and oxazole on  equal footing, a few conditions are considered.
The first one is a static consideration, where the target excited states are chosen to have a similar character.
In particular,  the selected states must contain a strong signature of the lower symmetry in oxazole, and
the effect of the electronegativity of the foreign atoms must be readily observable.
This can be achieved by exciting states that create nodal structure around the foreign atoms. 
Further, in order to populate a target excited state, it must be optically accessible. That is, the intensity of the transition from the ground to target excited state (proportional to the transition dipole moment $\mu^{2}$)
must be high. From a dynamical perspective, charge migration should be faster than the timescale of nuclear rearrangement, ideally in the sub-femtosecond regime, yet accessible with a few-femtosecond laser pulse. The target excited states satisfying both constraints will thus have similar energies and transition dipoles. Further, understanding the dynamics of the electronic density and EFD after such a pump pulse is greatly simplified if the wavepacket is composed of only a few components in the excited state manifold. This is achieved by exciting only a small fraction of electronic populations and optimizing the laser parameters (field intensity and carrier frequency), thereby leading to a controlled degree of charge migration.

\begin{figure}[b!]
\includegraphics[width = \linewidth]{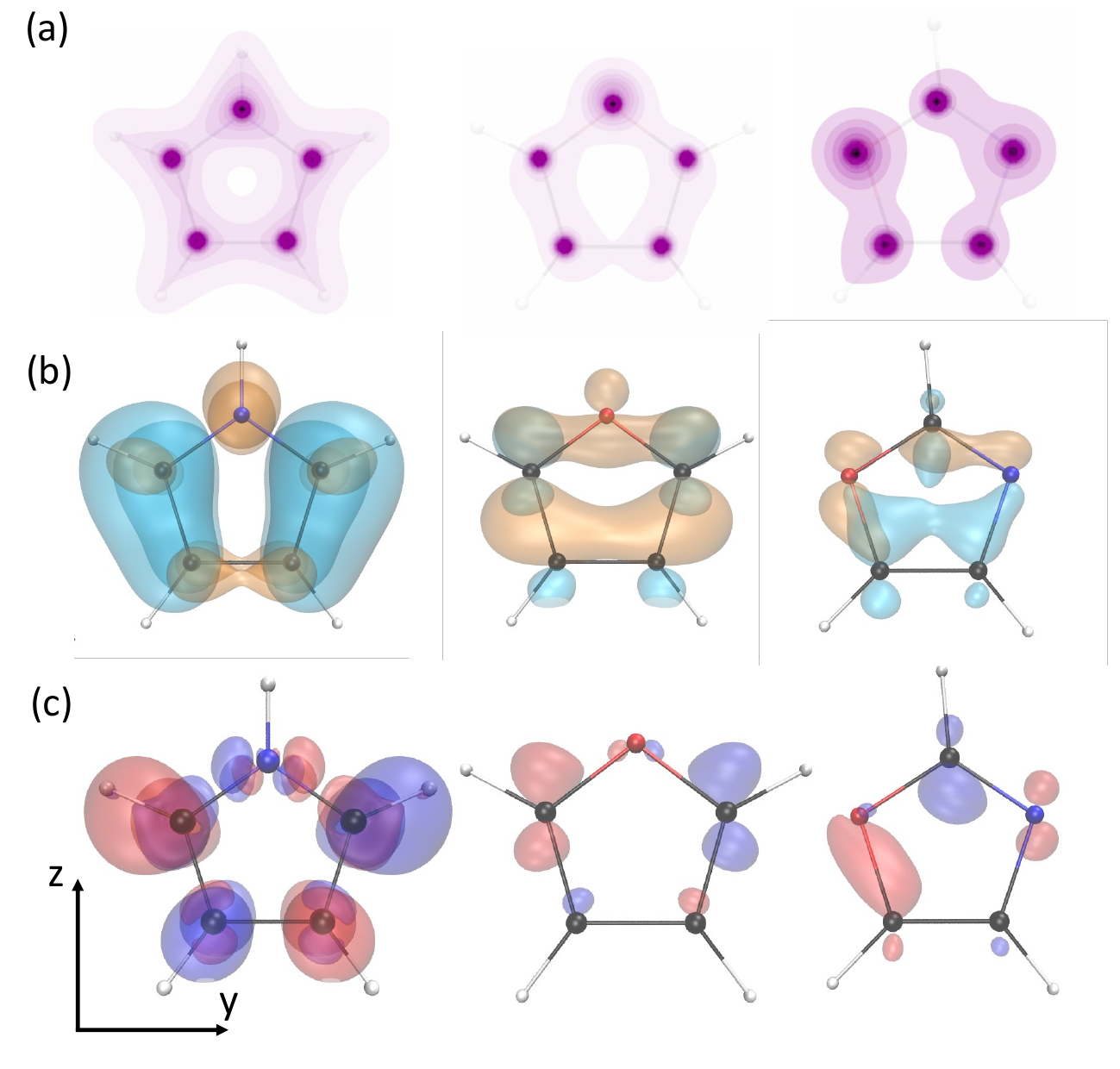}
\caption{Depiction of the electronic structure in the five-membered ring molecules. 
Pyrrole,  furan, and oxazole are represented from left to right. 
(a) Ground state electron density (isocontours equally separated by 0.001\,a.u.).
(b) Natural transition orbitals. Orange and blue surfaces represent the particle and hole densities, respectively. 
(c) Difference electron densities between target excited and ground electronic states, $\rho_{m}(\mathbf{r})-\rho_{0}(\mathbf{r})$. Red (blue) colour represents regions of density depletion (increase).
The isosurface values are $\pm0.005$, $\pm0.02$, and $\pm0.03$ for pyrrole, furan, and oxazole, respectively for both NTOs and transition densities.}
\label{fig31}
\end{figure}

The charge distribution corresponding to the ground electronic state of the three molecules is presented in Fig.~\ref{fig31}(a). 
All three molecules are projected in the $yz$ plane throughout this chapter. 
Pyrrole and furan are symmetric with respect to reflection about the $z$ axis that bisects molecule through the foreign atoms and the opposing CC bond, 
and the ground state densities both transform according to the totally symmetric point irreducible representation (IRREP) of the $\mathcal{C}_{\textrm{2v}}$ point group.
The presence of the nitrogen atom (blue) appears to favour delocalization of the electrons over the whole ring in the $\pi$-orbitals.
The more strongly electronegative oxygen atom (red) leads to a more compact ground state density in furan (central panel), which is more concentrated 
at the foreign atom.
Due to the presence of both oxygen and nitrogen atoms on the two sides of the pentagon, oxazole is not symmetric with respect to a reflection in the plane
and retains only the operations of the $\mathcal{C}_{\textrm{s}}$ point group. 

\begin{figure*}[tb!]
\includegraphics[width = \linewidth]{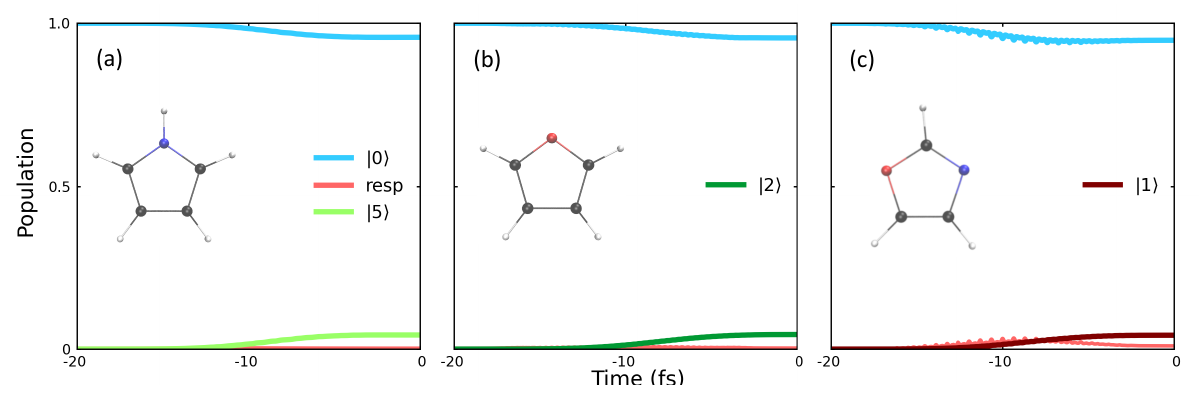}
\caption{Laser-induced population dynamics  in heterocyclic five-membered ring molecules
triggered by 20\,fs sine-squared  linearly polarized pulse along $y$ axis.
The label ``resp'' indicates  the residual population of all other electronic states, which acts as a non-linear response during the excitation.
The three molecules are shown in ball-and-stick representation, which are projected in the $yz$ plane throughout in this work.
White, grey, blue and red colours represent hydrogen, carbon, nitrogen and oxygen, respectively.
(a) Pyrrole: carrier frequency resonant with the energy of the 5$^{\text{th}}$ electronic state (6.26 eV); peak intensity of 3.7$\times$10$^{13}$ W/cm$^{2}$.
(b) Furan: carrier frequency resonant with the energy of the 2$^{\text{nd}}$ electronic state (6.26 eV); peak intensity of 2.8$\times$10$^{13}$ W/cm$^{2}$.
(c) Oxazole: carrier frequency resonant with the energy of the 1$^{\text{st}}$ electronic state (6.37 eV); peak intensity of 7$\times$10$^{13}$ W/cm$^{2}$.}
\label{population}
\end{figure*}

To identify interesting target state in each molecule, the natural transition orbitals (NTOs) of 
all states up to the ionisation threshold have been inspected.  
The NTOs of the lowest excited state accessible via a linearly polarised field along the y axis are shown in Fig.~\ref{fig31}(b). 
These correspond to the fifth, second and first excited states in pyrrole, furan, and oxazole, respectively. 
The transition intensities of these target excited states are much higher than the transition intensities from the ground to all other excited states of the respective molecule. 
NTOs offer a compact depiction of electronic transitions from a reference state in terms of particle and hole densities~\citep{martin2003natural}. 
In all three molecules, the excitation indeed creates nodal structure around the foreign atoms but the excitation characters are radically different.
The hole density (blue) in pyrrole is found to be delocalized on both sides of the ring on the carbon atoms, with a clear $\pi$ character and a node along the $y$ axis.
The particle density (orange) is more strongly localized close to the atoms of the ring, with a $\pi$ character around the nitrogen atom.
The structure is very different in furan, where the hole density is localized around the carbon atoms and exhibits a stronger 
in-plane bonding character (reminiscent of a $\sigma$ character, in opposition to an out-of-plane 
$\pi$-character).
The same character is found in the particle density, which is delocalized along the C-O-C fragment, as well as through space between the two C=C bonds.
Both excitations in pyrrole and furan retain the structure of the $\mathcal{C}_{\textrm{2v}}$ point group.
The NTO densities in oxazole are more similar to the furan case, although it is the hole density that has the through-space structure.

The difference densities between the ground and target states are also shown in Fig.~\ref{fig31}(c).
The strong $\pi$-type characters, i.e., the out-of-plane bonding characters of the excitation in pyrrole is noticed. 
In opposition to $\pi$ character, a clear nodal structure in the plane of the molecule analogous to
 the $\sigma$ character of the excitations in furan and oxazole
can be readily recognized. 
More importantly, the difference densities reveal the symmetry of the excited states,
which are found to transform according to the B\textsubscript{2} IRREP of the $\mathcal{C}_{\textrm{2v}}$ point group
for pyrrole and furan. These important differences should lead to markedly different charge migration dynamics after laser excitation.

Keeping in mind the dynamical requirements described above, we have employed a linearly polarized pulse along the $y$ direction to induce the charge migration.
The resultant population dynamics is presented in Fig.~\ref{population}.
The pulses for three molecules have comparable peak intensities, with similar carrier frequencies tuned at the respective target transitions.
The peak intensities of the pulses, in excess of 10$^{13}$W/cm$^2$, are tuned to obtain similar population in the target states,
hence leading to a similar contrast in the charge migration patterns.
As evident from the figure, the target excited states reach around $4 \%$ population in all three cases.
The values of the transition dipole ($\mu^{2}$) to the target excited states are 0.9898 $ea_{0}$, 
1.1071 $ea_{0}$, and 1.0048 $ea_{0}$ for pyrrole (fifth), furan (second), and oxygen (first), respectively. 
This compares favourably to all other transitions with values lower than 0.25 $ea_{0}$ in all three molecules.

The population in all off-resonant electronic states are termed as ``resp'' in the figure, as they mediate the non-linear response of the 
electron density to the strong external laser pulse.
Electronic response is found to be more important in oxazole than in either pyrrole or furan.
During the pulse, the population of these non-resonant states is exactly zero for pyrrole and furan, but
non-zero  in oxazole.
This is because of the strong transition dipole moments to other excited states in oxazole, which stems from its different chemical nature compared to the other two molecules.
The field-free charge migration sets into motion at time zero, after the pulse. 
Since the target excited states in the three systems have similar excitation energies with respect to the ground electronic state,
the charge migration will have a similar oscillation period around 650\,as.

\begin{figure*}[t!]
\includegraphics[width =  \linewidth]{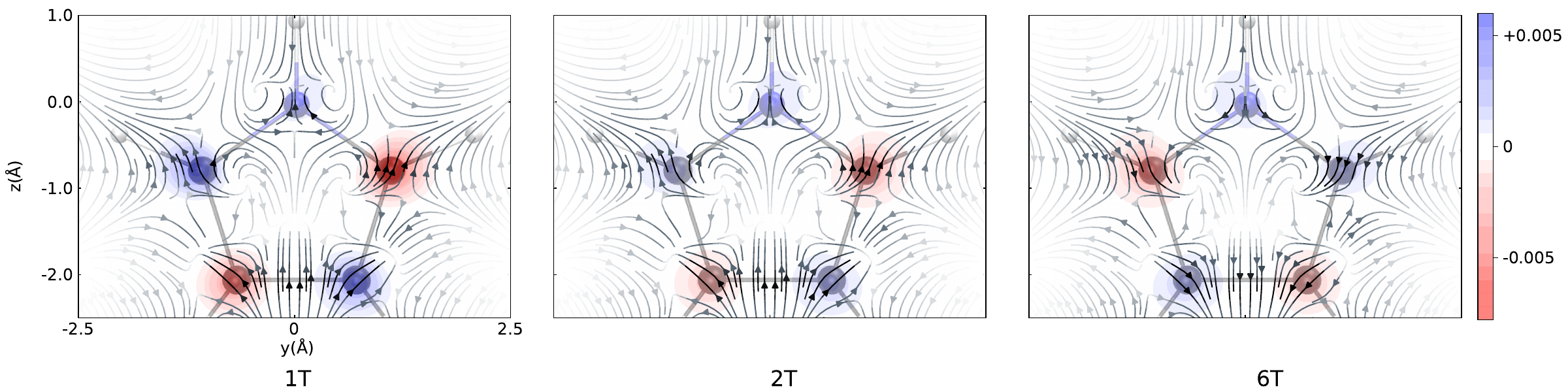}
\caption{Time-dependent difference electron and associated flux densities during 
charge migration in pyrrole at 1{\sffamily{T}}, 2{\sffamily{T}} and 6{\sffamily{T}}.
Here, {\sffamily{T}} = ${\tau}/8$ with  
$\tau =  660$ attoseconds  as  the characteristic timescale of the charge migration in pyrrole.
Ground-state electron density is subtracted  in the time-dependent difference electron density at each subsequent time step. 
The colour of the streamline arrows varies from white to black according to their increasing intensity. 
The red and blue contours present the depletion and the enhancement of the difference electronic density with respect to the ground state density. 
For better visualisation, the projection of pyrrole in the $yz$ plane is superposed in ball-stick representation.}
\label{pyrrole_stream}
\end{figure*}

To shed light on the effect of symmetry and electronegativity on field-free charge migration on its characteristic timescale,
we analyze snapshots of the time-dependent difference electron density and the associated EFD.
For the former, the contribution from the electronic ground state is subtracted from the wavepacket at each time step.

Figure~\ref{pyrrole_stream} presents the the difference density and the EFD at three time steps for pyrrole. 
The overall behaviour of the difference densities at all times reflects that they are mostly located  
around the atoms, which indicates charge localization.
This corresponds well to the picture offered by the density difference in the left panel of Fig.~\ref{fig31}(c).
The strong charge localisation around the atoms should not affect strongly the bonding properties in pyrrole. Consequently, it is unlikely that the excitation chosen would lead to any bond breaking, 
ring-opening or other photochemical reaction.
Moreover, the difference densities  are positive and negative in alternate carbon atoms, whereas  
nitrogen is always surrounded by the positive electron density. 
At 1{\sffamily{T}} and 2{\sffamily{T}}, the positive contour around  nitrogen  is tilted towards the positive $y$ axis
as evident from the figure.  This situation is reversed at 6{\sffamily{T}}. 
The difference densities around carbon atoms are anti-symmetric about $y$ axis at all time. 
The overall intensity of the difference density reduces from 1{\sffamily{T}} to  2{\sffamily{T}}. 
There is a phase shift  of ${\pi}/{2}$ from  2{\sffamily{T}} to 6{\sffamily{T}}, which is in accordance with the 
half time-period  of the charge migration as evident from the figure. 
The characteristic timescale for the  charge migration in pyrrole is $\tau=  660$ attoseconds.
%for the specific superposition state created here. 

\begin{figure*}[t!]
\includegraphics[width = \linewidth]{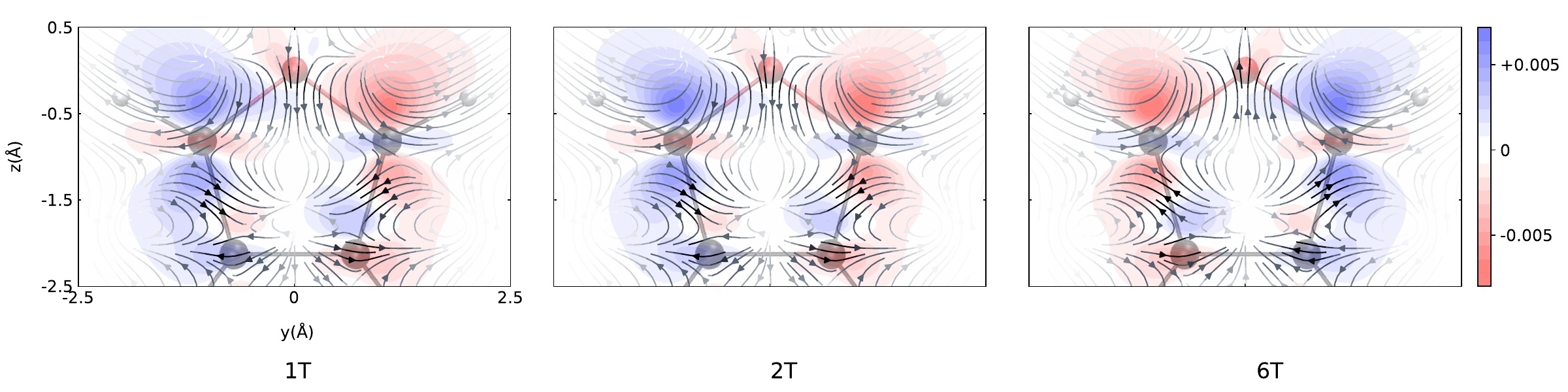}
\caption{Same as Fig.~\ref{pyrrole_stream} for furan  with $\tau =  660$ attoseconds  as  the characteristic timescale.} 
\label{furan_stream}
\end{figure*}

The electron flow can be understood better in terms of the EFD. 
As evident from Fig.~\ref{pyrrole_stream}, the direction of the electron flow is mostly around the atoms. 
The streamlines start from the atomic positions and terminate around the bonds at all times.
This reflects the atom-to-bond charge migration in pyrrole, which corresponds well to the picture offered by the NTO in Fig.~\ref{fig31}(b).
Indeed, the particle density is observed to be strongly localized around the atoms.
Consequently, the streamlines appear as piecewise disjoint lines of flow around each atom.  
From a symmetry point of view, there is no general electron flow from left to right or from upward to downward.
At 1{\sffamily{T}}, the flow around the bottom two carbon atoms is upward inside and downward outside of the pyrrole ring. 
Similar to the difference densities, the streamlines follow the same phase reversal from 2{\sffamily{T}} to 6{\sffamily{T}}. 
The EFD transforms according to the the totally symmetric IRREP of the  $\mathcal{C}_{\textrm{2v}}$ point group at all times.

At this juncture it is natural to ask how the above findings will alter if one replaces nitrogen by a more electronegative atom, say oxygen. 
To explore the question further, let us analyze the difference density and the EFD  for furan  at three time steps as presented in Fig.~\ref{furan_stream}. 
The electronic ground state density and the NTO of the target transition in Fig.~\ref{fig31} already indicate  
a significant change in the charge migration dynamics.
Unlike for pyrrole, the positive and negative contours of the difference electron densities are distributed not only around the atoms but also around the bonds,
and the latter is higher in intensity. 
They are anti-symmetric about the $y$ axis except around the oxygen atom. 
These findings are in agreement with the density difference profile observed in  Fig.~\ref{fig31}(c).

The bond strengthening (weakening) shown by this increase (decrease) in the electronic densities 
around the bonds is primarily visible on the two sides of the oxygen atom. In contrast to pyrrole, these will affect the bond strength and potentially cause bonds to break, resulting in ring opening on either side of the foreign atom. 
Exciting furan with short, intense laser pulses of the form used in this work is therefore more likely to trigger photochemical processes on longer timescales.
Interestingly, the difference density around the oxygen is negative at all times, reflecting the larger electronegativity of oxygen compared to the carbon atoms of the ring.
Since the oxygen substitution preserves all operations of the $\mathcal{C}_{\textrm{2v}}$ point group, the density difference upon charge migration in furan
retains the same B\textsubscript{2} IRREP as for pyrrole.

From the EFD for furan displayed in Fig.~\ref{furan_stream},  
a swirling type of motion between the atoms of the ring is evident.
The streamlines show a flow downward through the bonds between the oxygen and its nearest carbon atoms, and going upward to the hydrogen atoms at time steps  1{\sffamily{T}} and  2{\sffamily{T}}. 
The direction of the flow is completely reversed at  6{\sffamily{T}}. 
These patterns are consistent with the NTO picture in  Fig.~\ref{fig31}(b),
which show a stronger in-plane  bonding character and more nodal structure in the plane of the molecule for the target excitation.
Despite the apparent complexity of the flow patterns, they transform according to the the totally symmetric IRREP of the  $\mathcal{C}_{\textrm{2v}}$ point group at all times.
It can be concluded that the stronger in-plane  bonding 
character of the target excitation in a ring bearing a more electronegative atom does not
affect the symmetry of the charge migration.

\begin{figure*}[hbt!]
\includegraphics[width = \linewidth]{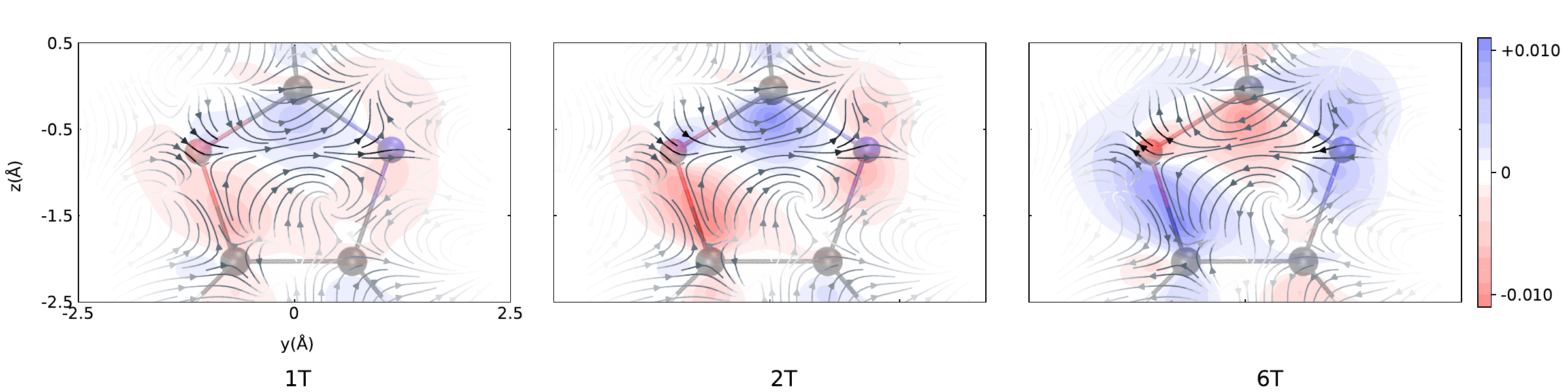}
\caption{Same as Fig.~\ref{pyrrole_stream} for oxazole with $\tau =  650$ attoseconds  as  the characteristic timescale.} 
\label{oxazole_stream}
\end{figure*}
For the charge migration in pyrrole and furan, both molecules belong to $\mathcal{C}_{\textrm{2v}}$ point group.  
The symmetry of the five-membered ring can be reduced by substitution of two carbon atoms by both nitrogen and oxygen.
To understand this further symmetry reduction, let us explore the charge migration in oxazole, which has 
$\mathcal{C}_{\textrm{s}}$ symmetry. 

As a result of the symmetry reduction, 
the charge migration in oxazole  changes drastically as two carbon atoms are replaced by one nitrogen and one oxygen atoms on the opposite sides of the pentagon ring (see Fig.~\ref{oxazole_stream}).
Due to the presence of the different atoms in the ring, 
the difference electron densities are entirely different in comparison to  pyrrole and furan, 
and do not exhibit any symmetric behaviour upon rotation with respect to the  $z$ axis or reflection in the $xz$ plane.
The regions of the density increase/depletion are not specific to the bonds or atoms. 
The dumbbell structure observed at all time step in the difference electron density around nitrogen atom hints at a strong contribution
of a $p$-type orbital in the molecular plane. 
At  1{\sffamily{T}},   the negative contours are present around the bonds and positive contours are situated in the space between nitrogen, oxygen and  carbon atoms.
In general, there is no strong correlation between the difference density plots during charge migration and the ones reported in Fig.~\,\ref{fig31}(c).

To have a better understanding of the charge migration mechanism in oxazole,
let us analyze the time-resolved EFD after laser excitation, as depicted in Fig.~\ref{oxazole_stream}. 
The multiple directions of the flux densities provide a picture of the charge migration that is consistent with the NTO densities depicted in Fig.~\,\ref{fig31}(b).
A swirling motion is observed, separated in bottom and top contributions that barely exchange electrons at all times.
This is intriguing, since the excitation using a $y$ polarized pulse would likely create nodal structure along this direction.
As such, a left-right separation of the charge migration should be expected, as was the case for pyrrole and furan.
From the NTOs in Fig.~\,\ref{fig31}(b), it appears that the top part of the particle density only migrates towards the hole density
located between the two foreign atoms. We attribute this feature to their electronegativity, that prefer to retain the excited electron
localized in this fragment.
On the other hand, the bottom part of the charge migration sees a transfer of the particle localized on the CO bond to the CC bond across the ring.
This part of the charge migration operates through space rather than following the bonds, in stark contrast with the more symmetrical pyrrole and furan.
The bonds connecting to the oxygen atom experience the strongest bond strength oscillations, 
whereas bonds connected to the nitrogen atom see less pronounced fluctuations. 
Much like the case of furan, it can be inferred from the electronic density and EFD snapshots that oxazole could probably undergo further photochemical reactions.

\section{Summary}  
In summary, we have investigated the role of symmetry reduction and of electronegativity on adiabatic attosecond charge migration in 
selected heterocyclic five-membered aromatic ring molecules.
Time-dependent transition electron  and  flux densities were used to unravel the mechanism of charge migration
induced by selective laser pulses. 
To compare the charge migration on three molecules on equal footing, transition with similar energetics were used to reach target excited states with high transition intensities and nodal
structure close to the foreign atoms.
It has been found that laser-induced charge migration and the corresponding EFD are significantly different in the three molecules in spite of these constraints.
Moreover, variations in electronic densities and EFDs reveal bond strengthening or weakening upon excitation by light, which is useful to infer photochemical potential in these molecules. 
For the laser pulses used in this chapter, 
it was inferred that pyrrole is less likely to undergo ring-opening reactions than furan and oxazole.

The presence of the nitrogen in pyrrole and oxygen in furan make the charge migration in both molecules significantly different. 
In pyrrole, a signature of charge  localization is observed during charge migration  as the difference electron 
densities are positioned  around the atoms.
In contrast, electron density depletion/increase upon light-induced charge migration in pyrrole are delocalized around the atoms and around the bonds,
which indicate bond-to-atom charge transfer in furan. The charge migration patterns observed in both cases are totally symmetric.
For oxazole, the difference electron densities documenting the charge migration dynamics display no symmetry outside of the $yz$ plane
as a result of the lower symmetry, from $\mathcal{C}_{\textrm{2v}}$ in furan and pyrrole to $\mathcal{C}_{\textrm{s}}$. 
In this case, charge migration takes the form of  a swirling motion through space. 

We believe that our results on attosecond charge migration in heterocyclic five-membered rings will motivate
further theoretical investigations in other types of molecular systems, where studies of structure-migration relationships
are still few and far apart.
Further, state-of-the-art emerging experimental techniques, such as attosecond transient-absorption spectroscopy~\citep{kraus2015measurement, calegari2014ultrafast}, 
high-harmonic generation spectroscopy~\citep{folorunso2021molecular, he2022filming, dixit2018control, chandra2019experimental}, 
or time-resolved x-ray diffraction~\citep{dixit2012imaging, dixit2014theory}, to name but a few,
will provide potent probes for our findings in heterocyclic five-membered ring molecules in coming future.

\cleardoublepage
\chapter{Role of Structural Saddling on Charge Migration}\label{Chapter4}

Technological advancements in recent years have allowed  us to control the synthesis of different types of corroles with interesting coordination chemistry. 
The properties of the central metal in metal-corroles are always the focus of ample attention. 
Due to the intriguing coordination chemistry and its compatibility with a variety of transition metals~\citep{nardis2019metal}, 
metal-corroles have demonstrated potential for applications as 
photosensitizers~\citep{jiang2019corrole, mahammed2019corroles} and 
catalysts~\citep{dogutan2011electocatalytic, gross2000epoxidation, mahammed2003aerobic}, to name but a few.
Copper corrole is one of the first systems bearing a transition metal in which copper was experimentally inserted in the N$_4$ coordination core of the corrole ring. 
Unlike other metal-corrole complexes, 
copper corrole is found to have a saddling in its equilibrium structure (see Fig.~\ref{fig41}), which makes this molecule very interesting for different applications~\citep{ghosh2000electronic, luobeznova2004electronic, broring2007revisiting, pierloot2010copper}. 
A combined study of x-ray absorption spectroscopy and TDDFT have been  performed 
to understand the electronic structure of copper corrole~\citep{lim2019x}. 
Moreover, the role of the  saddling on the electronic structure properties of low-lying electronic states in copper corrole
was explored by analysing the results of planar and saddled geometries~\citep{pierloot2010copper}. 
Furthermore, static x-ray diffraction with  density functional analysis confirmed the  saddling 
feature of copper corroles~\citep{alemayehu2009copper, broring2007revisiting}. 
Several theoretical works  have been carried out to understand the saddling in copper corroles~\citep{ghosh2000electronic,luobeznova2004electronic, alemayehu2009copper, lim2019x}. 

\begin{figure}[h!]
\includegraphics[width = \linewidth]{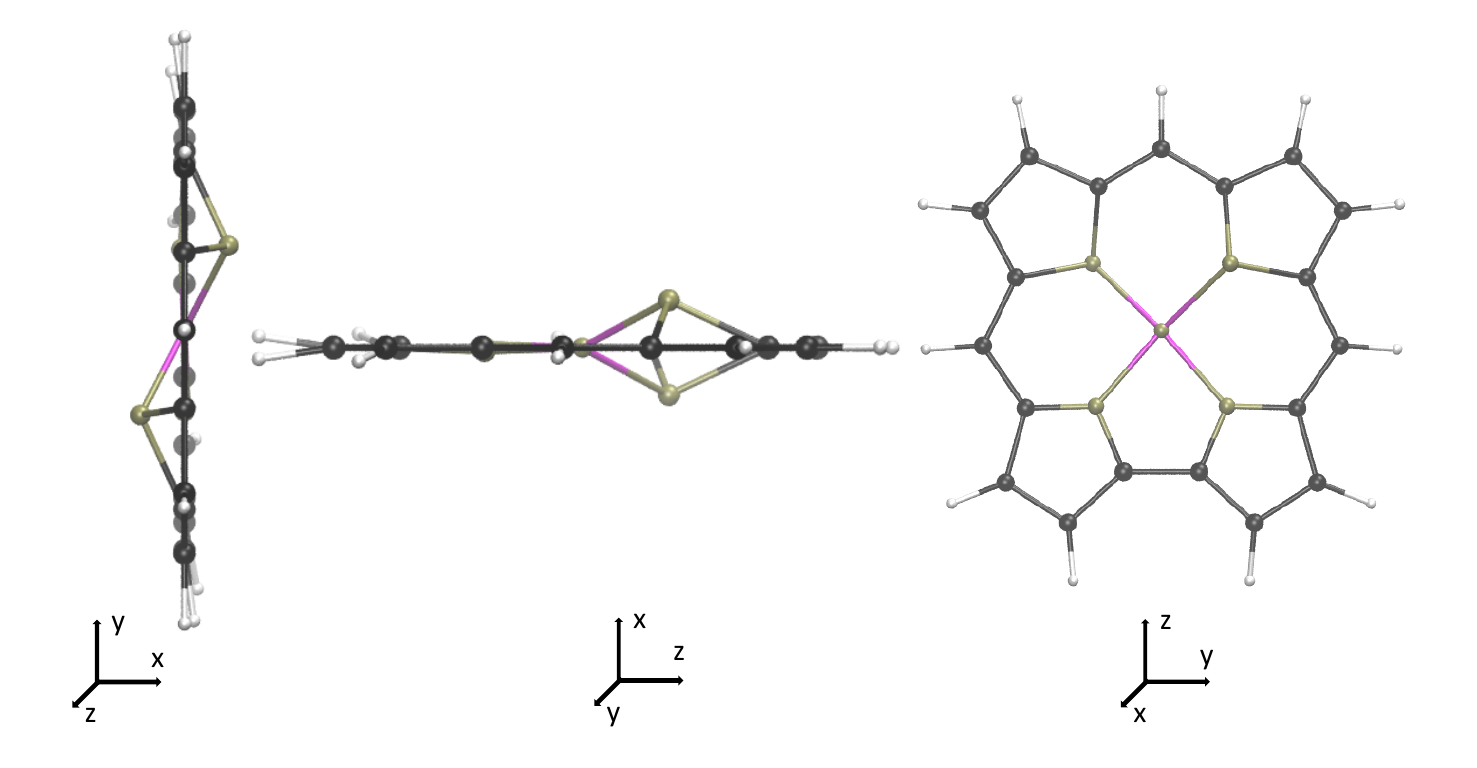}
\caption{Ball-stick representation of unsubstituted  copper corrole in different orientations, with 
the molecule  in the $yz$ plane.
Magenta, tan, black, and white spheres represent copper, nitrogen, carbon, and hydrogen atoms, respectively.} \label{fig41}
\end{figure}

{\it A priori}, it is not obvious how the  saddling in a molecular structure will affect the charge migration on attosecond timescale, during which the effect of nuclear vibrations is insignificant.
Moreover, what would be the signature of the structural saddling in any experimental probe signal, if any? 
The main aim of this chapter is to investigate such crucial questions. 
In this chapter, we discuss time-resolved imaging of charge migration in an unsubstituted copper corrole within a pump-probe configuration. 
An ultrashort pump pulse induces charge migration in copper corrole, which is imaged by TRXD 
at various pump-probe delay times.  
 Additionally, we will analyse time-dependent EFDs
to understand the mechanistic details of the charge migration associated with the induced dynamics. The time-dependent  EFDs provide an additional information 
about the direction of the electron flow during charge migration in copper corrole. 

In this chapter,  $N_\textrm{states}=80$ lowest-lying excited states below the ionization threshold are used to achieve the convergence of the 
excitation dynamics induced by the pump pulse in copper corrole. 
The states are computed using the CAM-B3LYP functional~\citep{yanai2004new} 
and aug-cc-pVDZ basis sets~\citep{dunning1989gaussian} on all atoms,
as implemented in Gaussian16~\citep{frisch2016gaussian}.

\section{Results and Discussion}

The molecular structure of the unsubstituted copper corrole 
in ball-stick representation is shown in Fig.~\ref{fig41}.
All axes and coordinates are described in the molecular frame of reference,
and it is assumed that the molecule is oriented in the $yz$ plane of the laboratory frame. 
From the projections in the two other planes, the saddling in copper corrole is evident, 
which makes this particular type of corrole  appealing for detailed investigations. 

To trigger the charge migration, 
a 10 fs cosine-squared  linearly polarized pulse along the $y$ axis is employed.
The wavelength and the peak intensity of the pump pulse are chosen to 360 \,nm  and 
3.5$\times$10$^{13}$ W/cm$^{2}$, respectively.
%For an ionization potential of about 6.78\,eV, the Keldysh parameter is found to be 2.8, such that electron loss to the ionization channel via multiphoton processes is not dominant. 
It has been experimentally demonstrated that a coherent transfer of population from ground to excited states is possible with such intensity in molecular systems~\citep{prokhorenko2005coherent}.
The resulting population dynamics of the electronic states during the pump pulse is presented in Fig.~\ref{fig42}. 

\begin{figure}[h!]
\includegraphics[width= \linewidth]{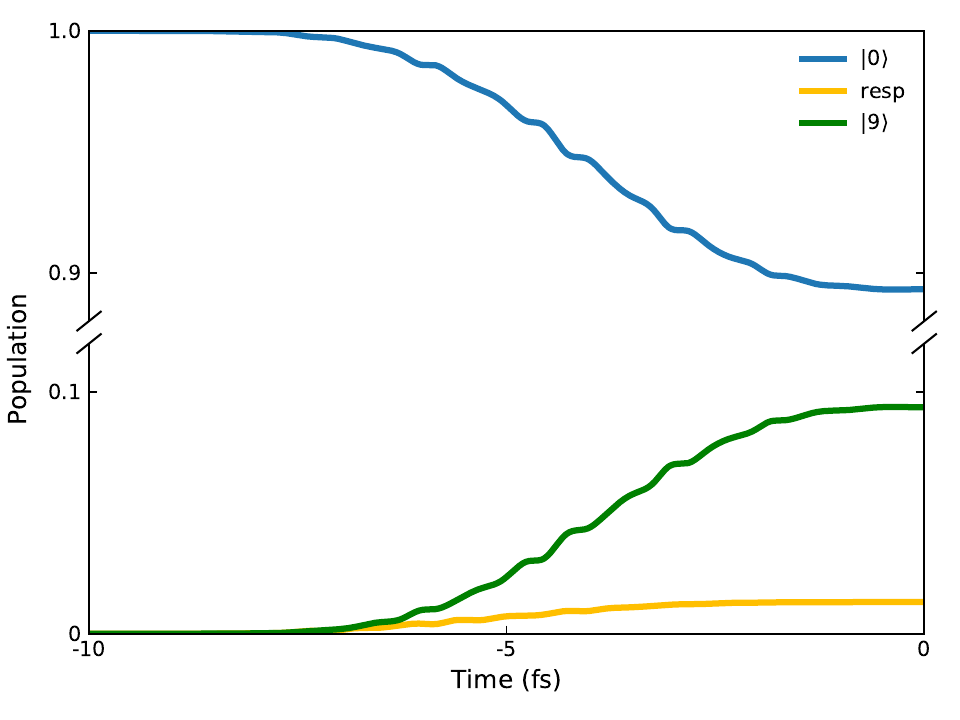}
\caption{Population dynamics of selected electronic states of copper corrole. 
 $\vert 0 \rangle$ represents the ground electronic state, and $\vert 9 \rangle$ represents 
the target electronic state with energy $E_{9}$ = 3.45 eV.
The orange line labeled ``resp'' represents the electronic response in the presence of the field, leading to the 
residual population of the rest of the electronic states after the excitation.
A cosine-squared linearly polarized pulse of 360 \,nm wavelength with peak intensities of 
3.5$\times$10$^{13}$ W/cm$^{2}$ is used to induce the population dynamics.  
The pulse is 10\,fs short and polarized along the $y$ axis.  
Time zero defines the onset of field-free electron dynamics.} \label{fig42}
\end{figure}

Transfer of population from the electronic ground state to other excited states starts as the pump pulse interacts  and 
the electron dynamics in  copper corrole sets into motion.  
As evident from Fig.~\ref{fig42}, only a small amount of population around $9.4\%$ is transferred to the ninth  electronic excited state,
and most of the population remains in the ground state, i.e., $89\%$ at the end of the pulse. 
Small population transfer from the ground to excited states is common in various  excitation schemes during experiments~\citep{liu2018attosecond, glownia2016self}, 
as it can be achieved in a more controlled manner with lower field intensities.
The rest of the electronic populations, termed as ``resp'' in Fig.~\ref{fig42}, is distributed among other excited states with insignificant probability. 
The insignificant contributions from other states involved in the electronic response during the excitation can be neglected for further analysis.
The energy  difference between the ground  and the ninth excited states is $\Delta E = 3.45$\,eV. 
Focusing only on the dominantly populated state allows us to estimate a characteristic timescale for the electron dynamics as
$\tau = \hbar/ \Delta E$= 1.2\,fs. %, which is used for discussions. 
Time zero in Fig.~\ref{fig42} represents the onset of field-free charge migration, i.e., after the pump pulse ended. 

\begin{figure}[h!]
\includegraphics[width= \linewidth]{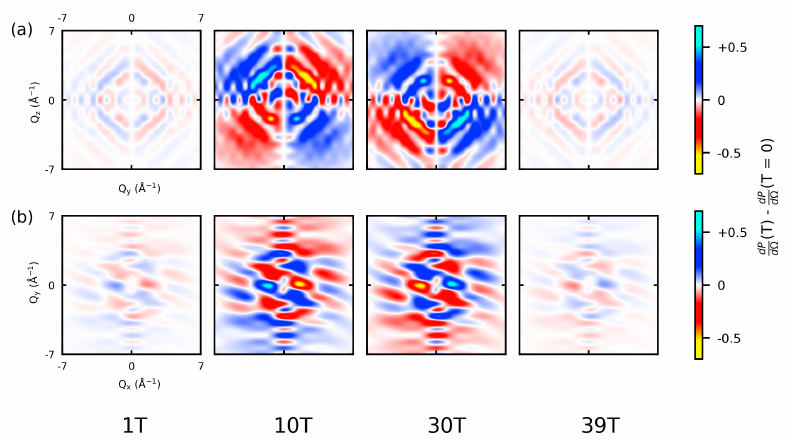}
\caption{Time-resolved difference diffraction signals for copper corrole in (a) $Q_{y}-Q_{z}$ and 
(b) $Q_{x}-Q_{y}$ planes at different pump-probe delay times during field-free charge migration.
Here,  {\sffamily{T}}  = ${\tau}/{40}$ is chosen with $\tau = 1.2$ fs as  the characteristic timescale of the charge migration.
The intensity of the diffraction patterns is presented in units of  ${dP_{e}}/{d \Omega}$.  
The time-independent diffraction signal at zero delay time is subtracted at all subsequent delay times.} \label{fig43}
\end{figure}

After triggering the charge migration by a short, intense pump pulse, we employ
TRXD in a pump-probe configuration to probe the dynamics using a probe pulse much shorter than 
{\sffamily{T}}  = 30 as.
Figure \ref{fig43} presents time-resolved diffraction signals at different pump-probe time delays during the field-free charge migration dynamics.
As the saddling is present in both the  $zx$ and $xy$ planes, while 
the molecule lies in the $yz$ plane without any saddling (see Fig.~\ref{fig41}),
the diffraction signals  are presented  in $Q_{y}-Q_{z}$ and $Q_{x}-Q_{y}$ planes, which  are  
used to emphasize the role of saddling during the charge migration.
For representation purpose, the total signal at the zero pump-probe delay time is subtracted from the total signals at different subsequent delay times.
Ground and the 10 lowest-lying excited states are used to simulate the TRXD signal, 
i.e., $f = [0,10]$ in Eq.\,\eqref{eq:DSP_final}.

At a glance, it seems that the overall intensity variation of the diffraction signals in  the
$Q_{y}-Q_{z}$ [Fig.\,\ref{fig43}(a)]  and $Q_{x}-Q_{y}$  [Fig.\,\ref{fig43}(b)]   planes is approximately  
the same at all delay times, up to the phase and the signal magnitude. 
The global maxima and smaller local features migrate up and down in the $Q_{y}-Q_{z}$ plane.
In the upper panels, the signal is found to be antisymmetric with respect to reflection about the $Q_{y} = 0$ line.
At the beginning of the charge migration, i.e., at delay time {\sffamily{T}}, the signal also exhibits reflection symmetry with respect to the $Q_{z} = 0$ line around low $\mathbf{Q}$ region.  
This feature in the signal is also true around the characteristic timescale $\tau = 1.2$ fs = 40{\sffamily{T}}, 
which is clearly visible from the signal at 39{\sffamily{T}}  [the last figure of  Fig.\,\ref{fig43}(a)].
Note that the total signal, which is not the difference signal,  is strictly anti-symmetric at $\tau = 0$ and 
40{\sffamily{T}} , and the snapshots at {\sffamily{T}}  and 39{\sffamily{T}} 
are chosen at the onset of symmetry reduction.
This scenario changes significantly at the two intermediate delay times, 10{\sffamily{T}}  and 30{\sffamily{T}}. 
The intensity of the signal exhibits an extremum close to the center, with local maxima at around values of $\pm 1$ in $Q_{y}$ and $Q_{z}$.
The extremum is initially more intense and diffused in the upper half of the plane, i.e., at positive values along the $Q_{z}$ axis at delay time 10{\sffamily{T}}.
The symmetry reduction can be understood using a simplified model of the charge migration as a superposition of $|0\rangle$ and $|9\rangle$ electronic states.
As derived from Eq.\,\eqref{eq:DSP_final} [also see  SI of Ref.\,\citep{hermann2020probing}], the time-dependent part of the
TRXD signal in such superposition state reads
\begin{equation}\label{tls}
\frac{dP({\mathsf{T}})}{d\Omega} \propto  \cos(\omega_{09}{\mathsf{T}})\,\text{Re}\big[{\mathcal L}_{09}(\mathbf{Q})\big]
-\sin(\omega_{09}{\mathsf{T}})\,\text{Im}\big[ {\mathcal L}_{09}(\mathbf{Q})\big],
\end{equation}
where $\omega_{09}=(E_9-E_0)/\hbar$ is the transition frequency between the $|0\rangle$ and $|9\rangle$  states, and
\begin{equation}
{\mathcal L}_{09}(\mathbf{Q}) = \sum_{f}~ \left[ \int \mathrm{d}\mathbf{r}  \int \mathrm{d}\mathbf{r}^{\prime} ~ \langle \Phi_{0}  | \hat{\rho}(\mathbf{r}) | \Phi_{f} \rangle \langle \Phi_{f}| \hat{\rho}(\mathbf{r}^{\prime})  | \Phi_9 \rangle~ e^{i \mathbf{Q} \cdot (\mathbf{r} -\mathbf{r}^{\prime})} \right].
\end{equation}
As evident from the middle column of Fig.~\ref{fig47}, the real and imaginary parts of ${\mathcal L}_{09}(\mathbf{Q})$ in the $Q_{y}-Q_{z}$ plane have different symmetry under reflection over the $Q_{z}$ axis.
Thus, symmetry reduction occurs by oscillation between these two $\text{Re}\big[{\mathcal L}_{09}(\mathbf{Q})\big]$ and $\text{Im}\big[{\mathcal L}_{09}(\mathbf{Q})\big]$ terms.

\begin{figure}[h!]
\includegraphics[width= \linewidth]{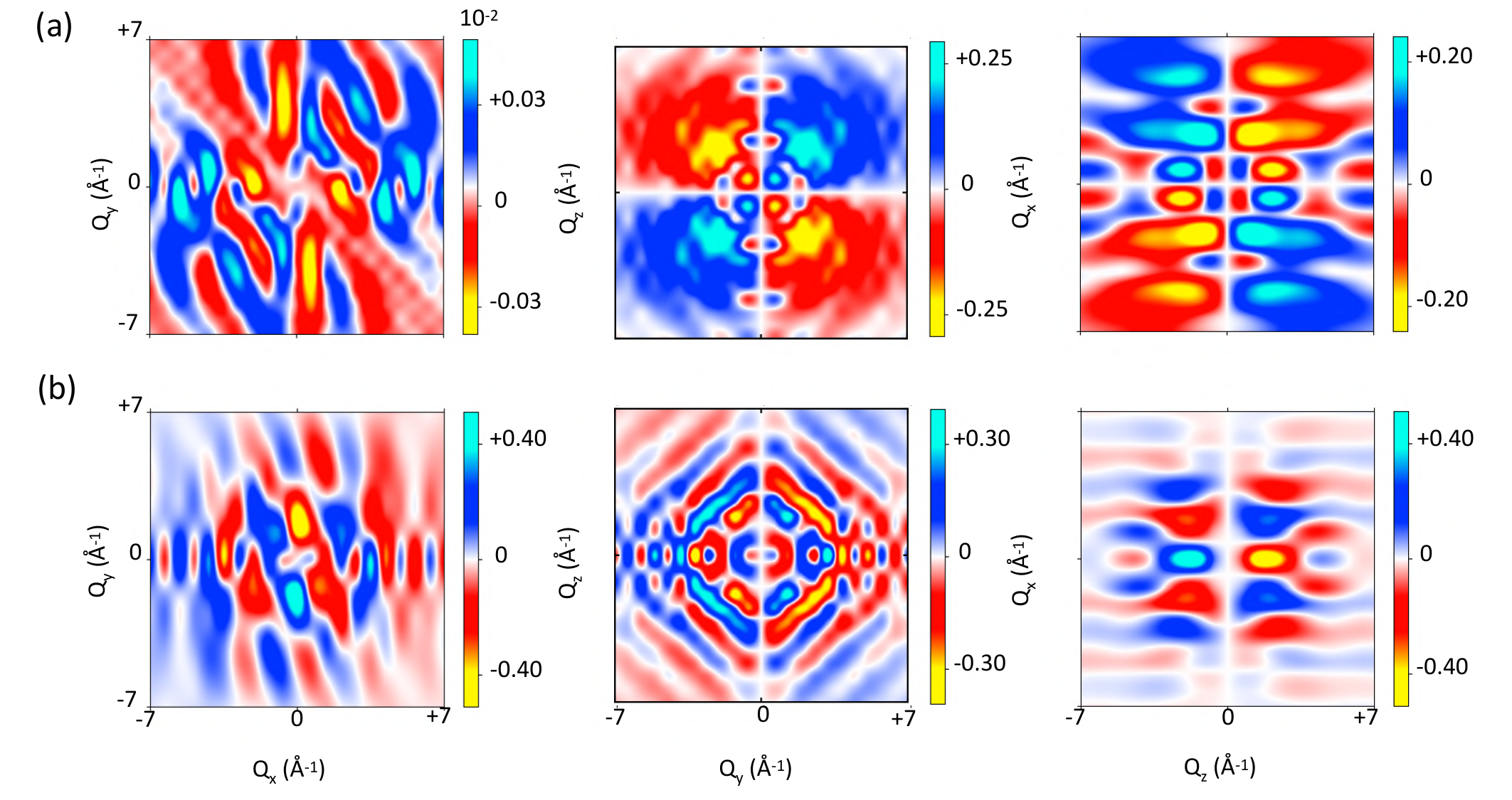}
\caption{(a) Real and (b) imaginary part of  $\mathcal{L}_{90}(\mathbf{Q})$ in  the $Q_{x}-Q_{y}$, $Q_{y}-Q_{z}$, and $Q_{z}-Q_{x}$ planes, respectively (from left to right).} \label{fig47}
\end{figure}

Let us analyse the signal at 30{\sffamily{T}}, which deserves a separate discussion. 
Signal depletion is observed in the $Q_{z} = Q_{y}$ direction at both delay times as documented in the 
top panels of Fig.\,\ref{fig43}.
In real space, this direction includes the nitrogen atom that buckles {\it above} the molecular ($yz$) plane.
Regions of signal enhancements (in blue and cyan) are found along the $Q_{z} = -Q_{y}$ direction, coinciding with the $z = -y$ direction, 
which contains the nitrogen atom  buckling {\it below} the molecular ($yz$) plane.
Left-right reflection about the $y=0$ line is the 
only symmetry element remaining in the projection of the molecule in the $yz$ plane, 
which gives rise to the pattern observed in the TRXD signal.
Hence, interference effects due to electronic coherences lead to the symmetry reduction of 
 the TRXD signal.
The asymmetry of the maxima with respect to  
the $Q_z=0$ line observed at both delay times, 10{\sffamily{T}}  and 30{\sffamily{T}},
correlates in both cases with signal depletion for the nitrogen above the corrole plane
or with signal enhancement for the nitrogen below the plane.

The signal in the $Q_{y}-Q_{z}$ plane at the beginning and end of the  characteristic timescale transforms 
according to the B\textsubscript{1} irreducible representation (IRREP) of the $\mathcal{C}$\textsubscript{2v} point group. 
However, at other delay times, the  signal transforms according to the A'' IRREP 
of the $\mathcal{C}$\textsubscript{s} point group [see the second and third figures 
of  Fig.\,\ref{fig43}(a)].
Incidentally, this is also the point group of the molecule projected in the $yz$ plane;  see right panel of Fig.\,\ref{fig41}. 
Let us understand this symmetry alteration  by analysing 
the time-dependent difference density during field-free charge migration, as shown in
Fig.\,\ref{fig44}. 
As reflected from the top panels, the difference density in the $yz$ plane
is found to transform according to the A'' IRREP of the $\mathcal{C}$\textsubscript{s} point group. 
This IRREP belongs to
the point group of the plane projection of the molecule. 
Also,  it is known that the Fourier transform of a transition density belonging to the 
$\mathcal{C}$\textsubscript{s} point group
corresponds to the $\mathcal{C}$\textsubscript{2v} point group~\citep{defranceschi1990symmetry}.

The temporal evolution of the signals in the $Q_{x}-Q_{y}$ plane for different delay times 
is significantly different than the signals in the $Q_{y}-Q_{z}$ plane [see Figs.~\ref{fig43}(a) and \ref{fig43}(b)]. 
The presence of the twisted structures in the diffraction signals  seems a signature   
of the saddled structure of copper corrole during charge migration. 
Both the TRXD signal [Fig.\,\ref{fig43}(b)] and the difference density [Fig.\,\ref{fig44}(b)] 
retain exactly the same structure at all times, 
albeit with phase reversal and intensity variations during the dynamics.
Both the signal and the difference density 
transform according to the A\textsubscript{u} IRREP of the $\mathcal{C}$\textsubscript{i} point group,  to which
belongs the projection of the molecule in the $xy$ plane. 
It was shown that the Fourier transform of a transition density 
belonging to the $\mathcal{C}$\textsubscript{i} point group will belong to the same group~\citep{defranceschi1990symmetry}.

\begin{figure}[h!]
\includegraphics[width =  \linewidth]{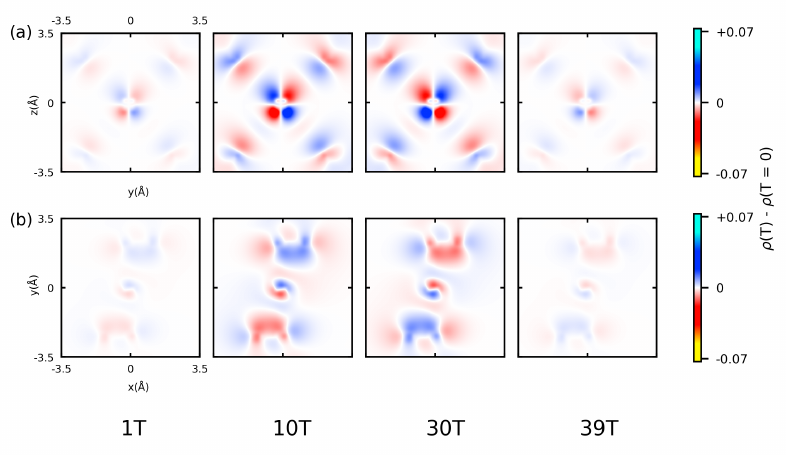}
\caption{Time evolution of the electron  density of the wavepacket 
during field-free charge migration in copper corrole.
The electron density at zero time delay is subtracted at all subsequent delay 
times and the difference density is represented
in the (a)  $yz$  and (b) $xy$ planes.
As in Fig.\,\ref{fig43},  {\sffamily{T}}  = ${\tau}/{40}$ is chosen with $\tau = 1.2$ fs as  the characteristic timescale of the electron dynamics.} \label{fig44}
\end{figure}

To better understand  the connection between the results shown in Figs.~\ref{fig43} and \ref{fig44}, let us 
analyse the key expression for TRXD, which reveals that the difference diffraction signal  encodes the Fourier transform of the transition electron density $\int \mathrm{d}\mathbf{r} ~\langle \Phi_{f}  | \hat{\rho}(\mathbf{r}) | \Phi_{k}  \rangle~ e^{-i \mathbf{Q} \cdot \mathbf{r}}$ [see Eq.~\eqref{eq:DSP_final}]. 
Also, the difference density of an electronic wavepacket consists of several terms of transition electron density,  
i.e., $\langle \Phi_{f}  | \hat{\rho}(\mathbf{r}) | \Phi_{k}  \rangle$.  
Note that it is  well established that the TRXD is  simply not related to the 
Fourier transform of the instantaneous electron density of the wavepacket, 
$\int \mathrm{d}\mathbf{r} ~\langle \Psi ({\mathsf{T}})  | \hat{\rho}(\mathbf{r}) | \Psi ({\mathsf{T}})  \rangle~ e^{-i \mathbf{Q} \cdot \mathbf{r}}$; and 
electronic coherences and the transition electron density  play  crucial 
roles in TRXD~ \citep{dixit2012imaging}. 
From the top panels of Fig.~\ref{fig43}, it is evident that the electronic coherences do not destroy
the symmetry relations between the transition density and its Fourier transform at early and later delay  times, but they  induce symmetry reduction in the TRXD signal at intermediate delay times during the dynamics.
This finding is specific to this particular example and the choice of exciting field is likely to affect the times at which  symmetry reduction occurs due to interferences.

\begin{figure}[h!]
\includegraphics[width=  \linewidth]{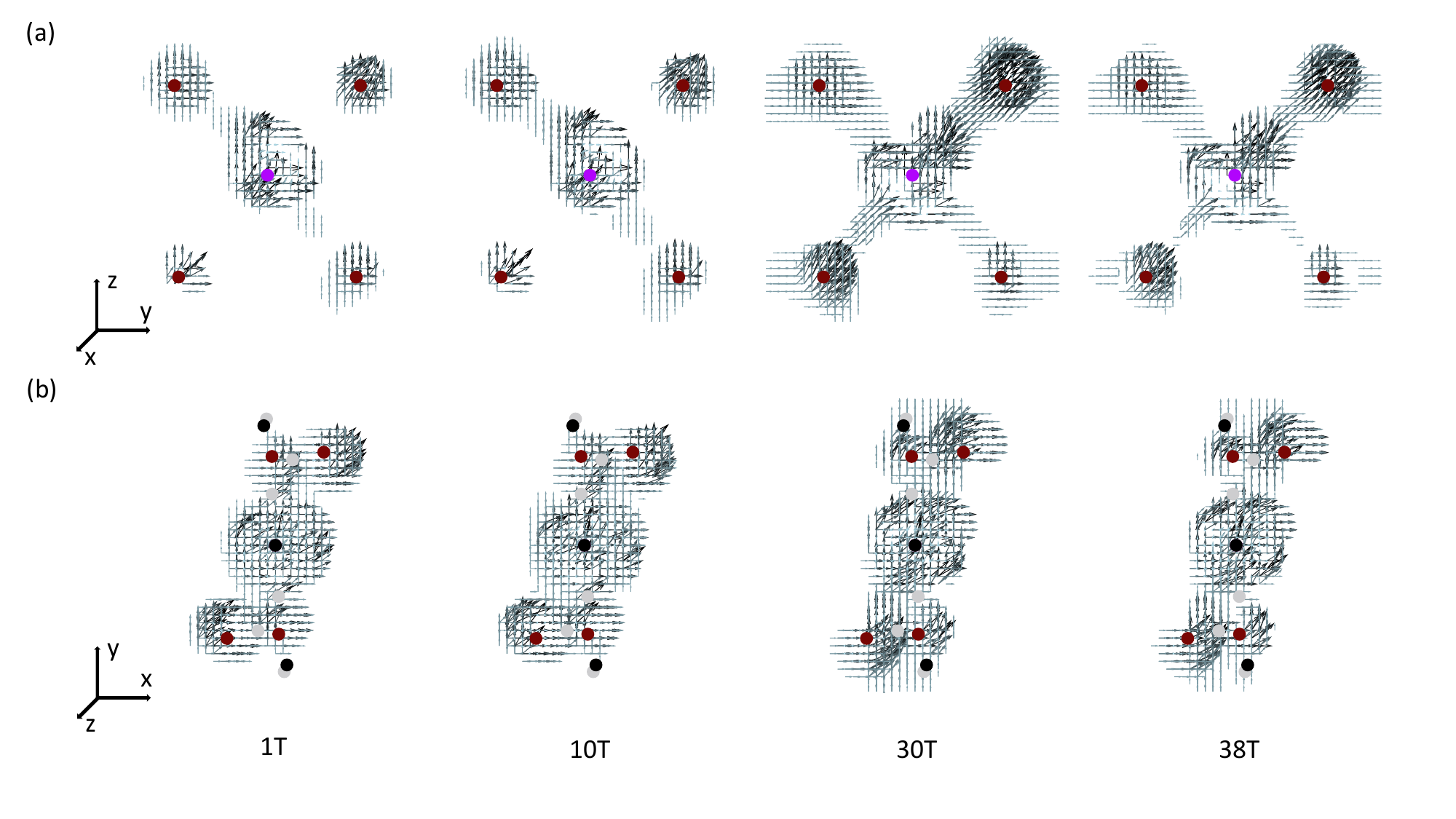}
\caption{Time-dependent electronic flux densities for copper corrole in the
(a) $yz$ and, (b) $xy$ planes at different pump-probe delay times during field-free charge migration.
The laser excitation parameters are the same as in Fig.\,\ref{fig42}.
The period {\sffamily{T}} = ${\tau}/{40}$ is chosen as the oscillation period $\tau = 1.2$ fs of the electron dynamics.
The black, violet, maroon, and gray dots  represent carbon, copper, nitrogen, and hydrogen atoms, respectively.} \label{fig45}
\end{figure}

Let us explore the charge migration  in real space to complement the mechanistic picture of the 
charge migration, which could provide  a deeper understanding of the time-resolved diffraction signal in detail. 
Knowledge of the dominant components of the electronic wavepacket allows one to 
analyse the charge migration at different instances in real space.
Recently, it has been discussed that the analysis  of the transient EFD provides a detailed 
understanding  of the electron dynamics as the  flux density maps the
direction of the electron flow in real space~\citep{hermann2020probing, tremblay2021time}.

Figure~\ref{fig45} presents the EFDs associated with the laser-induced charge migration  
at the same instances as in Fig.~\ref{fig43}. 
The central part of the corrole ring containing the nitrogens and the copper is zoomed in
to emphasize  the dominant contribution to the flux densities.  
It is evident from Fig.~\ref{fig45}(a)  that 
the charge migration  is taking place between the nitrogen atoms via the copper atom. 
Most of the charge seems to be displaced from the in-plane nitrogen in the bottom left to the nitrogen atom saddled
above the surface in the top right, which corresponds to the $Q_z = Q_y$ line in momentum space.  
As reflected from Fig.\,\ref{fig43}, the TRXD signals decrease along the $Q_z = Q_y$ line, so 
we assign this migration pattern to a hole displacement during the dynamics.
Synchronously, the nitrogen atom below the plane (top left) and the one in the bottom right appear to feed electrons to the copper atom,
which corresponds well to the region of the TRXD signal increase along the $Q_z = -Q_y$ line in the TRXD signal;  see Fig.~\ref{fig43}(a). 
The view of the flux densities in the $xy$ plane  provides complementary information. 
As evident from Fig.~\ref{fig45}(b), out-of-plane nitrogen atoms are also connected to each other via the copper atom.
This is revealed by the synchronous changes in the direction of the rotation of the  
flux densities around the nitrogen atoms.
At {\sffamily{T}}  and 10{\sffamily{T}}, the flux densities in the upper two nitrogen atoms rotate anti-clockwise, 
whereas they rotate clockwise for the bottom ones. 
This picture is reversed in the last two time steps at 30{\sffamily{T}} and 39{\sffamily{T}}.

\begin{figure}[h!]
\includegraphics[width=  \linewidth]{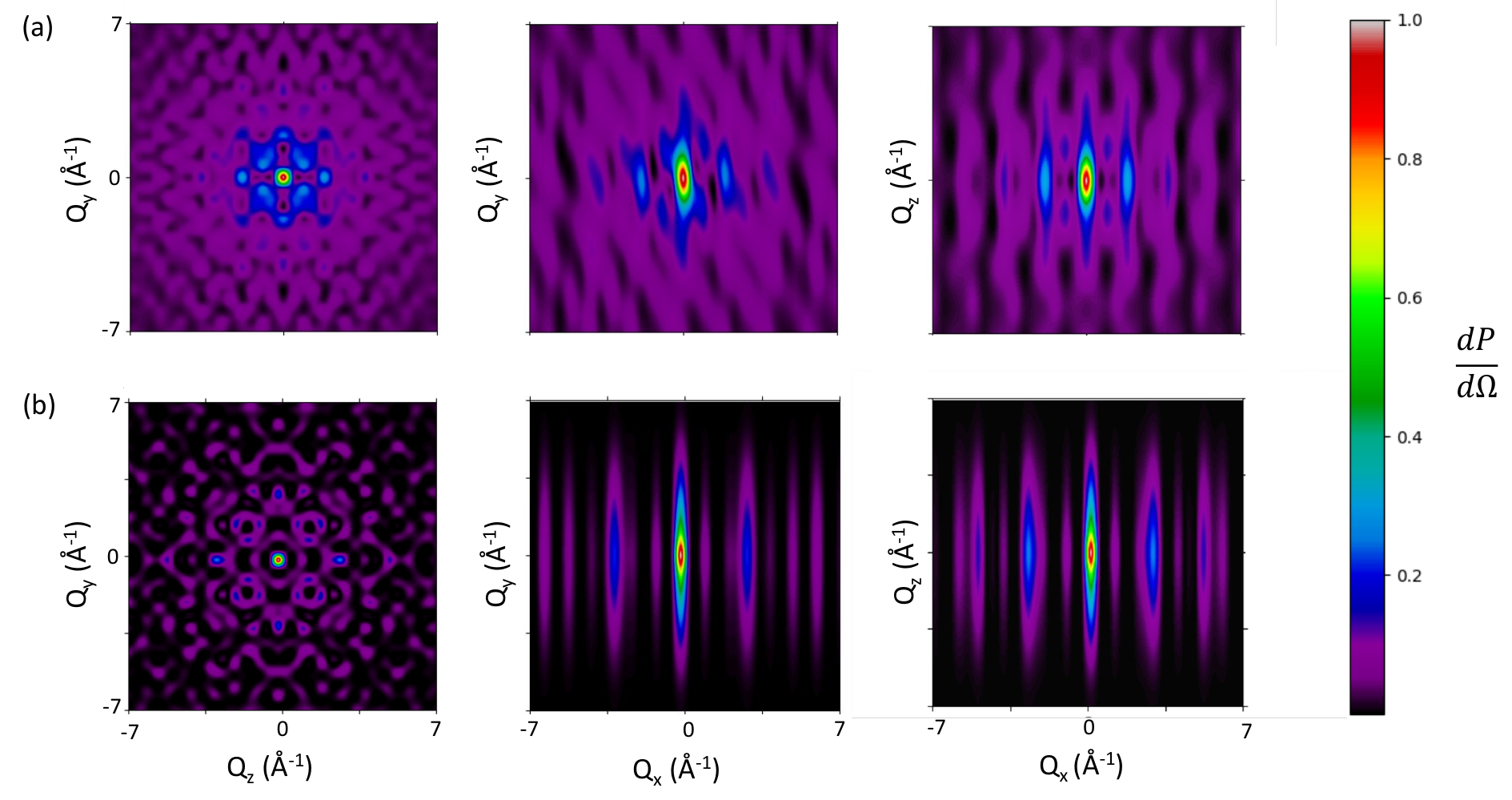}
\caption{The static diffraction signals corresponding to  the ground state for (a) copper corrole and (b) copper
porphyrin in the $Q_{y}-Q_{z}$, $Q_{y}-Q_{x}$ and $Q_{z}-Q_{x}$ planes (from left to right). 
The signals are normalized with respect to their maximum values.} \label{fig46}
\end{figure}

Analysis of the EFDs in  Fig.~\ref{fig45} also shows that 
the flux densities in the diagonal nitrogen atoms are different in magnitudes and directions. 
The present findings are in stark contrast with similar laser-induced charge migration dynamics in planar molecules,  such as benzene and porphyrin in which the flux densities and the charge migration dynamics
are observed to be symmetric at all times~\citep{hermann2020probing, barth2006periodic, barth2006unidirectional}.
In this sense, the asymmetry in the EFDs appears to be a signature of the saddling in copper corrole during charge migration dynamics,
which could be measured as a symmetry reduction in the time-resolved diffraction signals.
Although the distortion of the molecular structure due to the saddling occurs in the $xy$ plane, 
the signature of this symmetry reduction would rather be observed in the projection in the plane of the molecule, i.e., the $yz$ plane.
On the other hand, the connection between nitrogen atoms diametrically opposite of the copper atom  
is present in both the real-space and momentum-space views of the electron dynamics. 

To further confirm our claim that the diagonal symmetry is related to the saddled 
structure of the copper corrole, we simulate the static diffraction signal of copper porphyrin in 
the ground state, which exhibits a non-saddled structure.
For copper porphyrin in the ground state, the static diffraction signals are perfectly symmetric along 
the $Q_x$ = 0 and $Q_y$ = 0 planes. Moreover, owing to the non-saddled planar structure of copper porphyrin,
the static diffraction signals in the $Q_{x}-Q_{y}$ and $Q_{x}-Q_{z}$ planes are identical, as reflected 
in Fig.~\ref{fig46}(b). This is not the case for the signal of the saddled, 
nonplanar copper corrole shown in Fig.~\ref{fig46}(a).

\section{Summary}  
In summary, the present chapter discussed a first step towards understanding the interplay between
ultrafast charge migration, structural deformation, and symmetry reduction in  time-resolved x-ray imaging.
We investigated laser-induced dynamics in copper corrole, which has an interesting saddled geometry with only slightly reduced symmetry.
A linearly-polarized pump pulse is used to trigger electron dynamics, which is imaged by TRXD with atomic-scale spatiotemporal resolution. 
We find that the difference diffraction signals are sensitive to the saddled structure, and 
this asymmetry is reflected in the EFDs. 
For the studied excitation, the saddled nitrogen atoms in copper corrole are found to
facilitate the coherent charge migration between nitrogen and copper atoms.
We believe that our results on imaging electron dynamics in symmetry-reduced systems will motivate further
theoretical and experimental research, in particular on light-induced ultrafast processes in various metal-corroles.

\cleardoublepage
\chapter{Imaging Charge Migration in Chiral Molecules}\label{Chapter3}

Understanding the chirality is essential for a broad range of sciences including the origin of homochirality 
on earth. A molecule without inversion symmetry or a symmetry plane which is not superimposable with its own mirror image is known as a chiral molecule. 
A pair of such chiral molecules is known as an enantiomers. 
Enantiomers exhibit identical physical properties, but show a strong
enantiomeric preference during chemical and biological reactions. 
Discerning the enantiomeric excess and handedness of chiral molecules 
is crucial in chemistry, biology, and pharmaceuticals.

Experimental methods based on chiral light-matter interaction, such as  
Coulomb-explosion imaging~\citep{pitzer2013direct, pitzer2016absolute, herwig2013imaging}, microwave spectroscopy~\citep{patterson2013enantiomer, eibenberger2017enantiomer}, Raman optical activity~\citep{barron1973raman, barron2009molecular} and laser-induced mass spectrometry~\citep{bornschlegl2007investigation, li2006linear} have become practice to discern enantiomers in gas phase. Moreover, ionization based photoelectron circular dichroism approaches~\citep{harding2005photoelectron, bowering2001asymmetry, nahon2006determination, janssen2014detecting, ritchie1976theory, powis2000photoelectron} 
not only allow probing chirality in the  multiphoton~\citep{lux2012circular, lehmann2013imaging, lux2015photoelectron, beaulieu2016probing} and strong-field regimes~\citep{dreissigacker2014photoelectron, beaulieu2016universality, rozen2019controlling}, but also help us to understand molecular relaxation dynamics~\citep{comby2016relaxation} and photoionization time delay~\citep{beaulieu2017attosecond}. Analogously, laser-induced photoelectron circular dichroism was recently used to obtain time-resolved chiral signal~\citep{beaulieu2018photoexcitation, harvey2018general}. 
In particular, Beaulieu et al. have employed time-resolved vibronic dynamics associated with a photoexcited  electronic wavepacket to explain time-resolved chiral signal~\citep{beaulieu2018photoexcitation}. Recently, the signature of chirality  is also probed by chiral high-harmonic generation~\citep{cireasa2015probing, smirnova2015opportunities, wang2017high, harada2018circular, baykusheva2018chiral, neufeld2019ultrasensitive, baykusheva2019real}.
In most of the aforementioned methods, left- and right-circularly polarized light were used to probe 
 chirality in gas phase for randomly orientated chiral molecules.
An alternate method based on  pairs of linearly polarized laser pulses with skewed mutual polarization was proposed recently~\citep{yachmenev2016detecting}. 

\begin{figure}[]
\includegraphics[width = \linewidth]{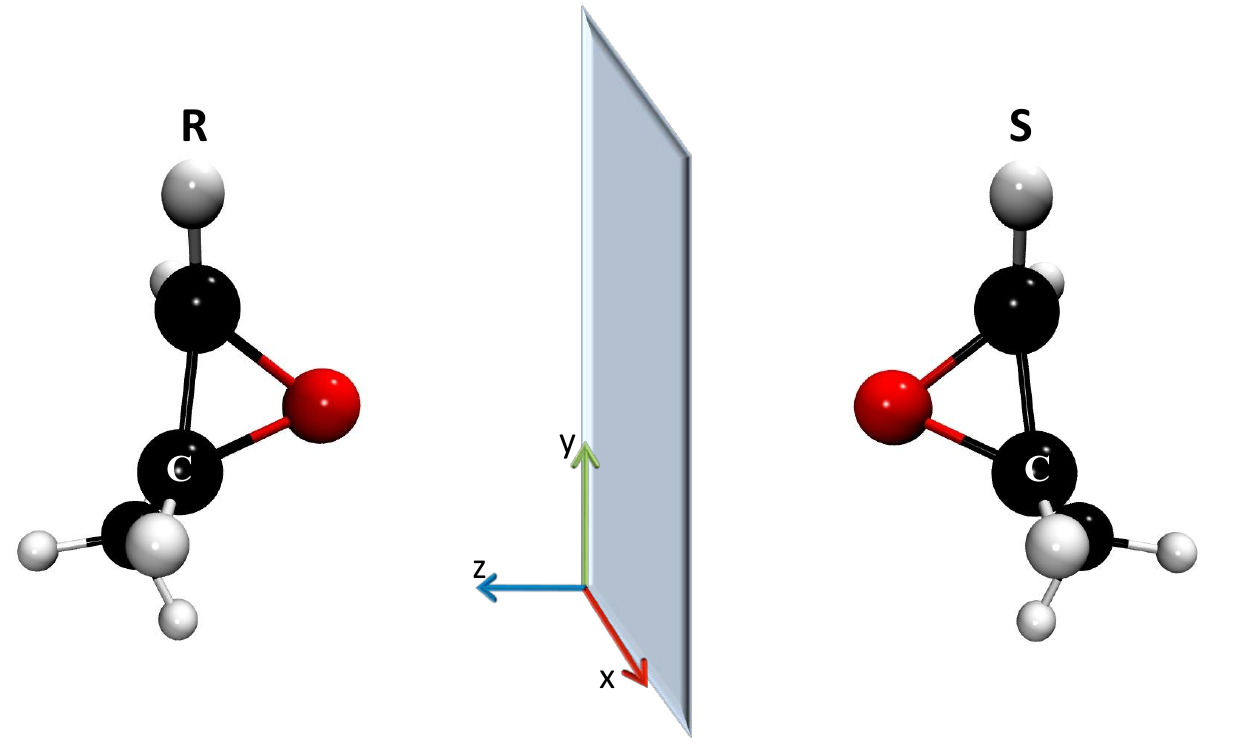}
\caption{R- and S-enantiomers of epoxypropane in the molecular fixed frame.
Gray, black and red spheres represent hydrogen, carbon and oxygen  atoms, respectively. The 
chiral carbon is labelled by C. 
The $xy$ plane is the mirror plane and the normal axis lies along the $z$ direction.} 
\label{fig0}
\end{figure}

Recently, it has been demonstrated experimentally  and theoretically that chiral molecules can be oriented in a specific direction~\citep{tutunnikov2018selective, tutunnikov2020observation}. When a linearly polarized laser pulse along $x$ direction is employed, the chiral molecules  
try to align themselves in such a way that the most polarizable axis in molecules will be along the $x$ direction. There will be an equal probability of alignment along the $\pm x$ direction and molecules can rotate freely in the perpendicular plane, i.e., a chiral molecule would lie along this alignment axis and rotate about it. Orientational averaging would imply that observed signals would present no signature of chirality. A time-delayed second laser pulse with skewed polarization, linearly polarized in the $xy$ plane at $45^{\circ}$ between both axes, induces a dipole in the molecule. It is given by $d_{i} = \sum_{j} \alpha_{ij} E_{j}$, where $\alpha_{ij}$ is the polarizability component and $E_{j}$ is the electric field component
of the laser pulse. As a result of the induced dipole, a torque $\vec{\tau} = \vec{d} \times \vec{E}$ is induced along the aligned molecular axis. The analysis of the components of the torque yields a nonzero value of the average of the time derivative of the torque along $z$ direction. The torque inducing this new alignment is proportional to the off-diagonal elements of the polarizability tensor. For nonchiral molecules, the off-diagonal elements have the same sign, so all molecules will experience the same torque. For chiral molecules, the off-diagonal elements of the polarizability tensor have opposite signs for different enantiomers. This implies that enantiomers will experience a torque in opposite directions, leading to opposite orientations, related by reflection symmetry (see Fig.~\ref{fig0}).

On the other hand, if a circularly polarized laser pulse is applied, the interaction with the chiral molecules will be different. For $E = E_{0}[\cos(\omega t) \hat{x} + \sin(\omega t) \hat{y}]$, the components of the induced torque will be $\tau_{x} = -E_{0}^{2}[\alpha_{zx} \cos(\omega t) + \alpha_{zy} \sin(\omega t)] \sin(\omega t)$, $\tau_{y} = E_{0}^{2}[\alpha_{zx} \cos(\omega t) + \alpha_{zy} \sin(\omega t)] \cos(\omega t)$, and $\tau_{z} = -E_{0}^{2}[\alpha_{xy}$ $\sin(\omega t) + \alpha_{yx} \cos(\omega t)]$. The average value of the $z$ component yields zero as we analyse the above expressions. For the other components of the induced torque, even if the components are different for the enantiomers as the polarizability tensors change sign, there is no simple $\pi$-phase change relation for the enantiomers. Therefore, the interaction with a circularly polarized laser pulse can break the symmetry of the enantiomers, but unidirectional orientation cannot be achieved.

%%%%%%%%%%%%%%%%%%%%%%%%%%%%

In this chapter, 
we explore charge migration in chiral molecules by studying ultrafast  charge migration induced by a 
linearly polarized intense laser pulse for a specific orientation of molecules.
It is not immediately obvious whether the charge migration in  a space-fixed excited enantiomer pair should remain same or not.
The answer is encoded in the EFDs, which maps the direction of electron flow.
The time-dependent behaviour of  EFDs in chiral molecules out-of-equilibrium 
remains uncharted territory.  This is one of the main focuses of the present chapter.
As will be discussed below, the EFD can become markedly different in an enantiomer pair
under specific laser-excitation conditions, which would lead to different experimental signals.
To illustrate the role of time-resolved EFD in chiral molecules,
epoxypropane (1,2-propylene oxide, see Fig.~\ref{fig0}) is used as a chiral molecule.
Epoxypropane has been used in chiral high-harmonic generation~\citep{cireasa2015probing},
as well as being observed in interstellar media~\citep{mcguire2016discovery, bergantini2018combined}. 

Four-dimensional imaging of  pump-induced charge migration in epoxypropane is  done by TRXD. 
In this chapter, we will establish that TRXD and EFD analysis provide complementary information about  
the charge migration in epoxypropane driven by linearly polarized light. 
Moreover,  we demonstrate that TRXD within pump-probe configuration offers an alternative to
probe the handedness of chiral molecules. 
Note that confirming the handedness of chiral molecules by determining their spatial arrangement is 
known as absolute configuration determination~\citep{mcmorrow2011elements, bijvoet1951determination, santoro2020absolute}. 
X-ray crystallography is the one of the most reliable methods to determine  
the absolute configuration~\citep{flack1999absolute}. 

In this chapter, all $N_\textrm{states}=31$ lowest-lying excited states below the ionization threshold are used to 
obtain convergence, checked by repeating the laser-driven dynamics simulations with $N_\textrm{states}=\{16, 21, 26, 31\}$.
The states are computed using the CAM-B3LYP functional~\citep{yanai2004new} 
and aug-cc-pVTZ basis sets~\citep{dunning1989gaussian} on all atoms,  as implemented in Gaussian16~\citep{frisch2016gaussian}. 
%The motion of nuclei  are typically much slower in comparison to the electronic motion, and 
%therefore,  nuclei are treated as frozen in this work. 

%%%%%%%%%%%%%%%%%%%%%%%%
\section{Results and Discussion}
\begin{figure}[]
\includegraphics[width= \linewidth]{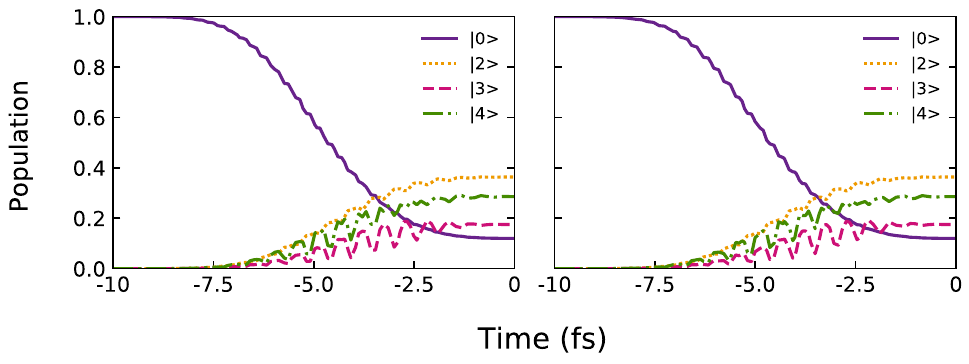}
\caption{Population dynamics of selected electronic states for (a) R- and 
(b) S-epoxypropane. A sine-squared 
(carrier frequency: 7.67\,eV; duration: 10\,fs; peak intensity:  10.1$\times$10$^{14}$ W/cm$^{2}$)
linearly polarized pulse along the $x$ axis is used to excited the first optically accessible band
in both enantiomers. $| 0 \rangle$ represents the ground electronic state. 
Only the excited states at energies $E_2=7.51$\,eV, $E_3=7.56$\,eV, and $E_4=7.73$\,eV are significantly populated throughout the dynamics.
The time origin, {\sffamily{T}} = 0, defines the onset of field-free charge migration.}
\label{fig1}
\end{figure} 

To reveal the charge migration of enantiomers in an external laser pulse,
it suffices to drive the system out-of-equilibrium, which transfers
some population from ground electronic to a few selected excited states.
In Fig.~\ref{fig1}, the time-evolution of the many-electron state populations in the two enantiomers of
epoxypropane is shown for
a 10\,fs sine-squared  linearly polarized pulse along the $x$ axis, with a 162\,nm wavelength (7.67\,eV)
and 10.1$\times$10$^{14}$ W/cm$^{2}$ peak intensity.
As evident from the figure,  the state populations are identical for both enantiomers at all times.
The small oscillations in the  population of individual electronic states are due to the permanent
dipole moments of these states, which interact with the laser pulse.
These oscillations do not affect the population transfer dynamics.
At the end of the pulse, i.e., at {\sffamily{T}} = 0, approximately 83$\%$ population is transferred from 
the initial ground electronic state to a coherent superposition of the 2$^{\text{nd}}$ (P$_2=36\%$),
3$^{\text{rd}}$ (P$_3 = 18\%$), and 4$^{\text{th}}$ (P$_4 = 29\%$) excited states.
The timescales associated with different charge migration processes 
can be estimated from the energy difference between the different populated states, $\tau=h/\Delta E$.
Due to significant population in the ground electronic state after the pulse, 
the largest energy difference of $\Delta E=7.89$\,eV leads to the fastest timescale for charge migration,
within the attosecond regime (524\,as).
However, as most of the electronic population is transferred to the excited states,
the dominant contribution to the charge migration will stem from interference effects
among these three excited states.
The timescale associated with these periodic processes range from 12.5\,fs to 82.7\,fs,
and the charge migration is thus predominantly happening on the femtosecond timescale. 

\begin{figure}[]
\includegraphics[width= \linewidth]{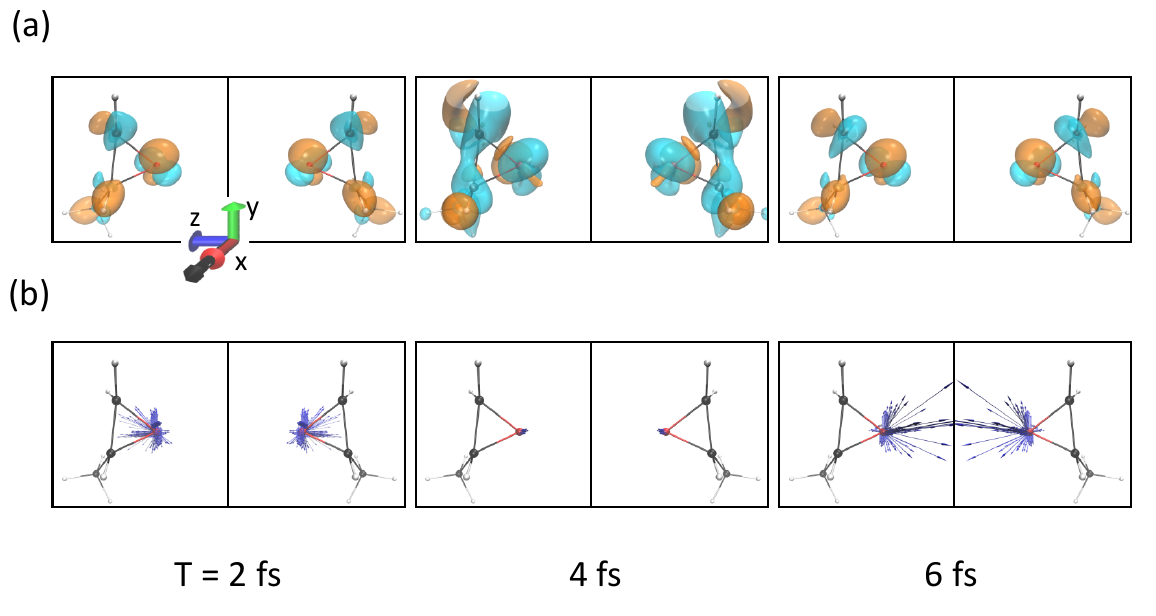}
\caption{(a) Electronic charge distribution difference, $\rho(\mathbf{r}, t)-\rho(\mathbf{r}, 0)$, and
	 (b) corresponding flux densities (blue arrows) for both  R- and S-enantiomers 
	 during field-free charge migration, at different times after the onset of the field-free charge migration.  	 	The field parameters are defined in the caption of Fig. \ref{fig1}.
	 Orange and blue colours represent isosurface values of  -0.0015 and +0.0015, respectively.}
\label{fig2}
\end{figure}

Although the populations are identical, it is not 
straightforward to infer whether the time-dependent charge distributions and
associated EFDs [$ \mathbf{j}(\mathbf{r},t)$ in Eq.~(\ref{eq:continuity})] corresponding to an electronic wavepacket
are identical or not for both the enantiomers. 
Figure~\ref{fig2}(a) presents the time-dependent charge distribution differences 
for both the enantiomers  at three different times after laser excitation. 
The system is prepared using the pulse defined in Fig.\,\ref{fig1}
in a superposition state consisting majority of the ground state
and of the 2$^{\text{nd}}$, 3$^{\text{rd}}$ and 4$^{\text{th}}$ excited states.
The charge distribution at the onset of field-free charge migration (i.e., at {\sffamily{T}} = 0)
is subtracted from the charge distributions at later times to reveal the charge migration.
At all times, the charge distribution differences for the enantiomers form mirror images,
and the chirality of the enantiomer pair is preserved (see Fig.~\ref{fig2}).
The charge distribution at time {\sffamily{T}} = 6\,fs is also found to be approximately 
similar to the distribution at {\sffamily{T}} = 2\,fs.
This is an indicative of a partial recurrence in the dynamics.
Because the electronic wavepacket is a superposition of many electronic states
with incommensurate energy differences, complete recurrence is not possible.
Interestingly, the oscillations in the charge distribution do not appear to involve
the chiral carbon atom (see also Fig.~\ref{fig0}). This is due to the fact that all states excited by the chosen pulse 
do not involve strong reorganization of the electron density close to the chiral center.
Prominent charge migration is taking place  around the oxygen atom only. Negative isocontour values (orange), corresponding to electron depletion, are found at time {\sffamily{T}} = 2\,fs and time {\sffamily{T}} = 6\,fs, 
whereas they are changed to positive isocontour values (blue) at time {\sffamily{T}} = 4\,fs. There is a small change of the isocontour for other carbon atoms except chiral carbon.

The variations of the one-electron density document charge migration but it is the EFD,
$\mathbf{j}(\mathbf{r},t)$, that yields spatially-resolved mechanistic information about the processes 
hidden within [see Eq.\,\eqref{eq:continuity}].
These EFDs are shown as blue arrows in Fig.~\ref{fig2}(b) for the same selected snapshots as above.
At all times, most of the flux densities are localized around the oxygen atom.
This confirms that the chiral carbon centre is not involved in this particular charge migration dynamics,
since many-body excited states inducing a nodal structure at the chiral center are found at higher energies
in epoxypropane.
It is interesting to observe that the direction of the flux densities have opposite phases at 
{\sffamily{T}}  = 2\,fs and {\sffamily{T}}  = 6\,fs,  although the magnitudes are almost equal. 
This picture contrasts with the one offered by the charge distribution differences
at these two times, which are approximately similar in particular around the top carbon atom.
This phase reversal in the flux densities describes a change in the flow direction of the electrons,
and this information can not be obtained from the charge distribution dynamics. 
Approaching this partial recurrence in the charge migration process could lead to the observed
sign reversal in the flow of electrons.

\begin{figure}[htb!]
\includegraphics[width= \linewidth]{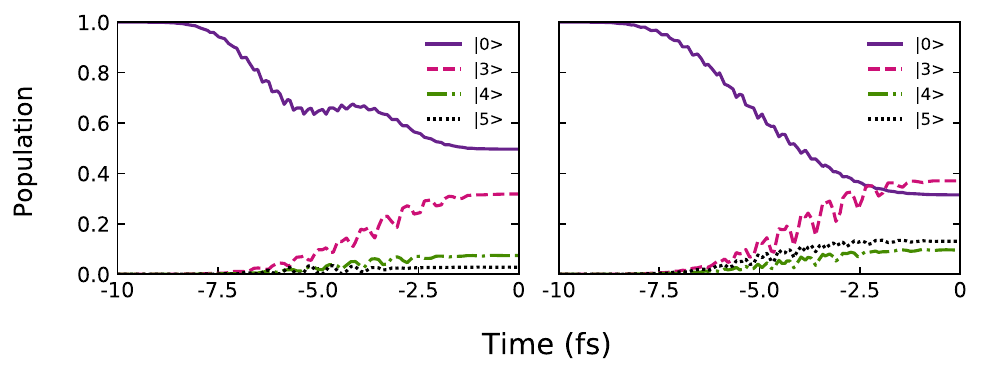} 
\caption{Population dynamics of selected electronic states for (a) R- and (b) S-epoxypropane. 
	A sine-squared 
(carrier frequency:  7.75\,eV; duration: 10\,fs; peak intensity:  9.1$\times$10$^{14}$ W/cm$^{2}$)
	linearly polarized pulse in the $yz$ plane is used to excite the first optically accessible band
	in both enantiomers. Only the excited states at energies $E_3=7.56$\,eV, $E_4=7.73$\,eV, and $E_5=7.84$\,eV
	are significantly populated throughout the dynamics.}
\label{fig3}
\end{figure} 

Note that the charge distributions and the EFDs, such as the ones depicted in Fig.\,\ref{fig2},
are related by the mirror reflection for both the enantiomers at all times. 
It is known that the mirror reflection of observables of one enantiomer along the mirror plane gives 
the same observable for other enantiomer due to the mirror symmetry of the chiral pair in the field.
A sign change upon reflection is a fundamental measure of chirality and it is evident from Fig.~\ref{fig2}.
The general trends discussed above remain unchanged for any other field-free charge migration process
in which the system is first prepared by laser pulses linearly polarized along the $z$ axis,  $y$ axis,
or for any laser polarization lying in the $xy$ plane.
In this case, the populations of electronic states will remain identical at all times for both the enantiomers,
whereas the time-dependent charge distributions and EFDs will retain the mirror symmetry
of the chiral pair.
This naturally brings up the question, how these results will change when we use 
linearly polarized pulses
along a plane including $z$ axis.

\begin{figure}[htb!]
\includegraphics[width= \linewidth]{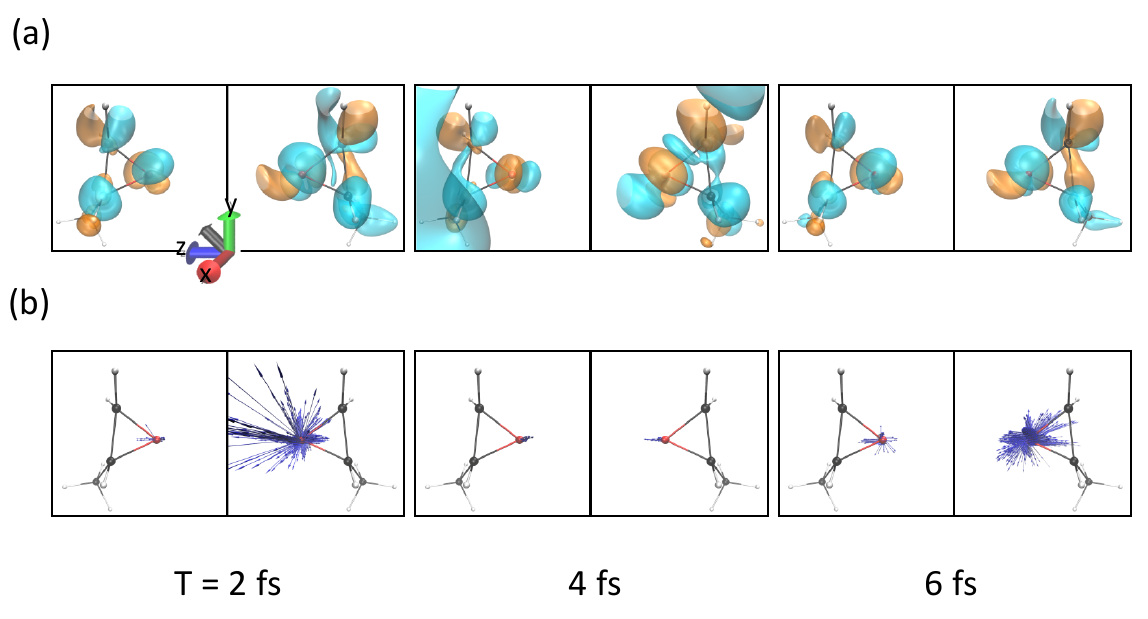}
\caption{(a) Electronic charge distribution difference and
	 (b) corresponding  flux densities (blue arrows) for both  R- and S-enantiomers 
	 during field-free charge migration, at different times after the laser pulse linearly polarized
	 in the $yz$ plane. The field parameters are defined in the caption of Fig. \ref{fig3}.
	 Orange and blue colours represent isosurface values of  -0.0015 and +0.0015, respectively.}
\label{fig4}
\end{figure} 

Figure~\ref{fig3} shows the population dynamics for selected electronic states
during a 10\,fs sine-squared pulse of 9.1$\times$10$^{14}$ W/cm$^{2}$ peak intensity and 160 nm wavelength (7.75\,eV).
The electric field is linearly polarized at $45^{\circ}$ between the axes in the $yz$ plane.
To achieve significant population among low-lying excited states, the pulse
parameters are tuned slightly compared to the previous case.
As opposed to polarization along the axes, the population dynamics for both enantiomers differ drastically.
For the R-enantiomer, $50\%$ population remains in the ground state at the end of the pulse,
whereas it is 70 $\%$ depleted for the S-enantiomer.
The dominant state at the end of the pulse is found to be the 3$^{\text{rd}}$ one (P$_3=32\%$ and $37\%$ for R- and S-enantiomers, respectively). The latter enantiomer populates also more efficiently
the 4$^{\text{th}}$ and 5$^{\text{th}}$ excited states (P$_4=10\%$ and P$_5=13\%$, respectively), providing a more democratic population distribution than in the R-enantiomer (P$_4=7\%$ and P$_5=2\%$).
It appears obvious that the field-free charge migration in the two enantiomers
following such linearly polarized excitation in the $yz$ plane will be radically different.

\begin{figure}[]
\includegraphics[width= \linewidth]{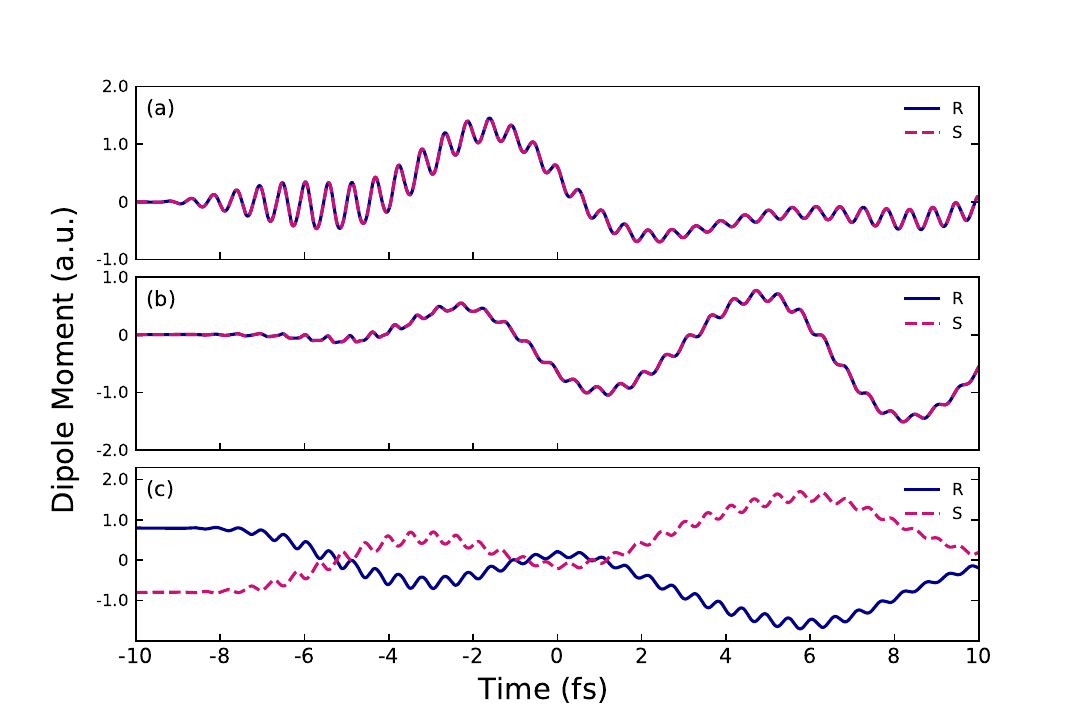}
\caption{Time evolution of (a) $x$-, (b) $y$- and (c) $z$-components of the total dipole moment
       during and after excitation by a 10\,fs sine-squared  linearly polarized pulse along the $x$ axis.
	The field parameters are the same as in Fig.\,\ref{fig1}.}
\label{fig5}
\end{figure} 

The charge distribution differences and EFDs for both the enantiomers
are shown in Fig.~\ref{fig4}.
In this case, only contributions to the electronic wavepacket
stemming from the 3$^{\text{rd}}$, 4$^{\text{th}}$, and 5$^{\text{th}}$ excited states are considered.
Fast oscillating contributions from the ground state are removed.
As could be inferred from the population at the pulse end, the charge migration dynamics
differs drastically for the two enantiomers. 
Moreover, neither the charge distributions nor the EFDs
are mirror images for both enantiomers,
unlike in the case of excitation by linearly polarized pulses in the $xy$ plane.
Again, the EFDs are concentrated around the oxygen atom. 
This is confirmed by looking at the flux densities [blue arrows in Fig.\,\ref{fig4}(b)],
which reveals that no electrons are flowing around nor through the chiral center.
As in the previous excitation scenario, the charge distributions are similar for both the enantiomers 
at times {\sffamily{T}}  = 2\,fs and $6$\,fs, in particular around the oxygen atom (see Fig.~\ref{fig4}). 
Although the relation to the evolution of the coherences in the system is as clear as above,
the flux densities are also found to be in opposite directions in these two snapshots.
Apart from the significant difference in the magnitude of the flux densities in the enantiomers, 
the nature of the evolution of the EFDs is also very different. 
At {\sffamily{T}}  = 2\,fs, the EFDs are equally distributed along $z$ axis and $x$ axis 
but at {\sffamily{T}}  = 4\,fs all of them are directed towards $x$ axis and they are distributed 
along all the direction for the R-enantiomer at {\sffamily{T}}  = 6\,fs. 
In contrast, for the S-enantiomer, at {\sffamily{T}}  = 2\,fs and {\sffamily{T}}  = 6\,fs 
most of the EFDs are 
distributed in the $xy$ plane but they are directed towards the positive $z$ axis
at {\sffamily{T}}  = 2\,fs, whereas they are directed towards negative $z$ axis in later time.

\begin{figure}[]
\includegraphics[width=  \linewidth]{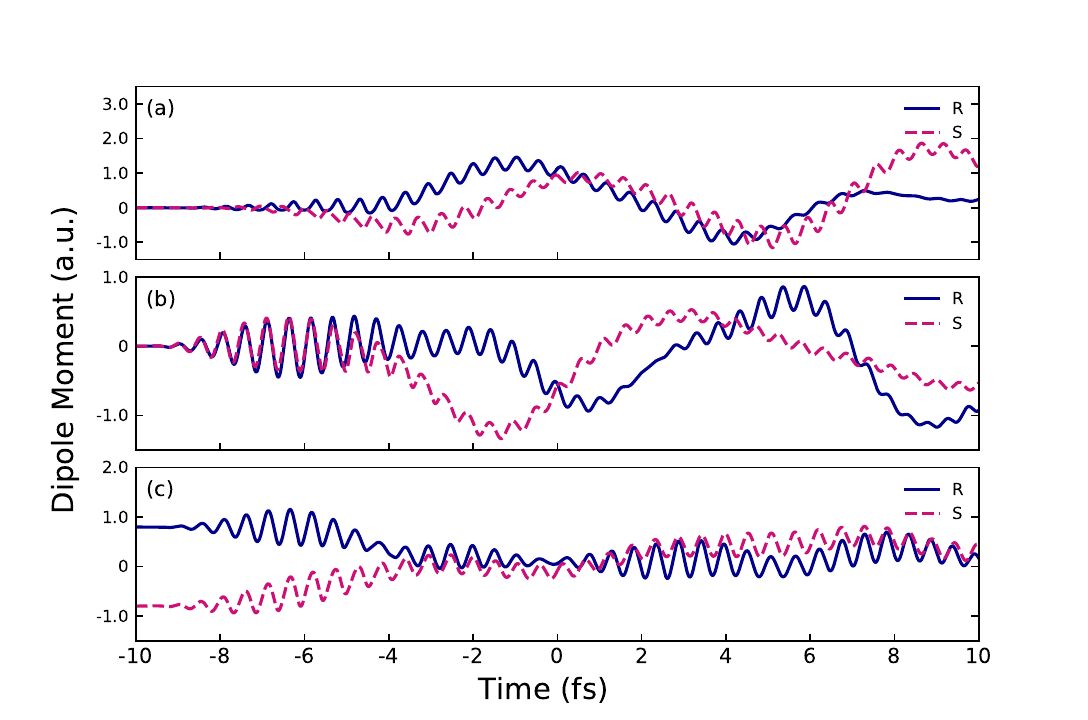}
\caption{Time evolution of the cartesian components of the dipole moment along
	(a) $x$-, (b) $y$- and (c) $z$-directions.
	The amplitude and phase along all the axes are different for both the enantiomers.
	The field parameters are the same as in Fig.\,\ref{fig3}.} 
\label{fig6}
\end{figure}

The previous simulations establish unequivocally that the field-free charge migration induced 
by linearly polarized pulse  behaves differently for laser polarization along an axis or 
in the $xy$ plane than for pulses polarized in a plane including the normal axis.
This laser-induced charge migration process is documented by charge distribution differences
and time-dependent EFDs, for a specific choice of molecular orientation.
Importantly, these observations remain generally valid for any orientation of the molecule,
as long as the molecule retains its orientation during the laser preparation phase.

These findings cannot be applied to aligned non-chiral molecules. For example, 
rotating the S-enantiomer by $180^\circ$ about the $y$ axis, 
the rotated S-enantiomer and R-enantiomer are not the same.
Consequently, the population dynamics driven by a linearly polarized pulse carefully chosen according to the 
prescriptions described below will be different.
To explain the physical origin of the radically different flux densities in enantiomer pairs,  
induced by linearly polarized light, 
time-evolution  of the dipole moment components for the enantiomers are calculated.
The enantiomers show mirror symmetry perpendicular to the $z$ axis in the absence of an external
field (see Fig.~\ref{fig0}).
It is the presence or the absence of a mirror symmetry for the enantiomer pair in the field that determines
the chirality of the charge migration, and this property is inherited by the subsequent field-free charge migration process.

The different Cartesian components of the total dipole moment are shown in Fig.~\ref{fig5}
during excitation using a sine-squared  linearly polarized pulse along the $x$ axis
and during subsequent field-free charge migration. The parameters of the pulse are chosen
as in Fig.\,\ref{fig1}.
It is evident from the figure that $x$- and $y$-components of the dipole moment are identical 
at all times for both the enantiomers.
On the contrary, the evolution of the $z$-component is opposite in sign for the R- and S-enantiomers  
 as the $z$ axis is the chiral axis in this case
(see Fig.~\ref{fig0}).
This mirror symmetry of the chiral pair can be seen by the opposite signs of the permanent dipole
for the R- and S-enantiomers prior to the laser excitation, i.e., at {\sffamily{T}}  = -10\,fs.
Upon excitation according to the first scenario, the pulse creates an electronic wavepacket, which follows 
 the mirror symmetry of the enantiomer pair.
The field generates a non-uniform time-dependent electronic distribution with different projections 
onto $z$ axis.
However, the electronic populations remain the same, as the orientation of the molecule relative to the 
$xy$ plane is arbitrary.
All linearly polarized pulses along any axis or in the $xy$ plane will
lead to a field-molecule interaction that preserves the mirror symmetry. 
Therefore, the $x$- and  $y$-components of the time-evolving
dipole moment will be identical for both enantiomers [see Figs.~\ref{fig4}(a) and \ref{fig4}(b)].
However, the projection of the $z$-component along the normal axis will remain
 the same in magnitude but opposite in phase by reflection symmetry among the enantiomers.
This statement is supported by our simulations, 
revealing an identical behaviour for the $z$-component of the dipole moment
at all times for any initial molecular orientation 
[see Fig.~\ref{fig4}(c) for a specific example]. 
Note that the $\pi$-phase variation in chiral molecules 
is a very widely known phenomenon, and is also the mechanism for 
photoelectron circular dichroism, photoexcitation circular dichroism, 
microwave three wave mixing, nonlinear based chiral
detection methods such as high-harmonic generation 
to detect chirality within the 
electric dipole approximation, etc.~\citep{harding2005photoelectron, nahon2006determination, powis2000photoelectron, yachmenev2016detecting, leibscher2019principles, beaulieu2018photoexcitation, cireasa2015probing, smirnova2015opportunities,  neufeld2019ultrasensitive}.

On the contrary, when the linear polarization of the pulse is rotated into 
a plane including the $z$ axis and any of the two other axes, 
the time-evolution of all components of the dipole moment is different in magnitude and phase (see Fig.~\ref{fig6}). 
For a pulse linearly polarized in the $yz$ plane, the pulse interacts coherently 
with the components of the dipole along both these axes simultaneously,
imposing a specific phase relation among them.
The relation between the $y$-component of the dipole moment is the same for both enantiomers,
while it is of opposite phase in the $z$-direction, therefore the torque experienced by the two enantiomers will be different.
This creates a different coherent non-uniform distribution of excitations in the R- and S-enantiomers.  
The different populations induced by linearly polarized pulse
lead to distinct charge migration patterns, which can be mapped
by the evolution of the charge distributions and by the time-dependent EFDs.

\subsection{Effect of imperfect orientation on electronic flux densities}
Till now, we have discussed the EFDs in an oriented chiral molecule, i.e.,  with 
fixed spatial orientation. In general, most of the gas-phase methods  deal with the 
ensembles of randomly oriented chiral molecules. 
Moreover, large and floppy chiral molecules are not easy to fix in the space. 
In the following, we consider the case where a chiral molecule is not fully fixed in the space
and has some uncertainty in its orientation. 
For this purpose, we will introduce the notion of average EFD,
which is incoherently averaged over different molecular orientations. 
To study the effect of orientation averaging on EFD 
for different orientations of the chiral molecule, the molecule is rotated by different angles. 
The absence of coherences implies an imperfect orientation in an ensemble of molecules,
where each molecule has a specific orientation which might be imperfectly aligned with the lab-frame axes.
In contrast, coherent averaging over a specific rotational quantum number would imply a complete lack of orientation of each individual molecule.

\begin{figure}[]
\begin{center}
\includegraphics[width=\linewidth]{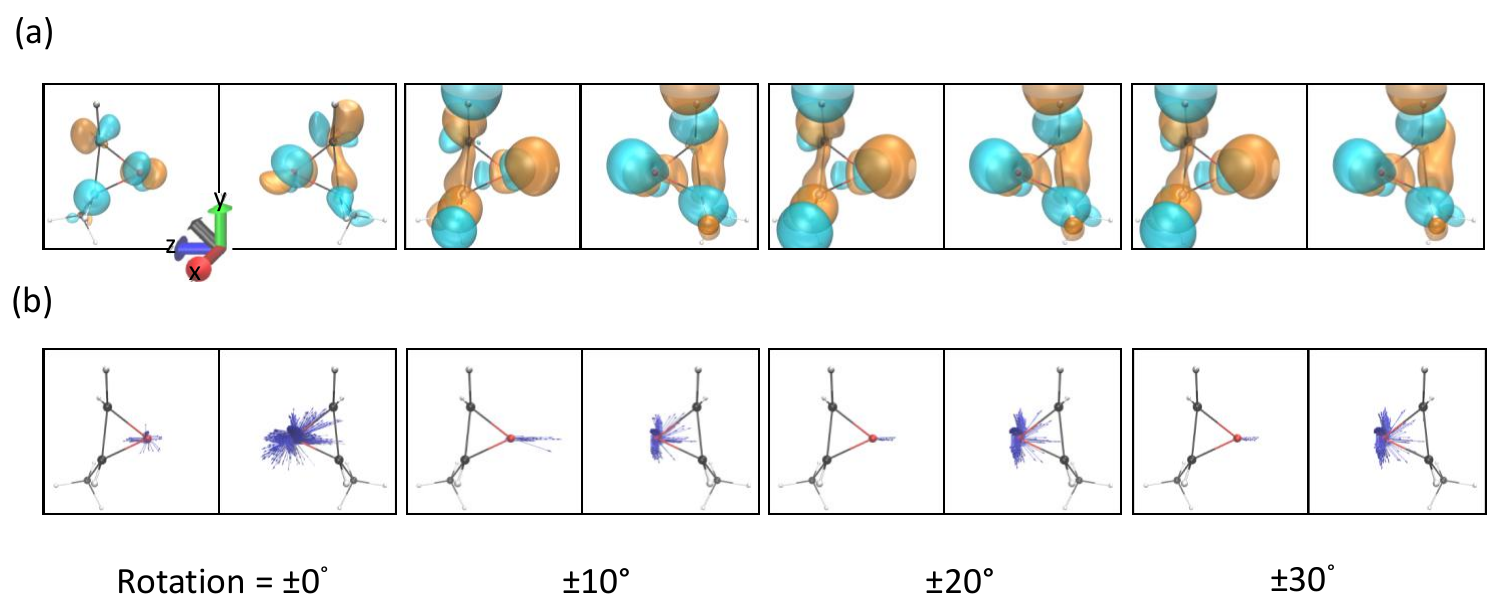}
\caption{(a) Electronic charge distribution difference and (b) corresponding  flux densities (blue arrows) for both  R- and S-enantiomers at different rotation angles with respect to 
$z$ axis, during field-free charge migration, at  {\sffamily{T}}  = 6\,fs for  linearly polarized laser pulse in 
$yz$ plane. Isocontours at +0.0020 (blue) and -0.0020 (orange) for the enantiomers.}
\label{plot_all}
\end{center}
\end{figure} 

To explore the impact of imperfect orientation on the EFDs, 
we calculate the flux densities at a particular time {\sffamily{T}}  = 6\,fs  after the end of the pulse with parameters specified in Fig.~\ref{fig3}. 
For this purpose, we investigate the charge migration driven by linearly polarized
laser in the $yz$ plane. In this scenario, we have flexibility to rotate the molecule in planes containing the normal axis, i.e., $yz$  and $xz$ planes as $z$ axis is our normal axis. 
The EFDs are calculated to interpret the effect of different molecular orientations. 
Fig.~\ref{plot_all}(a) presents the charge distributions  for the molecule rotated by
$\pm 10^\circ$,  $\pm 20^\circ$ and  $\pm 30^\circ$ along $z$ axis  at   
{\sffamily{T}}  = 6\,fs  during field-free charge migration.
As reflected from the figure, the charge distributions are  similar for the different values of rotational averaging.  Note that the populations in the excited enantiomers changes as we rotate the chiral molecule in space.

The average flux densities for larger-angle rotations 
are obtained by taking the average over the flux densities at all smaller-angle rotations.
For example, the average flux densities for $ \pm 30^\circ$ rotations are obtained from the 
flux densities averaged  incoherently over 
 $0^\circ$,  $ \pm 10^\circ$, $ \pm 20^\circ$ and $ \pm 30^\circ$ rotations. 
The average flux densities for both the enantiomers are presented in Fig.~\ref{plot_all}(b). 
It is evident from the figure that the average flux densities for $ \pm 10^\circ$, $ \pm 20^\circ$ and $ \pm 30^\circ$ rotations are 
very similar,
while they differ quantitatively from the flux densities for the space-fixed case, i.e., $0^\circ$ rotation. 
If we consider the flux densities around  individual atoms separately, say around the oxygen atom, there is a marked changes in 
the magnitude of the flux densities.
The presence of a noticeable amount of flux densities around chiral carbon  and upper carbon atoms  are significantly different from the 
flux densities  associated with the space-fixed molecule. 
The physical origin of these changes in the flux densities is the variations in the phases
due to the different orientations. 
Nonetheless, the main findings  remain unchanged:  linearly polarized pulse in the $yz$ plane induces 
charge migration and the associated time-dependent EFDs are found to be different
for enantiomer pairs, unrelated by mirror reflection.

%%%%%%%%%%%%%%%%%%%%%
\subsection{Imaging charge migration by TRXD}
As we have seen that the linearly  polarized pulse  in the $yz$ plane induces charge migration, which is
drastically different for the enantiomers (see Fig.~\ref{fig3}). This makes  them to study in further detail. 
We employ TRXD to image the induced charge migration in this particular case. 
Moreover, to make a connection between the EFD and TRXD signal, EFDs are projected in two different planes as shown in Fig.~\ref{fig01}. 
Thus, let us discuss the projected flux densities in detail before we discuss the results of TRXD.

The EFDs accompanying the charge migration for both enantiomers are presented in Fig.~\ref{fig01}. 
To avoid the fast oscillating contribution stemming from the 
coherent interaction with the ground state in the charge migration, 
only contributions from the excites states are considered in the reconstructed signal.
Snapshots are presented for $2$ fs, $6$ fs, $10$ fs, and $14$ fs, and the 2D representation in the $xz$ plane is obtained
by integrating over the $y$ coordinate.
The arrows of the vector field are color-coded according to their magnitude.
In both enantiomers, almost all the flux density is concentrated around the oxygen atom (in maroon).
It is important to mention that, maybe surprizingly, the chiral carbon atom is not involved in the dynamics,
as revealed by the transient flux densities. 
The most striking difference lies in the magnitude of the fluxes.
The strength of the flux densities for R-enantiomer is weak in comparison  to S-enantiomer during the charge migration. 
At all times, they both exhibit a similar circular pattern around the oxygen atom.
The direction of the circular charge migration changes as the dynamics progresses.

\begin{figure}[]
\includegraphics[width =  \linewidth]{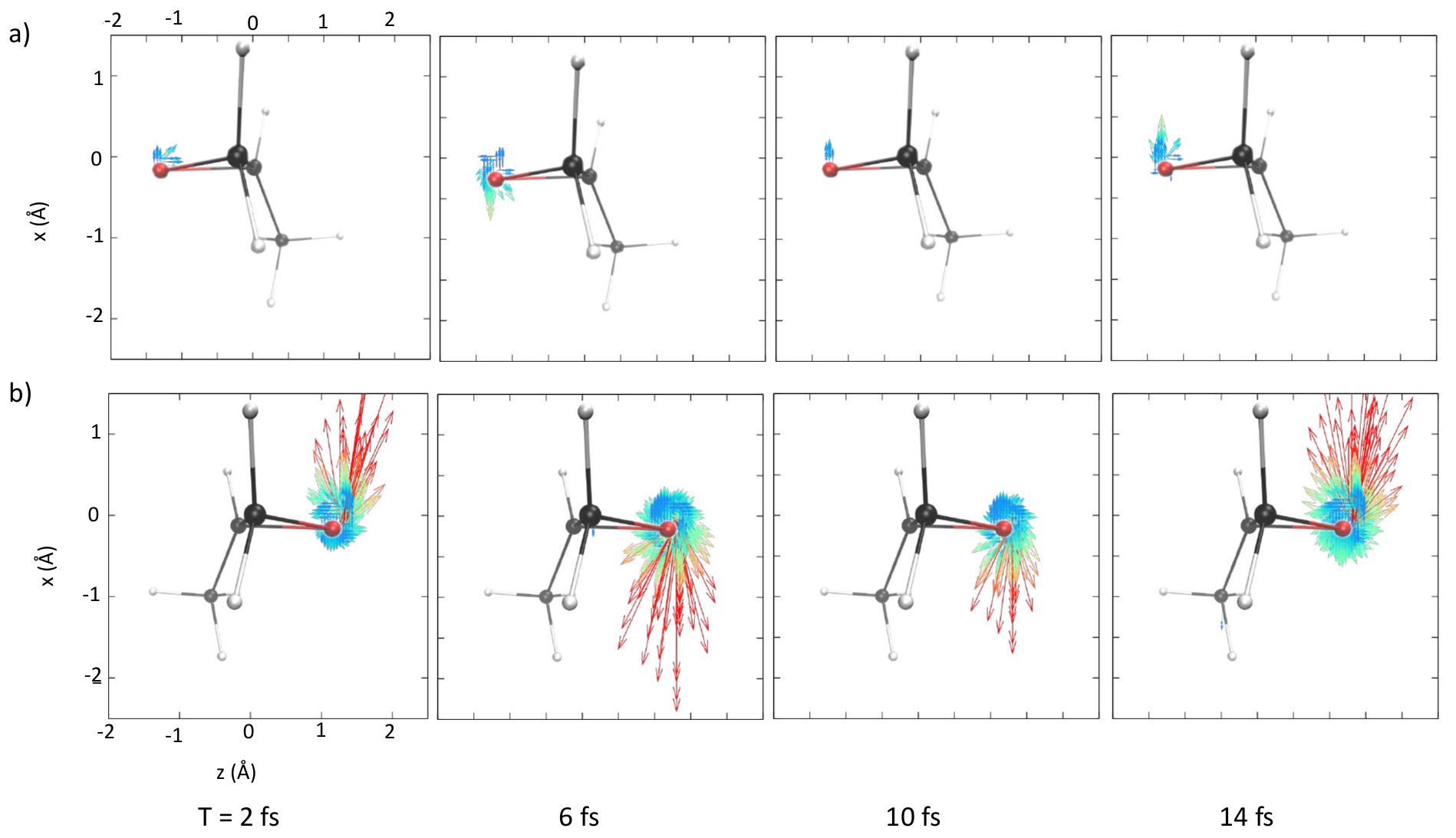}
\caption{Time-dependent electronic flux densities for (a) R- and, (b) S-enantiomers of epoxypropane in the 
$xz$ plane at $2$ fs, $6$ fs, $10$ fs, and $14$ fs
after the pump pulse.
Violet, maroon and gray colors are representing carbon, oxygen and hydrogen atoms, respectively. 
Here, $z$ axis is defined as the chiral axis in the molecular frame of reference.} \label{fig01}
\end{figure}

Since the direction of the electron flow is more prominent for the S-enantiomer, as reflected from Fig.~\ref{fig01}(b),
it can therefore be more readily described.
Arrows corresponding to the flux densities are emerging from the oxygen atom along all directions,
corresponding to an approximately circular flux density.
Yet, the longest arrows in red represent the most probable direction for the flow of electrons,
and these are changing with time.  
At $2$ fs, arrows are mostly emerging along the positive $x$-direction and it  is almost reversed in the next snapshot, i.e., at $6$ fs.
The flux densities for the S-enantiomer are along the same direction at $2$ fs and  $14$ fs, which indicate  
the partial recurrence. 
Similar observations can be made for the circular component of the flux 
density in the R-enantiomer, where the 
direction of the circular flux is opposite to that of the S-enantiomer.
On the contrary, the flux component with the largest magnitude points in the same direction for both enantiomers at all times.
A tentative explanation could be that the circular component stems from the phase transferred to the molecule by the laser pulse. 
This leads to opposite relative phases among the three excited states.
On the other hand, the dominant component of the flux density could be associated with the response of the electron cloud
in the laser pulse during the preparation phase. This electronic response behaviour would then be inherited by the subsequent
charge migration dynamics, explaining that both enantiomers react with the same phase.

\begin{figure}[]
\includegraphics[width=\linewidth]{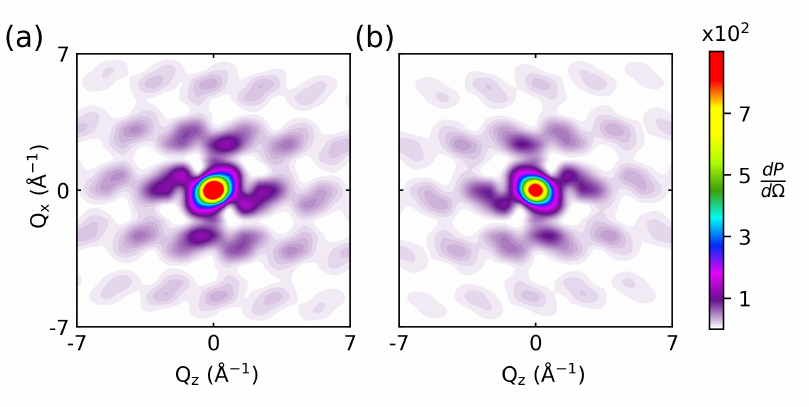}
\caption{Time-independent diffraction patterns for  (a) R- and (b) S- enantiomers of epoxypropane on left and 
right panels, respectively. The  diffraction patterns are presented in  $Q_{z}-Q_{x}$ plane in units of  
${dP_{e}}/{d \Omega}$. An ultrashort x-ray pulse of 8 keV mean energy is aimed to obtain 1.55~$\AA$~spatial resolution in these simulations. 
All diffracted photons up to 60$^\circ$ are collected in the detector.} \label{fig02}
\end{figure}

The different timescale of the charge migration can be understood by analysing 
the characteristic timescales  involved in the system, i.e., $\tau = \textit{h} / \Delta E$ where $\Delta E$ 
is the energy difference between two eigenstates. The characteristic timescales corresponding to 
$E_3$ = 7.56 eV, $E_4$ = 7.73 eV, and $E_5$ = 7.84 eV are $\tau_{43} = 24.32$ fs, $\tau_{53} = 14.77$ fs and $\tau_{54} = 37.59$ fs.
Different contributions of the excited states and their respective interference including these timescales are the reason for non-identical charge migration
in the two excited enantiomers.

After the pump pulse  and in the absence of an external field, the time-dependent coefficients in 
Eq.\,\eqref{eq:TD_wavefunction} simplify to the analytical form containing  amplitude and energy-dependent phase.
At {\sffamily{T}}  = 0, Eq.\,\eqref{eq:DSP_final} reduces to time-independent diffraction signal as 
\begin{equation}\label{eq:DSP_T0}
\frac{dP}{d\Omega} = 
\frac{dP_{e}}{d\Omega} \sum_{f}  \left|  \sum_{k} C_{k} 
\int d\mathbf{r} ~\langle \Phi_{f}  | \hat{\rho}(\mathbf{r}) | \Phi_{k} \rangle~ e^{-i \mathbf{Q} \cdot \mathbf{r}}\right|^{2}. 
\end{equation}
As evident from the above equation,  the time-independent diffraction signal is emerging from the 
joint contribution of the elastic and time-independent inelastic parts  of the total diffraction signal. This term 
serves as a constant background in TRXD.  

\begin{figure}[]
\includegraphics[width= \linewidth]{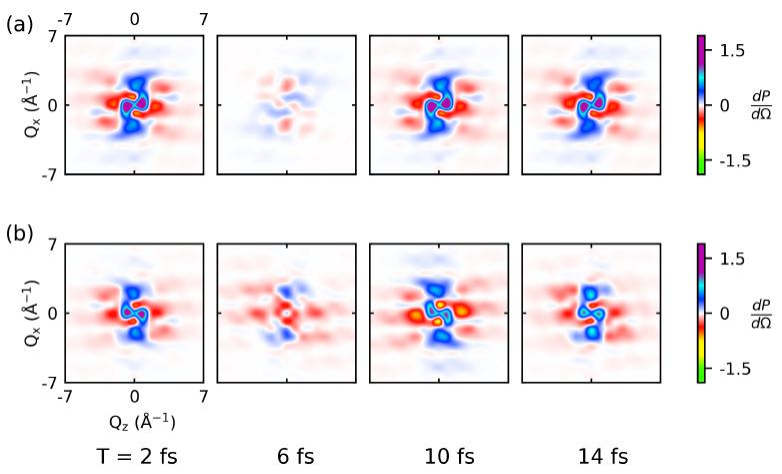}
\caption{Time-resolved diffraction patterns for (a) R- and, 
(b) S-enantiomers of epoxypropane 
in $Q_{z}-Q_{x}$ plane at  2 fs, 6 fs, 10 fs, and 14 fs. 
Intensity of the diffraction patterns is presented in units of  ${dP_{e}}/{d \Omega}$. 
Here, time-dependent diffraction signal at {\sffamily{T}}  = 0 fs is subtracted 
to the subsequent delay time.  An ultrashort x-ray pulse of 8 keV mean energy is aimed to obtain 1.55~$\AA$~spatial resolution in these simulations. 
All diffracted photons up to 60$^\circ$ are collected in the detector.} \label{fig03}
\end{figure}

The time-independent contributions to the total diffraction signal for both the enantiomers after the excitation are presented in Fig.~\ref{fig02}.  These signals are simulated using Eq.~(\ref{eq:DSP_T0}). 
From the figure, one can see that the time-independent diffraction signals 
are different for both enantiomers. First, the intensity of  the 
time-independent  signal  is stronger for the R-enantiomer in comparison to the S-enantiomer.   This is especially true in the low-momentum region.
Also, the signals are not mirror images of each other, which could allow to discriminate between the enantiomers by simple inspection of the two diffraction signals.
Since these diffraction signals 
are static in nature, the dynamical information about the charge migration is lost.

To image the charge migration, we  present  
time-dependent difference diffraction signals  where the total diffraction signal at zero delay time is subtracted 
from the subsequent delay times.  To simulate time-resolved diffraction signal,  the DSP  is computed according to Eq.\eqref{eq:DSP_final}.
The six lowest-lying excited eigenstates (i.e., $f = [0, 6]$) were sufficient to obtain the converged diffraction 
signals within the assumption that the sum-over-states expression truncation due to the detector response.

Time-resolved diffraction signals at selected delay times  for both the enantiomers are presented in Fig.~\ref{fig03}.
The total diffraction signal at {\sffamily{T}}  = 0\,fs is subtracted from the signal at subsequent delay times.
It is observed that the time-independent signal is three orders of magnitude stronger than this time-dependent signal. 
Similar observations have been reported earlier in the case of TRXD for different systems~\citep{carrascosa2021mapping, keefer2021imaging}.  
In spite of having low magnitude, the information about ultrafast dynamics of charge migration 
is imprinted  in the difference  diffraction signals as they are significantly different for 
 both enantiomers at different delay times. 
The central parts as well as the higher momenta signals 
are changing in a very distinct way for both enantiomers. 
Note that $\langle \Phi_{i} \vert \Phi_{j} \rangle$ is not exactly zero in our 
case due to numerical limitations, which impose 
a small but finite value with threshold value of 10$^{-5}$ 
on the LR-TDDFT coefficients. 
This, together with the numerical coefficient re-normalization, explain the small but non-zero value of the diffraction signal at the centre in Fig.~\ref{fig03}.

By comparing the time-dependent diffraction pattern of the enantiomers at 2 fs, it can be seen that the overall signals are not very different for both enantiomers.
There are two lobes of positive signal around the centre and they are of almost same intensity with different orientations for the respective enantiomers.
The negative part of the signal is scattered at higher momenta. There is a mismatch in the intensity at the central part of the lobes for two enantiomers.
 
At the next time step, i.e., at 6 fs, the diffraction signal reduces significantly for the R-enantiomer.
In contrast, the signal is just reversed in the central part, dominated by negative extrema  and spreading to higher momenta for S-enantiomer. 
The signal at 10 fs is comparable with the signal at 6 fs, but the orientation of the signal is different. 
On the contrary, for the R-enantiomer, the signal at 6 fs is similar to the one at 2 fs, and there is a slight decrease  in diffraction intensity at 14 fs. 
 The time-period of the time-dependent signal correlates
well with the timescale of the charge migration found in the flux density for the enantiomers as shown in Fig.~\ref{fig01}.

In order to understand how the time-dependent diffraction signals are related to the flux density, 
time-derivative of electron density in momentum space,  
$\vert \partial_t\rho({\mathbf{Q}})\vert$, at selected delay times are presented  in Fig.~\ref{fig04}. $\vert \partial_t\rho({\mathbf{Q}})\vert$ is zero at {\sffamily{T}}  = 0 fs [see Eq. (S7) in Ref.~\citep{hermann2020probing}]. 
Therefore, subtracting $\vert \partial_t\rho({\mathbf{Q}})\vert$ 
at {\sffamily{T}}  = 0 fs is not required.
$\vert \partial_t\rho({\mathbf{Q}})\vert$ is related to the flux densities, and it 
can be obtained by Fourier transform of the electronic continuity equation [see Eq.\,\eqref{eq:continuity}].

For both the enantiomers, the overall signal of $\vert \partial_t\rho({\mathbf{Q}})\vert$ decreases drastically from 2 fs to 6 fs. 
There is a partial revival of the signal at 10 fs for the R-enantiomer as observed in Figs.~\ref{fig01} and ~\ref{fig03}. 
The intensity of $\vert \partial_t\rho({\mathbf{Q}})\vert$  at 14 fs increases compared to that in 10 fs for R-enantiomer. In contrast, $\vert \partial_t\rho({\mathbf{Q}})\vert$  for S-enantiomer at 14 fs is approximately similar to the one at 2 fs.

As evident from Figs.~\ref{fig03} and \ref{fig04}, it is not straightforward to establish  one-to-one  
visual correspondent between DSP and $\vert \partial_t\rho({\mathbf{Q}}) \vert$, which is in contrast to the
earlier work discussed in Ref.~\citep{hermann2020probing}). 
There are two simple reasons behind this: the first one is that more than two electronic states 
are contributing to the electronic wavepacket in epoxypropane. 
As a result, time-dependent terms yield more than one sine term in the expression of 
$\vert \partial_t\rho({\mathbf{Q}}) \vert$ [see Eq. (S7) in Ref.~\citep{hermann2020probing}]. 
The other reason can be attributed to significant contribution stemming  from sine square term in comparison to the sine  term in the expression of $ \Delta {dP}/{d \Omega}$ [see Eq. (S12) in Ref.~\citep{hermann2020probing}]. These findings are in contrast to earlier case, 
where the wavepacket was composed of only two electronic states,
and the sine square term was negligibly small in comparison to the sin term~\citep{hermann2020probing}. 
Therefore, the comparison between DSP and $\vert \partial_t\rho({\mathbf{Q}})\vert$ can only be qualitative.

\begin{figure}[]
\includegraphics[width= \linewidth]{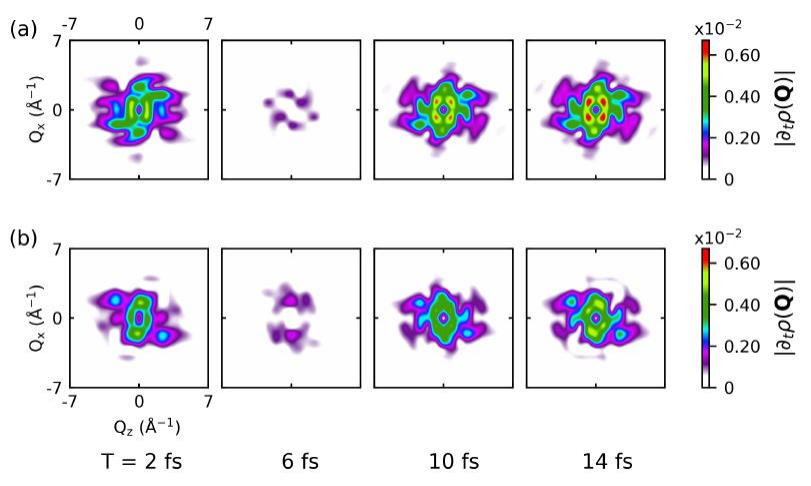}
\caption{Time-derivative of electron density in momentum space, 
$\vert \partial_t\rho({\mathbf{Q}})\vert$, for  (a) R- and, b) S-enantiomers  
of epoxypropane in $Q_{z}-Q_{x}$ plane at 2 fs, 6 fs, 10 fs, and 14 fs.} \label{fig04}
\end{figure}

As the time-resolved diffraction signals for two enantiomers are different, 
we can define the asymmetry parameter  as 
\begin{equation}\label{eq4} 
Y =  \frac{\sigma_R - \sigma_S}{\sigma_R + \sigma_S}, 
\end{equation} 
where $\sigma_R$ and $\sigma_S$ are the DSP  of R- and S-enantiomers, respectively. 
Here, the DSP is represented by $\sigma$ as a compact notation. 
Figures~\ref{fig05}(a)  and ~\ref{fig05}(b) present, respectively,  
the asymmetry parameter  along $Q_{x}$ and $Q_{z}$ axes as a function of time.
The domination of the red part over the blue is an indication that the signal for the S-enantiomer is stronger than the R-enantiomer along $Q_{x}$ axis. 
If we look only around the $Q_{x} = 0$ line, the signal for R-enantiomer is dominating over S-enantiomer for all the time steps. The case is almost reversed for $Q_{x} \neq 0$.  

\begin{figure*}[bht!]
\includegraphics[width=\linewidth]{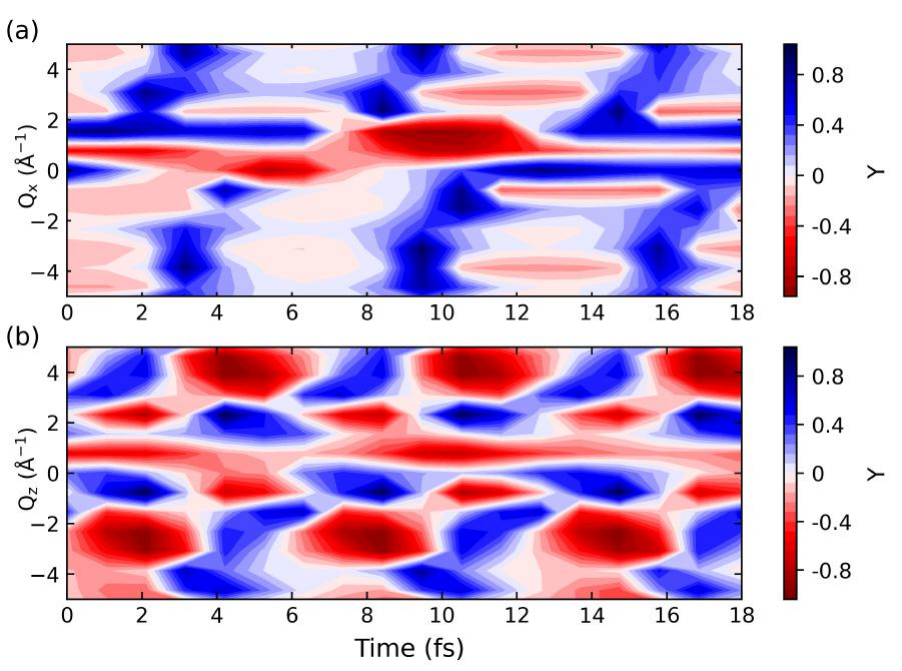}
\caption{Asymmetry parameter $ Y $ along (a) $Q_{x}$ and, (b) $Q_{z}$ axes as a function of  the delay time. The 1D TRXD signal is obtained from Eq.\,\eqref{eq:DSP_final} and integrated along the other axes.} \label{fig05}
\end{figure*}

The signals are more interesting along the chiral axis, i.e., along $Q_{z}$ axis. 
For a specific point in momentum space, the signal is changing in a periodic fashion. 
The same is true if we fix the time delay and walk along the $Q_{z}$ axis. 
These signals encode important information about the dynamical evolution of the system, which
is unique for a given choice of laser tagging parameters.
The asymmetry parameter can be used to determine the enantiomeric ratio in a sample of
unknown concentrations $C_R$ and $C_S$ in the R and S enantiomers, respectively.
In this case, the measured signal of the unknown mixture is given as the weighted sum of the two signals, 
$\sigma_\textrm{sample} = C_R\sigma_R + C_S\sigma_S$.
By subtracting from the TRXD signal of a pure enantiomer, say $\sigma_R$, this difference
\begin{equation}\label{eq5} \begin{aligned}
\Delta\sigma &=  \sigma_R - (C_R\sigma_R + C_S\sigma_S)\\
&= (1-C_R)\sigma_R - C_S\sigma_S= C_S(\sigma_R - \sigma_S)\\
\end{aligned}\end{equation} 
can be simply normalized and related to the asymmetry parameter
\begin{equation}\label{eq6} 
\frac{\Delta\sigma}{\sigma_R + \sigma_S}
=  C_S\left(\frac{\sigma_R - \sigma_S}{\sigma_R + \sigma_S}\right) = C_SY, 
\end{equation} 
Hence, from the knowledge of the TRXD signals of both enantiomers, 
the concentration of an enantiomer can be directly obtained by comparing the intensity of the difference
signal with that of the asymmetry parameter. As the dynamics becomes more intricate at longer times,
the time-dependent monitoring of the difference signal would potentially lead to a more 
precise determination of the concentrations.

\section{Summary}
In summary, we have explored how time-resolved electron densities and associated EFDs
behave when chiral molecules are driven by linearly polarized pulses. 
Moreover, it is shown by means of numerical simulations 
that  the signature of chirality in molecules can be studied by the choice of orientation in linearly polarized laser excitation,
which determines how enantiomers are driven out-of-equilibrium.
Our findings are valid for molecules with fixed spatial orientation as well as for floppy molecules with imperfect degree of orientation. 
It has been shown that two linearly polarized pulses can orient the chiral molecules in a specific direction~\citep{tutunnikov2018selective}. 
Particularly, epoxypropane and other chiral molecules can be oriented  in space for sufficiently  long times (of order of few picoseconds) using twisted polarization~\citep{milner2019controlled, tutunnikov2020observation}.
The enantiomers are oriented in such a manner to retain the mirror symmetry at the end of the second pulse. 
Our work, discussed in this chapter,  utilize such oriented enantiomers while driving them out of equilibrium and  opens new avenues
for imaging time-dependent 
EFD and  associated charge  distributions in enantiomers. 
In recent years, methods based on electric field-dipole interaction have gained importance 
as they  provide strong enantioselective signal~\citep{ordonez2018generalized}. 
Our method for generating EFD belongs to this family of novel field-dipole based methods,
and it is compatible with large, floppy chiral molecules.
Our method is different from  photoexcitation circular dichroism where 
enantiospecific bound electronic wavepacket in a randomly oriented molecules 
is launched by broadband circularly polarized pulse 
and linearly  polarized pulse is used to ionised the bound electronic wavepacket~\citep{beaulieu2018photoexcitation}.
In contrast to that,  in this chapter, linearly  polarized  pulse is used to prepare enantiospecific bound electronic wavepacket in an oriented chiral molecule.  
Moreover, we image the induced charge migration in epoxypropane using TRXD. 
The charge migration in an oriented epoxypropane is triggered by a linearly polarized pulse  at 45$^\circ$ in the 
$yz$ plane in the molecular frame of reference.  
For each enantiomer,  the induced charge migration is different, and  is imaged by TRXD. 
The dominating time-independent  as well as difference TRXD signals 
are analysed separately.
It is found that time-dependent difference signals are significantly different for both enantiomers.
We believe  that the present proof-of-principle results motivate to follow  laser-induced charge migration
in chiral molecules followed by TRXD. 
In particular, this could pave the way for a new, sensitive technique to determine the enantiomeric concentrations
in unknown samples.

\cleardoublepage
\chapter{Conclusions and Future Directions}
\section{Conclusions}
Present thesis is aimed to understand  the impact of various molecular symmetries in different molecules on attosecond charge migration. For this purpose, time-dependent electronic charge distributions and associated 
flux densities are used for spatiotemporal representation of the charge migration. 
Furthermore, TRXD is employed for four-dimensional  imaging of the charge migration in selected cases.

Starting with preparatory works about previously studied charge migration in ``highly symmetric'' benzene with  
$\mathcal{D}_{\textrm{6h}}$ symmetry, 
we have investigated how the charge migration will behave in different five-membered ring-shaped aromatic molecules. 
To this end, emphasis has been placed on those molecules that are neutral as opposed to ionic homocyclic and nonaromatic pentagon molecules.
In \textbf{Chapter3}, we have examined how charge migration in  
five-membered ring-shaped neutral molecules, such as pyrrole, furan, and oxazole, is influenced by symmetry reduction and electronegativity  of the foreign atom(s).
To compare the charge migration in these three molecules on equal footing, 
similar laser parameters are chosen to get the population in the target excited states with nodal structure adjacent to the foreign atom(s).
The direction of the charge migration  and its spatial distribution are understood by means of 
the time-dependent difference electronic charge and flux densities.   
The presence of nitrogen and oxygen causes significantly different charge migration  in pyrrole and furan, respectively; despite having identical  $\mathcal{C}_{\textrm{2v}}$ symmetry. 
The changes in the electron densities in pyrrole are primarily  positioned at atoms, whereas 
the difference electron densities are delocalized around the bonds in furan. 
Analysis of the flux densities unravel  atom-to-bond and bond-to-atom charge migration in pyrrole and furan, respectively.  
$\mathcal{C}_{\textrm{2v}}$ symmetry  is imprinted at different instances during  the attosecond charge migration in both molecules. 
As the symmetry is reduced from $\mathcal{C}_{\textrm{2v}}$ to $\mathcal{C}_{\textrm{s}}$ for oxazole, 
the time-dependent charge and flux densities display no reflection symmetry during charge migration. 
The difference electron densities are neither located at atoms nor at bonds in oxazole, which 
results a swirling motion originating from the lower left carbon to the lower right carbon through the space between the nitrogen and oxygen atom. 
Thus, it  is concluded that the charge migration is sensitive to the point group symmetry via the presence of foreign atom(s), such as nitrogen in pyrrole, oxygen in furan; and nitrogen and oxygen in oxazole. 
 
Following the charge migration in planar molecules, 
 \textbf{Chapter 4} is dedicated to understand the role of the structural  saddling on the charge migration  
 in copper corrole in  which  two nitrogen atoms are out-of-plane and exhibits $\mathcal{C}_{\textrm{2}}$  symmetry. 
To explore how the out-of-plane nitrogen atoms affect the charge migration, 
a linearly polarized pump pulse is used to initiate coherent electronic motion. 
Time-dependent charge and flux densities are used to understand the simulated results. 
Analysis of the electronic flux density reveals that the diagonal nitrogen atoms mediate coherent charge migration between them via a central copper atom.
In this chapter, TRXD is employed to image the charge migration. 
It is found that the signature of the saddling in copper corrole during charge migration  is 
imprinted   in time-resolved diffraction signals. 
To further assert the role of the saddling, 
a comparison of the static diffraction signals of nonsaddled planar copper porphyrin and saddled nonplanar copper corrole in their ground states is made.

After analyzing roles of different symmetries in molecules on the charge migration, 
\textbf{Chapter 5} is focused on  the charge migration in molecules without any symmetry. 
For this purpose, epoxypropane -- a chiral molecule with $\mathcal{C}_{\textrm{1}}$  symmetry, i.e., absence of molecular symmetry, is chosen.    
In general,  circularly polarized light is used to probe the chirality in molecules. 
In this chapter, different orientations of the linearly polarized pulses are used to induce charge migration in 
space-fixed R- and S-enantiomers of epoxypropane. 
It has been demonstrated  that the selective orientation of R- and S-enantiomers of epoxypropane in lab frame is possible 
with the help of two linearly polarized pulses~\citep{tutunnikov2018selective}.
After the orientation in lab frame, enantiomers exhibit  mirror symmetry. 
However, it is not {\it a priori} obvious whether  laser-induced charge migration in enantiomers will also display   mirror symmetry or not.  
It has been found that the electronic population dynamics and the associated flux densities are identical when 
the enantiomers are driven by a linearly polarized pulse  along  $x$ axis. 
These findings remain same for other linearly polarized pulses  along  $y$ and $z$ axes. 
On the other hand, if the enantiomers are driven by a linearly polarized pulse in  the $yz$ plane, the 
population dynamics and associated flux densities are drastically different for both  enantiomers. 
Our observations remain valid for  R- and S-enantiomers of epoxypropane with a fixed spatial orientation as well as for certain degree of imperfect orientation in lab frame. 

Time-resolved x-ray diffraction is used for four-dimensional imaging of the charge migration in oriented epoxypropane, induced by a linearly polarized pulse at  45$^\circ$ in the  $yz$ plane. 
Detailed analysis of the  x-ray diffraction signals reveals that the time-independent  diffraction signal is two orders in magnitude higher than the  time-dependent difference diffraction signals. 
Moreover, the time-independent diffraction signals for the two enantiomers differ remarkably from one another at higher momenta. 
Furthermore, a connection between TRXD and the electronic continuity equation is discussed by analysing the time-dependent diffraction signal and the time derivative of the total electron density in the momentum space. 
In this chapter, we have proposed a way to determine the concentration of an enantiomer in the racemic mixture by analyzing the asymmetry parameter stemming from the TRXD signal of a pure enantiomer. Our proposal 
could open an avenue to ascertain the enantiomeric contents in unidentified chiral molecules.  

\section{Future Directions} 
Attosecond charge migration is one of the important  topics of attosecond physics in present days. 
We believe that our findings on imaging charge migration in different molecules will catalyze further 
theoretical and experimental works, 
particularly on laser-induced ultrafast processes in various metal corroles and other symmetry-reduced molecular systems.  
%Numerous ultrafast chemical reactions need to be understood at a fundamental level on the theoretical side. Our research has the potential to contribute to that.

Present thesis focused on inducing and probing charge migration in chiral molecule with one chiral (stereo) 
center. 
It is known that there exists $2^{n}$ number of stereoisomers 
depending on the number of stereo centers, $n$, in a molecule. 
For example, there are two stereo centers in tartaric acid that generates four stereoisomers. 
The pairs of them which are connected by mirror reflection are called enantiomers  
and the pairs which are not connected by mirror reflection are known as diastereomers.
There are not much research  activities on  light-induced charge migration in diastereomers till date. 
It will be very interesting to study the charge migration in the stereoisomers where they are connected with (enantiomers) or without (diastereomers) mirror reflection.
Moreover,  the spatiotemporal imaging of the charge migration to understand the role of the multiple stereo centers can also be investigated using EFDs and TRXD. 

The origin of the  homochirality is an open question as 
it is still unclear why asymmetry is naturally preferred, 
resulting in the survival of only one type of enantiomer in living objects in earth.
Since light is the most frequent perturbation in nature, it might be conceivable to use either enantiomer transformation or enantiomer dissociation to understand why one enantiomer is preferred over another.
The transformation could  involve bond breaking and  formation,
which can be investigated using light-driven nuclear dynamics in enantiomer. This could potential contribute towards our understanding of homochirality.

Not only different molecules are interesting for various light-induced phenomena but also different properties of light can be harnessed to learn about various aspects of the charge migration in molecules.  
It is known that  light carry an extra degree of freedom: orbital angular momentum (OAM) -- associated with its spatial profile, apart from known spin angular momentum associated with light's polarization. 
The OAM carrying beam is also known as helical or twisted beam.
Investigating the interaction of twisted light with molecule is another  topic for future study.
The OAM of light has been extensively studied in the context of optical interferometry, quantum information processing,
 and nanoparticle manipulation, to name but a few.
It would be interesting  to explore how the helical laser pulse interacts with chiral molecules -- leading to an emerging concept of  time-resolved  helical dichroism. 
Moreover, how helical ultrashort x-ray pulses could be useful for TRXD will be  another open direction of research. 

Attosecond charge migration is actively studied in various molecular systems and  sheds 
light on various laser-driven processes in them. It would be interesting to extend such studies to  solid-state systems, which is an unexplored territory.   
Additionally, there is a plenty of room for theory to motivate and support the design of upcoming state-of-the-art experiments with suitable pump-probe methods  to probe charge migration. 
It would be interesting to apply attosecond transient-absorption spectroscopy, and time- and  angle-resolved photoionization spectroscopy within pump-probe configuration to image the EFDs associated with  the laser-induced charge migration in near future.   

%As technology develops, the generation of laser pulses and data processing enhance the field of charge migration and make it possible to image and perhaps influence dynamics at an ever-shorter timescale.

\cleardoublepage
%\include{Chapter7}
%\cleardoublepage
\end{spacing}
%%%%%% ADD SUPPLEMENtary results

\begin{spacing}{1.3}	
\addcontentsline{toc}{chapter}{Bibliography}
\bibliographystyle{abbrvnat.bst}
\bibliography{SG_thesis_final}

\cleardoublepage \addcontentsline{toc}{chapter}{List of Publications}
\include{Publication}\cleardoublepage
\cleardoublepage
\end{spacing}

\cleardoublepage \cleardoublepage
\addcontentsline{toc}{chapter}{List of Publications}
\markboth{List of Publications}{List of Publications}
\newpage
\thispagestyle{empty}
\begin{center}
\vspace*{-0.4cm} {\LARGE {\textbf{List of Publications}}}
\end{center}
{\setlength{\baselineskip}{8pt} \setlength{\parskip}{2pt}
\begin{spacing}{1.5}
\vspace*{.7cm} \noindent{\bf \large A. Part of this thesis} \vspace*{.7cm}
\begin{enumerate}
	\item {\bf Sucharita Giri}, Alexandra Maxi Dudzinski, Jean Christophe Tremblay, and Gopal Dixit: ``Time-dependent electronic current densities in chiral molecules'' \textbf{Physical Review A 102}, 063103 (2020).
	\item {\bf Sucharita Giri}, Jean Christophe Tremblay, and Gopal Dixit: ``Imaging charge migration in chiral molecules using time-resolved
x-ray diffraction'' \textbf{Physical Review A 104}, 053115 (2021). 
	\item {\bf Sucharita Giri}, Jean Christophe Tremblay, and Gopal Dixit: ``Probing the effect of molecular structure saddling on ultrafast
charge migration via time-resolved x-ray diffraction'' \textbf{Physical Review A 106}, 033120 (2022). 
     \item {\bf Sucharita Giri}, Gopal Dixit, and Jean Christophe Tremblay: ``Charge migration in heterocyclic five-membered rings'' (accepted in The European Physical Journal Special Topics).
\end{enumerate}

\vspace*{.7cm} \noindent{\bf \large B. Not part of this thesis} \vspace*{.7cm}
\begin{enumerate}
	\item {\bf Sucharita Giri}, Misha Ivanov, and Gopal Dixit: ``Signatures of the orbital angular momentum of an infrared light beam in the two-photon transition matrix element: A step toward attosecond chronoscopy of photoionization'' \textbf{Physical Review A 101}, 033412 (2020). 
	\item Nilanjan Roy, \textbf{Sucharita Giri}, and Partha Pratim Jana: ``Site preference and atomic ordering in the ternary Rh5Ga2As: first-principles calculations'' \textbf{Zeitschrift fuer Kristallographie-Crystalline Materials 236}, 147-54 (2021).
	\item  Jean Christophe Tremblay, Ambre Blanc, Pascal Krause, \textbf{Sucharita Giri}, and Gopal Dixit: 
	``Probing electronic symmetry reduction during charge migration via time-resolved x-ray diffraction'' 
	\textbf{Chem. Phys. Chem. e202200463} (2022).
\end{enumerate}

\newpage
\thispagestyle{empty}
\begin{center}
	\vspace*{-0.4cm} {\LARGE {\textbf{Conferences Attended}}}
\end{center}

\vspace*{.7cm} \noindent{\bf \large A. International}
\vspace*{.7cm}

\begin{enumerate}
	\item \textbf{ATTOCHEM2020}, Online (9 - 11 September 2020).
	\item \textbf{UXSS2021}, Online (14 - 18 June 2021).
	\item \textbf{UXSS2022}, Paris, France (10 - 14 October 2022).
	\item \textbf{Theory Meets XFELs Workshop}, Hamburg, Germany (2 - 4 November 2022).
\end{enumerate}

\vspace*{.7cm} \noindent{\bf \large B. National}
\vspace*{.7cm}

\begin{enumerate}
	\item \textbf{UFS 2017}, University of Hyderabad (02 - 04 November, 2017).
	\item \textbf{UFS 2019}, IIT Bombay (07 - 09 November, 2019).
	\item \textbf{8$^\textrm{th}$ TC2020}, IIT Roorkee (03 - 05 March, 2020).
\end{enumerate}

\end{spacing}

\cleardoublepage
\end{document}